\documentclass[acmsmall,screen,nonacm]{acmart}

\usepackage[]{hyperref}
\usepackage{xurl}
\usepackage{amsmath,amsfonts,amsthm}
\usepackage{algorithmic}
\usepackage{graphicx}
\usepackage{textcomp}
\usepackage{color}
\usepackage{wrapfig}
\usepackage{math-cmds}
\usepackage{bm}
\usepackage{xspace}
\usepackage{enumitem}
\usepackage{makecell}
\usepackage{multirow}
\usepackage{subcaption}
\usepackage{bussproofs} %
\usepackage{proof} %
\usepackage[ruled,vlined,linesnumbered]{algorithm2e}
\usepackage{listings}
\usepackage{lstlinebgrd}

\lstdefinestyle{pl_overview}{
  linebackgroundcolor={
    \ifnum\value{lstnumber}=2 \color{green!25}\fi
    \ifnum\value{lstnumber}=9 \color{green!25}\fi
    \ifnum\value{lstnumber}=10 \color{green!25}\fi
  }
}

\def\Green{\leavevmode\rlap{\hbox to \hsize{\color{green!50}\leaders\hrule height .8\baselineskip depth .5ex\hfill}}}

\newcommand{\sys}{CrypTorch\xspace}

\newcounter{obs_ctn}
\stepcounter{obs_ctn}
\newcommand{\obs}[1]{\noindent\fbox{\begin{minipage}{0.98\textwidth}
\emph{\textbf{Observation \arabic{obs_ctn}:}} #1
\end{minipage}}\stepcounter{obs_ctn}}

\newcommand\premSep{\hspace*{0.5em}}

\newcommand{\gInfer}[3]{\infer[\textsc{(#1)}]{#3}{#2}}

\newcommand{\type}[1]{\langle#1\rangle}
\newcommand{\round}[1]{\lfloor#1\rceil}

\DeclareMathOperator*{\encode}{\sigma}
\DeclareMathOperator*{\encrypt}{enc}
\DeclareMathOperator*{\secret}{\textsc{S}}
\DeclareMathOperator*{\pub}{\textsc{P}}
\DeclareMathOperator*{\add}{\oplus}
\DeclareMathOperator*{\mul}{\otimes}
\DeclareMathOperator*{\ltz}{ltz}

\newcommand{\btype}{bt}
\newcommand{\itype}{it}
\newcommand{\ftype}{ft}
\newcommand{\bool}{\textsc{Bool}}
\newcommand{\integer}{\textsc{I}}
\newcommand{\float}{\textsc{F}}

\newcommand{\defaultscale}{s_d}
\newcommand{\defaultint}{\integer_d}

\newcommand{\tightcolorbox}[2]{%
  {\setlength{\fboxsep}{1pt}\colorbox{#1}{#2}}%
}

\begin{document}
\title[\sys: PyTorch-based Auto-tuning Compiler for Machine Learning with Multi-party Computation]{\sys: Unified PyTorch-based Auto-tuning Compiler for Private Machine Learning Inference}

\author{Jinyu Liu}
\orcid{0009-0006-0069-4542}
\affiliation{%
   \institution{The Pennsylvania State University}
   \city{University Park}
   \state{PA}
   \country{USA}
}
\email{liu.jinyu@psu.edu}

\author{Gang Tan}
\orcid{0000-0001-6109-6091}
\affiliation{%
   \institution{The Pennsylvania State University}
   \city{University Park}
   \state{PA}
   \country{USA}
}
\email{gxt29@psu.edu}

\author{Kiwan Maeng}
\orcid{0000-0002-0321-8406}
\affiliation{%
   \institution{The Pennsylvania State University}
   \city{University Park}
   \state{PA}
   \country{USA}
}
\email{kvm6242@psu.edu}

\begin{abstract}

\emph{MPC-based ML} uses secure multi-party computation (MPC) to execute machine learning (ML) workloads across multiple parties without requiring any party to reveal its private data or model parameters.
Our characterization study finds that the performance of state-of-the-art MPC frameworks is often dominated by operator-approximation strategies, where ML operators not natively supported by MPC are replaced with iterative methods or polynomials.
These approximations are typically chosen heuristically by each framework and applied globally. As a result, they can be unnecessarily expensive in some cases, yet insufficiently accurate in others.
We find that the best approximation choice depends on the context, target framework, and application, making a single globally fixed approximation fundamentally suboptimal.
To address this problem, we design \sys, a multi-stage compiler for MPC-based ML that supports multiple MPC frameworks as backends.
At the core of \sys is its auto-tuning stage, which extracts cost estimates from the target MPC backend and quality estimates from a proxy PyTorch execution, and uses them to automatically select a set of approximations that maximizes performance while satisfying the given quality constraint.
We show that \sys's auto-tuner improves performance across multiple MPC backends by 1.14--1.60$\times$ without degrading accuracy, and by 1.17--1.92$\times$ with less than 3\% accuracy degradation.
\sys's unified design also improves programmability and usability by allowing multiple backends to share other common compiler functionality and optimizations. We further show that the auto-tuner extends to a related fully homomorphic encryption (FHE) backend, reducing the number of bootstrapping operations by 1.31--1.65$\times$.

\end{abstract}

\maketitle

\section{Introduction}

Machine learning (ML) models increasingly process sensitive data, such as X-ray images~\cite{xray}, microphone recordings~\cite{alexa, googlehome, fbportal}, and private user queries~\cite{yu2024privacy}.
At the same time, model owners are often reluctant to reveal their proprietary models.
This creates a fundamental dilemma over where to execute ML workloads: running the computation at either party requires the other party to reveal its private input or model weights.
\emph{MPC-based ML} addresses this problem by executing ML workloads across multiple parties using secure multi-party computation (MPC), protecting private data and models.
Numerous MPC-based ML frameworks have been developed in both academia~\cite{minionn, delphi, gazelle, ariann, bolt, falcon, securenn, charmeleon, astra, blaze, flash, trident, piranha, cheetah} and industry~\cite{crypten, cryptgpu, cryptflow, cryptflow2, spu, orca, sigma}.

MPC protocols cannot directly support several operations required by modern ML models.
Existing frameworks rely on \emph{operator approximations} that transform unsupported operations (\emph{e.g.}, $e^x$, inverse, or square root) into sequences of additions, multiplications, truncations, and comparisons.
Our characterization study (Section~\ref{sec:charac_approximation}) finds that most frameworks rely on heuristically selected approximations that are applied globally, which can either \emph{hurt performance} or \emph{degrade model accuracy}.
Moreover, these approximations are deeply embedded in their implementation, resulting in fragmented and duplicated efforts.
If a framework lacks a particular approximation, developers must reimplement it within that framework, often requiring non-negligible effort.

We introduce \sys, a multi-stage compiler that addresses these inefficiencies.
\sys decomposes common transformations in existing MPC frameworks into reusable compiler stages and \emph{shares a common frontend and middle-end across multiple backends}, reducing fragmentation and duplicated implementation effort.
At the heart of \sys is its operator-approximation stage, which determines which approximation to use for each unsupported operation in the computation graph.
\sys employs an \emph{auto-tuner} that estimates both the accuracy impact of each approximation and its cost on the target MPC runtime, then selects the configuration that maximizes performance while satisfying a user-specified quality constraint.
Our evaluation shows that the auto-tuner improves performance by 1.14--1.60$\times$ without degrading accuracy, and by 1.17--1.92$\times$ with less than 3\% accuracy degradation, across multiple MPC backends.
We further show that \sys's frontend and middle-end can also benefit fully homomorphic encryption (FHE) backends, another widely used approach to private inference that faces similar approximation challenges.
We summarize our contributions:

\begin{enumerate}[noitemsep, leftmargin=*, topsep=5pt]
    \item We study existing MPC-based ML frameworks~\cite{crypten, pigeon} and show that much of the performance overhead comes from operator approximations embedded in each framework.

    \item We propose \sys, a multi-stage compiler for MPC-based ML. At its core is an approximation auto-tuning stage that automatically selects the approximation for each operator to maximize performance while satisfying a user-specified quality constraint. This auto-tuning stage, together with the other front- and middle-end compiler stages, can be shared across and benefit multiple MPC backends.

    \item Our evaluation shows that the auto-tuner improves performance across multiple MPC backends by 1.14--1.60$\times$ without degrading accuracy, and by 1.17--1.92$\times$ with less than 3\% accuracy degradation. We also show that the same compiler front- and middle-end can extend to FHE backends and reduce expensive bootstrapping by 1.31--1.65$\times$, demonstrating the generality of our design to other non-MPC paradigms.

    \item We additionally demonstrate the practical programmability and usability benefits of \sys. By sharing a common front- and middle-end, \sys avoids duplicated implementation effort across MPC frameworks. We show that connecting existing frameworks as \sys backends can immediately extend them to models they previously did not support and allow them to benefit from shared optimizations.

    \item \sys is implemented as an extension to the PyTorch 2 compiler~\cite{pytorch2}, making it easy to extend using PyTorch 2's existing pass-writing interface~\cite{export_ir_pass_tutorial}.
    
\end{enumerate}

\section{Background and Motivation}

\subsection{MPC for Machine Learning Inference}
\label{sec:bg_mpc}

Multi-party computation (MPC) allows multiple parties to collaboratively run computations without revealing their secrets to each other. 
Using MPC for ML inference has been an active area of research recently, with numerous frameworks being developed from both academia and industry~\cite{falcon, hummingbird, aby3, securenn, charmeleon, astra, blaze, flash, trident, piranha, crypten, cryptgpu, cryptflow, spu, orca, sigma, ditto, cryptdough, pigeon, spu, smash}.
MPC protocols require participants to communicate frequently over the network~\cite{crypten, pigeon}, which makes them multiple-orders-of-magnitude slower compared to plaintext inference~\cite{pengzhi_ispass, jinyu_mpc_cal}. Their security guarantee relies on non-colluding assumptions; if $k$ out of $N$ participants (or more) collude, the confidentiality guarantee breaks (\emph{e.g.}, 2-out-of-2 for CrypTen~\cite{crypten}, 2-out-of-3 for Trio~\cite{hpmpc}).

Several MPC protocols exist, but we primarily focus on those based on \emph{arithmetic secret sharing}.
Arithmetic-sharing-based MPC protocols~\cite{crypten, falcon, pigeon, aby3} represent a secret using shares over an arithmetic ring or field and distribute these shares among the participating parties. The secret can be reconstructed only by appropriately combining the shares arithmetically, and no individual party can recover it from its own share alone.
Arithmetic sharing is particularly common in ML inference because it efficiently supports addition and multiplication, which dominate many ML workloads.
Accordingly, many MPC frameworks for ML primarily use arithmetic sharing and switch to other sharing schemes only for specific operations such as comparison (\emph{e.g.}, binary sharing~\cite{crypten, aby3}, Yao's sharing~\cite{aby3}, or function sharing~\cite{ariann, orca, sigma}).
Our work targets MPC frameworks that are primarily based on arithmetic sharing. Throughout this paper, we use the term MPC to refer to protocols that mainly rely on arithmetic sharing for addition and multiplication, while occasionally using other sharing schemes for selected operations.

\subsubsection{Limitations of MPC}
\label{sec:bg_mpc_limitations}
We do not discuss how popular MPC protocols work in detail, as this is not important for understanding the contribution of this paper. Instead, we highlight some of their key characteristics.
First, they can only support certain simple operations---addition, subtraction, multiplication, comparison, and truncation (integer division) by a known constant---which we call the \emph{supported set}.
These are sufficient to run certain ML operators like Conv2d, Linear, MaxPool, and ReLU, but are insufficient to run more complex operators, such as Softmax, GELU, LayerNorm, or SiLU, which require $e^x$, square root, reciprocal, and \emph{etc.} Thus, existing frameworks introduce several \textbf{\emph{approximations}} to the models~\cite{crypten, spu, bolt, sirnn, mpcformer}, using Newton-Raphson approximation, piecewise polynomials, and \emph{etc.}, to approximate these unsupported operators.
As we show in Section~\ref{sec:charac_approximation}, approximated operators dominate the cost in complex ML models.

Second, MPC operates over a \textbf{\emph{fixed-size integer ring}}. Since most ML models use floating-point, existing frameworks represent each floating-point value $x_f$ as a scaled integer ($x=\round{x_f\cdot s}$).
This is essentially fixed-point arithmetic, and therefore follows many of the same rules as conventional fixed-point systems~\cite{fixed_point_book}. For example, multiplying two scaled integers (\emph{e.g.}, $x=\round{x_f\cdot s}$ and $y=\round{y_f\cdot s}$) produces a value with an effective scaling factor of $s^2$ ($xy \approx \round{x_fy_f\cdot s^2}$), so a corresponding \emph{rescaling} operation (truncation by $s$) must follow.
Similarly, two values must have matching implicit scaling factors before they can be added.
However, MPC imposes stricter constraints. For example, changing bitwidths (accumulating at a larger bitwidth), using saturating addition, or temporarily converting values to floating point are all common in plaintext fixed-point systems, but are expensive in MPC and must be avoided~\cite{ditto}.
MPC frameworks implement their own rescaling rules, either explicitly or implicitly, and these sometimes contain subtle bugs, as we discuss in Section~\ref{sec:to_int}.

\subsection{FHE for Machine Learning Inference}
\label{sec:bg_fhe}

While \sys primarily targets MPC, parts of its compiler front- and middle-end can also benefit fully homomorphic encryption (FHE), another widely used approach for private ML inference~\cite{heax, f1, cheddar, orion, bts, cerium, thor, nexus, craterlake, ark, cheetah_fhe, 100x} (Section~\ref{sec:eval_fhe}). FHE allows certain computations to be performed directly on ciphertexts, even when the party performing the computation does not possess the secret key. This enables users to send encrypted data to an untrusted service provider and run ML inference without revealing the underlying inputs.
Like MPC, FHE-based ML inference is typically multiple orders of magnitude slower than plaintext inference~\cite{cerium, 100x, cheddar, pengzhi_ispass}, although the sources of overhead differ.
In FHE, multiplications consume a ciphertext's limited \emph{levels}; once too few levels remain, an expensive \emph{bootstrapping} operation is required to refresh the ciphertext, which often dominates the overall FHE overhead~\cite{orion}.
To balance computation and communication costs, several hybrid (F)HE--MPC systems have also been proposed~\cite{bolt, cheetah, gazelle, cryptflow2, cipherflow, delphi}.

\subsubsection{Limitations of FHE}
CKKS~\cite{ckks} is a popular FHE scheme that suits ML inference well due to its relatively fast arithmetic support.
We again do not discuss how CKKS works in detail, but briefly discuss the important limitations that make it relevant to \sys.
As in MPC, CKKS FHE can only support simple operations. It supports only addition, subtraction, and multiplication, and not even comparison or truncation.
As a result, they again \textbf{\emph{approximate}} functions they do not support but need for ML inference~\cite{cerium, nexus, cinnamon}.
As FHE does not even support comparison, operations like ReLU, which MPC can natively support, must be approximated with a high-degree polynomial~\cite{orion}.
MPC's approximations are expensive because the multiplication, truncation, and comparison used inside the approximations are by themselves expensive, whereas FHE's approximations are expensive because it requires more frequent bootstrapping, as evaluating the high-degree polynomials consumes many multiplicative levels.

Like MPC, FHE can also only operate on integers, and the result must be rescaled after every multiplication. However, FHE's rescaling is much more complex than MPC's, because it consumes a modulus level that must be replenished by expensive bootstrapping~\cite{orion}. Thus, placing rescaling and bootstrapping operations (along with other bookkeeping operations, such as relinearization and modulus switching) has been an important topic for FHE compilers~\cite{orion, hecate, elasm, dacapo, halo}.

\subsection{Threat Model}
This paper follows the standard threat models of the MPC-based ML frameworks on which it directly builds~\cite{crypten, pigeon}.
We protect only data values, while tensor shapes, model architectures, and computation patterns are considered public.
For MPC backends, we assume semi-honest parties, \emph{i.e.}, parties follow the prescribed protocol faithfully. 
Protecting against malicious parties that do not follow the protocol faithfully is out of scope.
We also assume that parties do not collude with each other. These are common assumptions in MPC literature.
We reuse low-level MPC protocols and kernel implementations from prior work and assume that these implementations are secure and free of correctness bugs.
The same assumption applies to our FHE experiments (Section~\ref{sec:eval_fhe}).
Security vulnerabilities arising from implementation bugs or flaws in the underlying cryptographic protocols are outside the scope of this work.

\section{Characterizing the Overhead of Approximation in MPC-based ML}
\label{sec:motivation}
\label{sec:charac_approximation}

Several MPC frameworks for ML have been developed and maintained independently.
As discussed in Section~\ref{sec:bg_mpc}, each framework implements its own approximations for operators that are not natively supported by MPC.
In most frameworks, these \emph{approximations are fixed and applied globally}, even when a cheaper approximation would suffice or a more accurate one is necessary.
Moreover, the implementation effort is fragmented and duplicated across frameworks.

\begin{figure}[t]
    \centering
    \includegraphics[width=\textwidth]{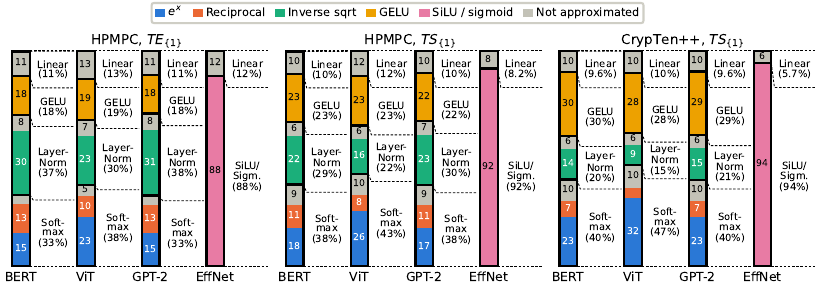}
    \caption{
    Latency breakdown of four models under three different MPC runtimes. Thick boxes are operators; colors split each into its major components.
    }
    \label{fig:breakdown_hpmpc}
\end{figure}

\subsection{Latency Breakdown}
\label{sec:charac_latency_breakdown}
Figure~\ref{fig:breakdown_hpmpc} shows the latency breakdown for four models (BERT-base, ViT, GPT-2, and EfficientNet) across three MPC configurations: (1) HPMPC~\cite{hpmpc} library using its Trio MPC protocol, configured to use exact truncation ($\mathit{TE}_{\{1\}}$); (2) the same HPMPC/Trio configuration, but using stochastic truncation ($\mathit{TS}_{\{1\}}$) instead of exact truncation (which has performance and potential security implications; see Section~\ref{sec:eval_setup_mpc}); and (3) CrypTen++~\cite{jinyu_mpc_cal}, an optimized variant of the CrypTen~\cite{crypten} library and protocol.
Details of the workloads, system configurations, and MPC frameworks are provided in Sections~\ref{sec:backend} and~\ref{sec:eval_setup_mpc}. All results assume a WAN connection between parties.

The gray portion shows operators that do not require approximation (\emph{e.g.}, Linear, Conv2d, or ReLU), while the remaining colors correspond to approximated operators (\emph{e.g.}, $e^x$, reciprocal, inverse square root, GELU, SiLU, and sigmoid).
The first observation is that \emph{most of the end-to-end latency comes from iterative or polynomial-based approximations}: 77--88\% for HPMPC with $\mathit{TE}_{\{1\}}$, 88--92\% for HPMPC with $\mathit{TS}_{\{1\}}$, and 90--94\% for CrypTen++.
This differs from earlier observations in the MPC literature, where linear layers~\cite{mpc_transformer_workshop, mpcformer, mpcpipe, mpc_transformer}, max operations~\cite{mpc_transformer, mpcformer}, or ReLU~\cite{hummingbird, snl, deepreduce} were often identified as the dominant costs.
The bottleneck has shifted because recent efforts have substantially reduced the cost of comparisons~\cite{hummingbird, aby3, jinyu_mpc_cal} and linear layers~\cite{piranha, pigeon, jinyu_mpc_cal} and increasingly target more complex models.
Even prior work that identified operators such as Softmax as expensive attributed most of the cost to the max operation during the approximation~\cite{mpcformer, mpc_transformer}. With recent optimizations that improved max~\cite{jinyu_mpc_cal, hummingbird, aby3}, we instead find that the remaining overhead is dominated by the iterative $e^x$ approximation (blue portion), rather than max (gray).

\obs{
Across MPC runtimes, iterative and polynomial-based approximations dominate end-to-end latency. This differs from earlier reports in the literature, reflecting how recent optimizations have shifted the performance bottleneck.
}

Another observation from Figure~\ref{fig:breakdown_hpmpc} is that \emph{no single operator is the sole bottleneck}.
Softmax (33--47\%), GELU (18--30\%), and LayerNorm (15--38\%) each account for a substantial fraction of the total latency.
This again differs from prior reports that emphasized a single dominant operator as the primary bottleneck~\cite{mpcformer, mpc_transformer}, while largely overlooking operators such as LayerNorm.
This shift is likewise due to recent advances and optimizations in the MPC community.
Moreover, the \emph{breakdown varies across frameworks and protocols}; with exact truncation, LayerNorm's relative overhead increases from 15--30\% to 30--38\%.
Several prior works focus on reducing the overhead of a single operator (\emph{e.g.}, Softmax~\cite{mpcformer, mpcvit, salvit}), implicitly assuming that doing so will broadly benefit MPC inference.
Our results show this is no longer universally true: multiple operators must be optimized jointly, and their relative importance depends on the underlying framework.

\obs{
Multiple operators contribute to end-to-end latency, and their relative importance depends on the framework. No single framework-agnostic dominating bottleneck exists.
}

\begin{figure}
    \centering
    \includegraphics[width=0.9\textwidth]{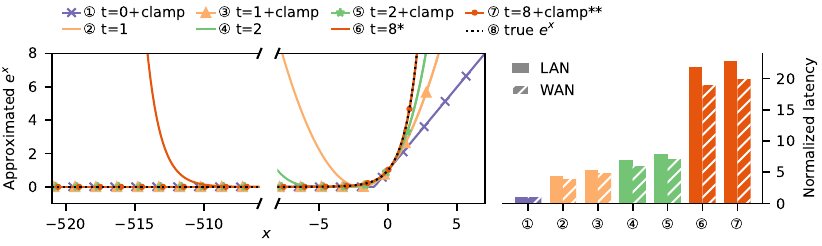}
    \caption{
    Functional behavior (left) and the latency (right) for various MPC approximations for $e^x$ under HPMPC-Trio with $\mathit{TE}_{\{1\}}$.
    *Used in~\cite{crypten, spu}; **Used in~\cite{spu_hack, jinyu_mpc_cal}. 
    }
    \label{fig:approx_exp}
\end{figure}

\subsection{Detailed Analysis on the Approximations}
\label{sec:charac_approx_detail}

We examine these approximations in more detail.
For illustration, we use the exponential function ($e^x$) as a running example, but similar discussion applies to other operators that require approximation.
Most frameworks~\cite{crypten, spu, bumblebee, nexus} approximate $e^x$ using $t$ iterative multiplications, with $e^x \approx (1 + \frac{x}{2^t})^{2^t}$, following the definition $e^x=\lim\limits_{n\to\infty}(1+\frac{x}{n})^n$.
This approximation diverges rapidly when $x < -2^t$, so some systems~\cite{spu_hack, jinyu_mpc_cal} clamp the output to zero in this problematic region.
Figure~\ref{fig:approx_exp} shows how the approximation behaves as $t$ varies, with and without clamping (left), together with the corresponding latencies (right).
The clamped variants (lines with markers) overlap almost exactly with their unclamped counterparts until the unclamped versions begin to diverge.

Approximations expose a wide range of performance--accuracy trade-offs.
In some cases, \emph{an approximation that appears reasonable and is widely used can still significantly degrade accuracy}.
Many frameworks~\cite{crypten, spu} use $t$=8 (red line) by default, which only diverges noticeably for inputs near $x < -512$.
However, such seemingly unlikely values do occur in practice and can completely destroy BERT-base accuracy.
Framework designers often choose approximations by minimizing error over a fixed input range~\cite{bolt}. Such manually designed approximations can therefore fail badly when real inputs fall outside the range considered during design.

In other cases, \emph{an approximation is unnecessarily precise and incurs substantial overhead with little or no accuracy benefit}.
As shown in Figure~\ref{fig:approx_exp}, $t$=2 and $t$=2+clamp, which still approximate $e^x$ reasonably well over a modest input range, are 3.1--3.2$\times$ and 2.8--2.9$\times$ cheaper than the $t$=8 and $t$=8+clamp variants used by many existing frameworks~\cite{crypten, jinyu_mpc_cal, spu}.
In more aggressive cases, even $t$=1 or $t$=0+clamp, which are 5.1--5.2$\times$ and 20--23$\times$ cheaper than $t$=8+clamp, may still be sufficiently accurate.
Prior work on MPC-friendly Transformers~\cite{mpcformer, mpcvit, salvit} has similarly shown that $e^x$ in Softmax can sometimes be replaced by a quadratic function of the form $(x+c)^2$ or by ReLU with little accuracy loss.
These choices closely correspond to the highly approximate $t$=1 and $t$=0+clamp.

\obs{
Sometimes, even widely used approximations are not accurate enough and significantly hurt model quality.
Other times, they are more accurate than necessary and incur avoidable performance overhead.
Existing frameworks typically choose one approximation heuristically and apply it globally~\cite{crypten, spu, bolt}.
}

Even for the same operator, the \emph{tolerable approximation error depends strongly on where it appears in the model}.
For example, we found that using $t$=0+clamp for the $e^x$ in the first Softmax of ViT degrades accuracy by only 0.56\%, whereas using the same approximation for the third Softmax degrades accuracy by \textbf{8.7\%}.
Similarly, reducing the inverse-square-root approximation in ViT's first LayerNorm from five iterations to one causes a \textbf{13.4\%} accuracy drop, while the same reduction causes at most 0.1\% degradation in half of the model's other LayerNorms.

\obs{
The same operator can tolerate very different amounts of approximation error depending on where it appears in the model, making manual tuning difficult.
}

\subsubsection{Applicability to FHE}

While this paper primarily focuses on MPC, similar opportunities exist for FHE frameworks. FHE systems likewise approximate unsupported operators, often using similar or identical approximation techniques~\cite{encryptedllm, nexus}.
In fact, FHE systems must approximate even relatively simple functions such as ReLU (Section~\ref{sec:bg_fhe}), since they do not natively support comparison.
Thus, the compiler frontend and middle-end of \sys can, in principle, be applied to benefit FHE systems as well. We evaluate this extension in Section~\ref{sec:eval_fhe}.

\begin{figure*}
    \centering
    \includegraphics[width=0.95\textwidth]{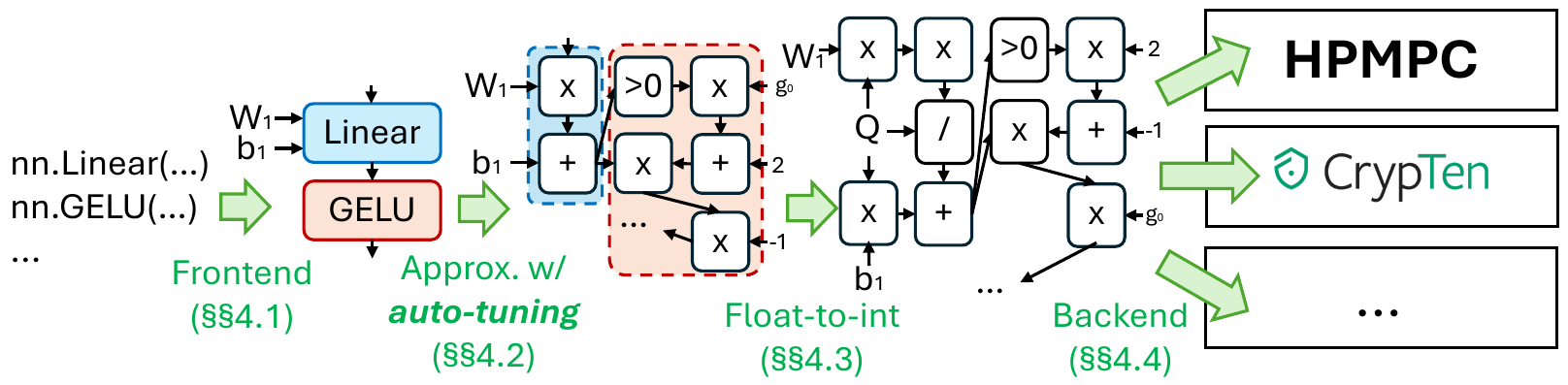}
    \caption{
    Overview of \sys. %
    }
    \label{fig:sys_overview}
\end{figure*}

\section{\sys Framework Design}

\sys is a multi-stage compiler that implements common functionality needed by MPC frameworks as reusable compiler stages shared across multiple MPC runtimes (Figure~\ref{fig:sys_overview}).

\begin{enumerate}[noitemsep, leftmargin=*, topsep=5pt]
\item The \emph{frontend} (Section~\ref{sec:frontend}) converts an annotated PyTorch program into \sys's IR, which is implemented as a lightweight extension of PyTorch 2's export IR~\cite{export_ir}.
\item The \emph{approximation stage} (Section~\ref{sec:approx}), which is the core of \sys, selects an approximation for each unsupported operator to maximize performance while satisfying the user-specified quality constraint.
Following the observations in Section~\ref{sec:charac_approximation}, we design the auto-tuner to account for each operator's cost on the target MPC runtime and its impact on overall model quality.
\item The \emph{float-to-int stage} (Section~\ref{sec:to_int}) converts floating-point computations to integer arithmetic and inserts the appropriate rescaling operations.
We also provide, to our knowledge, the first formal specification of this stage, whose behavior has largely been left implicit in prior work. This formalization helps avoid corner-case errors that have appeared in existing systems~\cite{crypten-conversion-bug}. We additionally implement a shared optimization~\cite{hummingbird} that benefits multiple runtimes.
\item The \emph{backend} (Section~\ref{sec:backend}) lowers the graph to the target runtime and applies backend-specific optimizations. Our HPMPC backend includes a new optimization (to our knowledge), which is an orthogonal contribution.
\end{enumerate}
\sys is implemented as an extension to PyTorch 2's compiler~\cite{pytorch2}. This offers practical benefits over existing frameworks implemented with custom C++ from scratch, as users familiar with PyTorch can easily extend the codebase using PyTorch's pass-writing interface~\cite{export_ir_pass_tutorial}.

\subsection{\sys Compiler Frontend}
\label{sec:frontend}
\sys's frontend (Figure~\ref{fig:sys_overview}, left) takes an annotated PyTorch program, compiles it into an IR, and deduces the owner of each intermediate tensor.
The output is an otherwise standard PyTorch export IR~\cite{export_ir} where tensors are annotated with owners.

\subsubsection{Programming Model and IR}
\label{sec:frontend_pl}
Figure~\ref{fig:pl_overview} shows an example program that runs an ML model (\texttt{net}), written in PyTorch, with \sys.
The highlighted lines (Lines 2, 9--10) are regular PyTorch code, while the rest is \sys-specific.
\sys reuses PyTorch's frontend and its IR (export IR~\cite{export_ir}), which can be generated by the \texttt{torch.export} function (Line 2).

\sys extends the export IR by adding and propagating each tensor's owner information (Line 3; Section~\ref{sec:front_ownership}).
In this example, party 0 owns the input \texttt{x}, and party 1 owns all other parameters (names returned by the helper function \texttt{get\_prms}).
Unless a tensor is owned by all parties, the compiler backend (Section~\ref{sec:backend}) encrypts it throughout.

To run approximation auto-tuning, the programmer first lists the approximation templates for each unsupported operator, which specify the forms of valid approximations that the auto-tuner may apply (\emph{i.e.}, \emph{passes}, Line~4). \sys already provides a range of templates derived from commonly used approximations, but users can add new ones through \sys's programming interface if needed (Section~\ref{sec:approx_pl}). The programmer then specifies the target MPC setup (Line~5), allowing the auto-tuner to accurately estimate the cost of each approximation under that setup, and invokes the auto-tuner with a tuning dataset and a quality threshold (Line~6).
In this example, the auto-tuner minimizes cost for the HPMPC runtime while ensuring that the KL divergence of the output logits changes by no more than 0.005 on the tuning dataset (Section~\ref{sec:approx_autotune}).

Finally, the graph is converted to integer-ring computation (Line~7; Section~\ref{sec:to_int}) and lowered to the target MPC runtime (Line~8; Section~\ref{sec:backend}).
For simplicity, these steps could be hidden behind a single library function, but our prototype exposes each stage explicitly to the programmer for full transparency. The output of every stage remains a lightly extended PyTorch export IR graph~\cite{export_ir} (\texttt{torch.fx.GraphModule}), allowing users to inspect the IR after each stage if desired. Before being lowered at Line~8, these graphs can also be executed like regular PyTorch programs on a single GPU for testing.
The final output is again a \texttt{torch.fx.GraphModule}, which can be invoked like a regular PyTorch model (Lines~9--10), but executes the computation using MPC.

\subsubsection{Ownership Propagation}
\label{sec:front_ownership}
The owner information for each tensor (Figure~\ref{fig:pl_overview}, Line 3) is recorded in the IR node's custom metadata field and propagated forward through the graph, using intersection when two flows meet.
For example, if tensor \texttt{x} is owned by party 0, \texttt{y} is owned by party 1, and \texttt{c} is owned by both (public), \texttt{x+c} is also owned by party 0, but \texttt{x+y} is owned by neither.
The compiler backend ensures that only the owner can recover a secret (Section~\ref{sec:backend}).
Ownership is not propagated across operators whose output does not depend on the values (\emph{e.g.}, the \texttt{zeros\_like} operator). 
Following existing frameworks, we allow the input data owner to decrypt the final result.

\subsection{Operator Approximation and Auto-tuning}
\label{sec:approx}

The operator-approximation stage (Figure~\ref{fig:sys_overview}, middle-left)  decomposes operators unsupported by MPC into a series of supported operators.
The output of this stage is an export IR that only contains the supported set of operators.
This stage also decomposes combined operators into their basic elements (\emph{e.g.}, Linear into MatMul and addition, ReLU into comparison and multiplication, \emph{etc.}), and rewrites all the comparisons into less-than-zero comparison (\emph{e.g.}, $a \ge b$ is converted to $1 - ((a - b) < 0)$).
Valid forms of approximations are expressed through a \emph{pass}.
\sys already provides a set of passes derived from approximations used in popular frameworks~\cite{crypten, spu, bolt, sirnn}, while also allowing users to introduce new approximation ideas through writing a custom pass (Section~\ref{sec:approx_pl}).
Given these choices, the auto-tuner selects a set of approximations that maximizes performance according to its search algorithm (Section~\ref{sec:approx_autotune}).

\begin{figure}
    \centering
    \hspace*{0.015\textwidth}%
    \begin{minipage}[t]{0.47\textwidth}
    \vspace{0pt}
    \begin{lstlisting}[
        language=Python,
        style=pl_overview,
        basicstyle=\ttfamily\scriptsize,
        breaklines=true,
        columns=fullflexible
    ]
import cryptorch as ct
mod = torch.export(net, args=(x,)).module()
ct.set_secret(mod, {"x":[0], get_prms(mod):[1]})
passes = [ct.ExpPass(), ct.GeluPass(), ...]
cost = ct.target_cost("hpmpc", passes, lan)
ct.autotune(mod, passes, cost, data, kl=0.005))
ct.to_integer(mod, ...)
ct.lower(mod, backend="hpmpc")
for x, y in test_loader:
    y_pred = mod(x)\end{lstlisting}
    \caption{User-level code for \sys (simplified).}
    \label{fig:pl_overview}
    \end{minipage}%
    \hspace{0.025\textwidth}%
    \begin{minipage}[t]{0.47\textwidth}
    \vspace{0pt}
    \begin{lstlisting}[
        language=Python,
        basicstyle=\ttfamily\scriptsize,
        breaklines=true,
        columns=fullflexible
    ]
class ExpPass(TunablePass):
    def get_patterns(self):
        return [lambda x: aten.exp(x)]
    def get_filters(self):
        return [lambda op: is_secret(op.args[0])]
    def get_replacement(self):
        y = 1+x/(2**<@\tightcolorbox{green!25}{self.t}@>)
        if <@\tightcolorbox{green!25}{self.clamp}@>:
            y = y * (x > -2 ** <@\tightcolorbox{green!25}{self.t}@>)
        for _ in range(<@\tightcolorbox{green!25}{self.t}@>):
            y *= y
        return y\end{lstlisting}
    \caption{Approximation with tunable knobs.}
    \label{fig:pass_lang}
    \end{minipage}
\end{figure}

\subsubsection{Expressing Approximations As Passes}
\label{sec:approx_pl}
A valid set of approximations for each operator is expressed as a pass (Figure~\ref{fig:pass_lang}).
Each pass specifies the target graph patterns to replace (Lines~2--3), a set of optional conditions under which the approximation applies (Lines~4--5), and the approximation choices to consider (Lines~7--12).
This example shows a pass that approximates $e^x$ (Line~3) using the iterative method from Figure~\ref{fig:approx_exp}, with optional clamping (Lines~7--12), when the input is secret (Line~5).
\texttt{self.t} and \texttt{self.clamp} correspond to the $t$ and clamp choices in Figure~\ref{fig:approx_exp}, and serve as tunable knobs that control the accuracy--performance trade-off.

For an approximation to be tunable, it must be written such that decreasing the tunable knobs (\emph{e.g.}, \texttt{self.t} and \texttt{self.clamp}) makes the approximation faster but less accurate.
The valid range of each knob must also be specified, although we omit it from the figure for brevity.
Because most approximations are iterative or polynomial-based, iteration count or polynomial degree can naturally serve as tuning knobs.
More generally, an arbitrary set of approximation choices can be exposed to the auto-tuner by wrapping them in \texttt{if}-\texttt{else} branches conditioned on the knob value.
\sys provides a default set of tunable approximation passes that span more- and less-accurate variants of the heuristically chosen approximations used in existing frameworks.
Users who wish to explore new approximation ideas can additionally implement their own passes.

\subsubsection{Auto-tuning Engine}
\label{sec:approx_autotune}
\label{sec:autotuner}

\begin{wrapfigure}{r}{0.47\textwidth}
    \centering
    \vspace{-20pt}
    \includegraphics[width=0.47\textwidth]{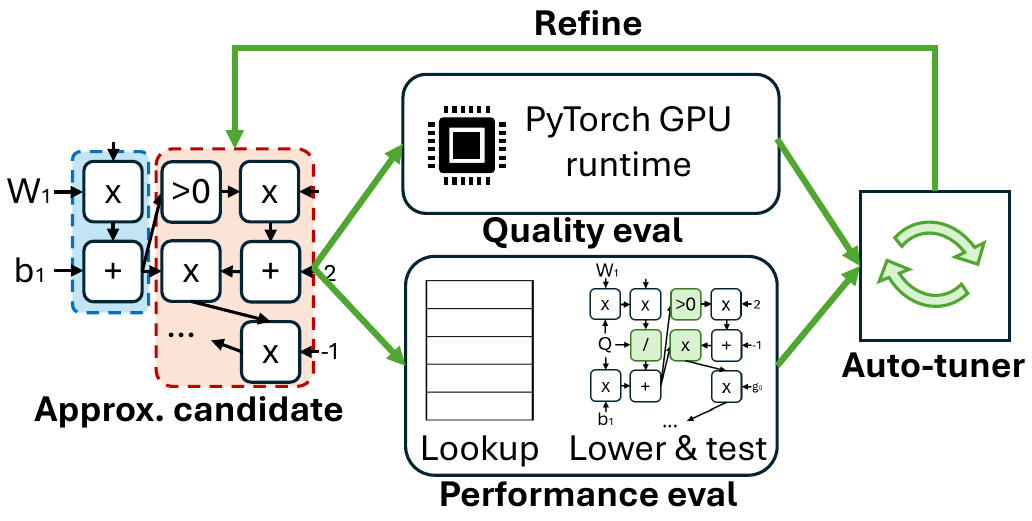}
    \vspace{-20pt}
    \caption{
    \sys's auto-tuner.
    }
    \vspace{-10pt}
    \label{fig:autotune}
\end{wrapfigure}

\sys's auto-tuner aims to automatically find, for each unsupported operator, a set of approximations that maximizes performance while minimizing quality loss in the target ML program.
The search engine starts with all tunable knobs at their maximum values and uses the selected search algorithm to find a knob configuration that improves performance while staying within the specified quality budget. Figure~\ref{fig:autotune} summarizes the auto-tuner.

\textbf{Search algorithms.}
\sys provides several basic search algorithms, and users can also implement their own. By default, the search engine uses a greedy hill-climbing algorithm, which repeatedly takes the step that yields the largest reduction in runtime overhead per unit loss of quality.
To specify the quality budget, users choose a quality metric (\emph{e.g.}, accuracy, loss, or KL divergence) and a threshold. For example, if the user sets the quality budget to a KL divergence of 0.005 (Figure~\ref{fig:pass_lang}, Line~6), the search engine reduces a knob only if the resulting logits have a KL divergence below 0.005 from those of the float model on the given tuning dataset. We empirically found that guiding the search while constraining the KL divergence works best as the quality metric because it provides a finer-grained signal than accuracy and does not cause overfitting.

\textbf{Efficient candidate evaluation.}
To evaluate different candidates, \sys must repeatedly modify the approximated IR graph. A naive implementation would regenerate the entire IR graph for every candidate, even though most of the graph remains unchanged across settings.
To avoid this overhead, \sys first identifies all tunable sites in the IR graph, \emph{i.e.}, sites matched by a tunable pass such as Figure~\ref{fig:pass_lang}, and replaces each with a submodule that can be modified independently.
The auto-tuner can then update only the affected submodules when evaluating a new candidate, while reusing the rest of the graph from previous evaluations.
After tuning completes, \sys inlines these submodules to produce a single unified IR graph.

\textbf{Estimating the quality.} 
\sys must estimate how each knob value affects overall ML inference quality. The easiest and most accurate way is to compile each candidate and run it on the target MPC runtime, but this adds prohibitive overhead to the auto-tuner. Instead, \sys measures the quality impact by running the candidate graph on a standard PyTorch GPU runtime in floating point. This is easy to implement because, as explained in Section~\ref{sec:frontend_pl}, every intermediate graph generated by \sys is PyTorch export IR, which can already run natively in PyTorch. We empirically find that while the result of floating-point execution is not exactly the same as running on the MPC runtime, this provides sufficient signal to guide the search effectively, while allowing multiple orders-of-magnitude faster search.

\textbf{Estimating the cost.} To make informed approximation choices, \sys needs to estimate the cost of each candidate knob setting. However, different backends have very different cost characteristics (Section~\ref{sec:charac_latency_breakdown}), and optimizations applied in later compiler stages (Sections~\ref{sec:hummingbird} and~\ref{sec:backend_additional_opts}) can further change these costs in non-trivial ways. As a result, an approximation that appears beneficial at this stage may interact with later optimizations and ultimately degrade performance.
To obtain an accurate estimate, \sys runs each candidate through the remaining compiler stages and evaluates its cost at the target backend.
Doing this for the entire network for every candidate would be prohibitively expensive, so \sys instead processes only the affected submodule through the remaining pipeline and caches the resulting cost in a lookup table, avoiding repeated evaluation of the same candidate.
This approach assumes that submodule costs compose additively, which may not hold perfectly when later stages optimize across submodule boundaries. Nevertheless, we empirically find that the resulting estimate is within 0.05\% of the true cost.

\subsection{Converting to Integer Ring}
\label{sec:to_int}

The output from the previous stage (Section~\ref{sec:approx}) holds only the supported set of operators (\emph{e.g.}, addition, multiplication, truncation, and comparison) but still operates on floating-point numbers.
The float-to-int stage (Figure~\ref{fig:sys_overview}, middle-right) translates the graph into one that operates only on integers. 
This stage also marks nodes that take secret inputs so the final backend stage (Section~\ref{sec:backend}) lowers them to appropriate MPC kernels.
Conversion to the integer ring mainly involves ensuring that the scaling factor $s$ matches between two values before addition, and that the output of a multiplication is rescaled properly (Section~\ref{sec:bg_mpc_limitations}).
The rule is similar to that of fixed point~\cite{fixed_point_book}, but not exactly the same (\emph{e.g.}, see the discussion around \textsc{Sec-PubFloat Add} below). 
Sharing this stage across backends eliminates the need for each MPC framework to reimplement the same conversion logic, which is sometimes error-prone~\cite{crypten_conversion_bug}.
We formally define the translation rules for this stage, which, to our knowledge, provides the first formal specification of this conversion process.

\subsubsection{Grammar of the Input/Output Graphs}
\label{sec:grammar}

\begin{figure}[t]
\centering
\begin{small}

\begin{subfigure}[t]{0.48\linewidth}
\centering
\[
\ensuremath{\displaystyle
\begin{array}{l@{\ \ }l@{\ }r@{\ }l}
    \text{Expr} & e &::= &x: t \mid \ltz(e) \mid \add(e, e) \mid \mul(e, e)\\
    \text{Add-like op} & \add &::= & \text{add} \mid \text{sub} \mid ...\\
    \text{Mul-like op} & \mul &::= & \text{mul} \mid \text{conv2d} \mid \text{matmul} \mid ...\\
    \text{Type} & t  &::= & \type{o, \btype}\\
    \text{Owner} & o &::= & \top \mid p \in \mathcal{O}\\
    \text{Base type} & \btype &::= & \itype \mid \ftype\\
    \text{Int type} & \itype &::= & \bool \mid \integer64 \mid \integer32 \mid \integer16 \mid ...\\
    \text{Float type} & \ftype &::= & \float64 \mid \float32 \mid ...
\end{array}
}
\]
\caption{Input graph $e$.}
\label{fig:e_syntax}
\end{subfigure}
\hfill
\begin{subfigure}[t]{0.48\linewidth}
\centering
\[
\ensuremath{\displaystyle
\begin{array}{l@{\ \ }l@{\ }r@{\ }l}
    \text{Type} & T &::= & \type{\ell, \btype, s}\\
    \text{Scale} & s &::= & 1 \mid 2 \mid ...\\
    \text{Security label} & \ell &::= & \pub \mid \secret\\
    \text{Expr} & E &::= & x \mid \add(E, E) \mid \mul(E, E)\\
    & & \mid & \ltz(E) \mid \encode(E, s) \\
    & & \mid & \add_{SEC}(E, E) \mid \mul_{SEC}(E, E)\\
    & & \mid & \ltz_{SEC}(E) \mid \encode_{SEC}(E, s)\\
    & & \mid &  \encrypt(E)
\end{array}
}
\]
\caption{Output graph $E$.}
\label{fig:E_syntax}
\end{subfigure}
\caption{Grammar for the input/output graphs of the integer ring conversion.}
\end{small}
\end{figure}

This stage translates the input IR, a graph $e$ containing a mix of plaintext integer and floating-point operations, into a graph $E$ with type $T$, which uses only integers and dedicated operators for computations involving secret inputs.

The judgment form is $\Gamma \vdash e \Rightarrow E: T$.
Figure~\ref{fig:e_syntax} summarizes the grammar for $e$, which only contains the supported set of operators: add-like operators (addition, subtraction, \emph{etc.}); mul-like operators (MatMul, Conv2d without bias, \emph{etc.}); and less-than-zero comparison ($\ltz$). The approximation stage rewrites every comparison into $\ltz$ (Section~\ref{sec:approx}).
Each tensor has a base type $\btype$, either integer or floating point. Together with its owner $o$ this forms the tensor's type $t$.
An owner is either $\top$, meaning the value is public and known to every party, or one of the input owners $o$.

Figure~\ref{fig:E_syntax} defines the grammar for the output $E$.
Its type $T$ records whether each expression is secret ($\secret$) or public ($\pub$) rather than its owner. It also records the implicit scaling factor $s$.
As explained in Section~\ref{sec:bg_mpc_limitations}, all MPC computation must be on an integer ring, so non-integers are first multiplied by a scaling factor and rounded ($\round{x_f \times s}$). The scaling factor propagates through operations, growing on multiplication and shrinking on truncation. $T$ tracks that information.
The scaling operator $\encode$ rescales $E$: if $E$ had scaling factor $s_{old}$, then $\encode(E,s_{new})=\round{E\times \frac{s_{new}}{s_{old}}}$. Depending on the sizes of $s_{old}$ and $s_{new}$, this can be either multiplication, truncation, or no-op, but we express it as $\encode$ for simplicity.
$E$ contains the normal (plaintext) operators ($\add$, $\mul$, $\ltz$, and $\encode$) and their secret-involving variants ($\add_{SEC}$, $\mul_{SEC}$, $\ltz_{SEC}$, and $\encode_{SEC}$), which are used when at least one operand is a secret.
It also has an $\encrypt$ operator, which encrypts plaintext values into MPC secret shares.

\subsubsection{Translation Rules}
\label{sec:transformation}
Now, we describe the translation, which (1) encrypts secret tensors, (2) inserts proper $\encode$ operators to manage the scaling factor, and (3) replaces operators with at least one secret input with their secret-involving variants.

\begin{figure}
\begin{small}
\[
\begin{aligned}
\gInfer{No-Enc}
{
}{
\Gamma \vdash x: \type{\top, \btype}
\Rightarrow x: \type{\pub, \btype, 1}
}
\qquad\qquad
\gInfer{Enc}
{
o \ne \top
}{
\Gamma \vdash x: \type{o, \btype}
\Rightarrow
\encrypt(\encode(x, \defaultscale)):
\type{\secret, \itype, \defaultscale}
}
\end{aligned}
\]
\end{small}
\caption{Rules for encryption.}
\label{fig:encryption_rule}
\end{figure}

\emph{Encryption} rule is shown in Figure~\ref{fig:encryption_rule}. If a tensor is owned by all parties ($o=\top$), $\ell = \pub$, and it remains unchanged (\textsc{No-Enc}).
If only one party owns the tensor, the tensor is converted to an integer and encrypted into secret shares (\textsc{Enc}).
During the encryption, a fixed ring size of $\defaultint$ and a scaling factor $\defaultscale$ is used
(\emph{e.g.}, CrypTen~\cite{crypten} uses $\defaultint=\integer64$ and $\defaultscale=2^{16}$).

\begin{figure}
\begin{small}
\[
\begin{array}{c}
\gInfer{Sec-Sec Add}
{
\Gamma \vdash e_1 \Rightarrow E_1: \type{\secret, \defaultint, s_1} \premSep \Gamma \vdash e_2 \Rightarrow E_2: \type{\secret, \defaultint, s_2} \premSep  \max(s_1, s_2) = s_{max}
}{
\Gamma \vdash \add(e_1, e_2) \Rightarrow \add_{SEC}(\encode_{SEC}(E_1, s_{max}), \encode_{SEC}(E_2, s_{max})): \type{\secret, \defaultint, s_{max}}
}
\\[4ex]
\gInfer{Sec-PubInt Add}
{
\Gamma \vdash e_1 \Rightarrow E_1: \type{\secret, \defaultint, s_1} \premSep \Gamma \vdash e_2 \Rightarrow E_2: \type{\pub, \itype, s_2} \premSep  \max(s_1, s_2) = s_{max}
}{
\Gamma \vdash \add(e_1, e_2) \Rightarrow \add_{SEC}(\encode_{SEC}(E_1, s_{max}), \encode(E_2, s_{max})): \type{\secret, \defaultint, s_{max}}
}
\\[4ex]
\gInfer{Sec-PubFloat Add}
{
\Gamma \vdash e_1 \Rightarrow E_1: \type{\secret, \defaultint, s} \premSep \Gamma \vdash e_2 \Rightarrow E_2: \type{\pub, \ftype, 1} \premSep \max(s, \defaultscale) = s_{max}
}{
\Gamma \vdash \add(e_1, e_2) \Rightarrow \add_{SEC}(\encode_{SEC}(E_1, s_{max}), \encode(E_2, s_{max})): \type{\secret, \defaultint, s_{max}}
}
\end{array}
\]
\end{small}
\caption{Rules for add-like operators.}
\label{fig:add_rules}
\end{figure}

\emph{Add-like operators} are translated through rules in Figure~\ref{fig:add_rules}. When adding secrets (\textsc{Sec-Sec Add}) or adding a secret with a public integer (\textsc{Sec-Int Add}), the scaling factors must match (Section~\ref{sec:bg_mpc_limitations}). If not, \sys inserts a scaling function ($\encode$ or $\encode_{SEC}$, depending on the secrecy of the input) to match the scaling factor to the larger ($s_{max}$) of the two. $\encode(E, s)$ is a no-op if $E$ is already of scale $s$.
When adding a secret and a public floating-point (\textsc{Sec-PubFloat Add}), the floating-point is first converted to an integer using a scaling factor of $\max(s, \defaultscale)=s_{max}$.
It is crucial to use $\max(s, \defaultscale)=s_{max}$ instead of the secret's scaling factor $s$ when converting the floating-point. Otherwise, a large conversion error may be introduced when $s$ is small (\emph{e.g.}, $s$=1).
We found that CrypTen simply uses $s$ in these conversions (the conversion occurs dynamically inside the library)~\cite{crypten_conversion_bug} and occasionally suffers from large conversion errors.
This rule is unique to MPC: in a standard fixed-point system, multiplication between a floating point and a fixed point with $s=1$ (normal integer) yields a floating-point output; in MPC, the output of a secret computation can never be a floating point.
We only show interesting cases, and rules with the operators switched (\emph{e.g.}, \textsc{PubInt-Sec Add}, which works similarly to \textsc{Sec-PubInt Add}) or trivial cases (addition between public values) are omitted.

\begin{figure}
\begin{small}
\[
\begin{array}{c}
\gInfer{Sec-Sec Mul}
{
\begin{array}{c}
\Gamma \vdash e_1 \Rightarrow E_1: \type{\secret, \defaultint, s_1} \premSep \Gamma \vdash e_2 \Rightarrow E_2: \type{\secret, \defaultint, s_2} \premSep \max(s_1, s_2) = s_{max}
\end{array}
}{
\Gamma \vdash \mul(e_1, e_2) \Rightarrow \encode_{SEC}(\mul_{SEC}(E_1, E_2), s_{max}): \type{\secret, \defaultint, s_{max}}
}
\\[2ex]
\gInfer{Sec-PubInt Mul}
{
\begin{array}{c}
\Gamma \vdash e_1 \Rightarrow E_1: \type{\secret, \defaultint, s_1} \premSep \Gamma \vdash e_2 \Rightarrow E_2: \type{\pub, \itype, s_2} \premSep \max(s_1, s_2) = s_{max}
\end{array}
}{
\Gamma \vdash \mul(e_1, e_2) \Rightarrow \encode_{SEC}(\mul_{SEC}(E_1, E_2), s_{max}): \type{\secret, \defaultint, s_{max}}
}
\\[2ex]
\gInfer{Sec-PubFloat Mul}
{
\begin{array}{c}
\Gamma \vdash e_1 \Rightarrow E_1: \type{\secret, \defaultint, s} \premSep \Gamma \vdash e_2 \Rightarrow E_2: \type{\pub, \ftype, 1} \premSep \max(s, \defaultscale) = s_{max}
\end{array}
}{
\Gamma \vdash \mul(e_1, e_2) \Rightarrow \encode_{SEC}(\mul_{SEC}(E_1, \encode(E_2, \defaultscale)), s_{max}): \type{\secret, \defaultint, s_{max}}
}
\end{array}
\]
\end{small}
\caption{Rules for mul-like operators.
}
\label{fig:mul_rules}
\end{figure}

\emph{Mul-like operators} follow rules in Figure~\ref{fig:mul_rules} (again, only interesting rules are shown). Similar to addition, $\encode_{SEC}$ and $\mul_{SEC}$ must be used when the input is a secret, and as explained in Section~\ref{sec:bg_mpc}, each multiplication must be followed by a rescaling operation ($\encode$).
\sys rescales the result to follow the larger of the two ($s_{max}$).
When a floating-point is involved (\textsc{Sec-PubFloat Mul}), as in \textsc{Sec-PubFloat Add}, the floating-point is first converted using a default scaling factor ($\encode(E_2, \defaultscale)$).
\emph{Less-than-zero operators} ($\ltz$) on a secret input is lowered to the MPC version. The resulting $\ltz_{SEC}(E)$ has a scaling factor of $s$=1, as the output is only 0 or 1, and using any other scaling factor is redundant. Below is the rule:
\[
\begin{array}{c}
\gInfer{LTZ}
{
\Gamma \vdash e \Rightarrow E: \type{\secret, \defaultint, s}
}{
\Gamma \vdash \ltz(e) \Rightarrow \ltz_{SEC}(E): \type{\secret, \defaultint, 1}
}
\end{array}
\]

\emph{Other operators} were omitted from the formalism for simplicity, but 
\sys supports other common PyTorch operators.
Operators that only change the tensor's shape and not the values, \emph{e.g.}, transpose, flatten, stack, permute, reshape \emph{etc.}, and operators that only use the tensor's shape and/or data types as an input, \emph{e.g.}, zeros\_like, full\_like, \emph{etc.}, can be trivially supported, because these do not depend on the secret value.
Non-binary add-like operators, \emph{e.g.}, sum, mean, AvgPool, AdaptiveAvgPool, \emph{etc.} are basically addition between secret values (plus truncation, if needed) and follows the \textsc{Sec-Sec Add} rule from Figure~\ref{fig:add_rules}.
Max and MaxPool operators can be decomposed into a series of addition, multiplication, and ltz.

\subsubsection{Additional Optimizations (HummingBird~\cite{hummingbird})}
\label{sec:hummingbird}

We can implement additional optimizations that operate on the integer graph and apply across multiple MPC runtimes at this stage.
For example, we implement the HummingBird~\cite{hummingbird} optimization here, which is currently implemented separately in several existing MPC frameworks~\cite{pigeon, hummingbird, jinyu_mpc_cal}. This allows a single implementation to be shared across multiple compatible backends.
HummingBird accelerates secret comparison by estimating the input range and performing the comparison using only the number of bits required to represent that range.
Because this optimization requires knowledge of the exact integer representation and bitwidth, it cannot be applied in earlier stages.

Existing frameworks implement different variants of HummingBird. The original work selects a separate bitwidth for each comparison by searching an exponential configuration space~\cite{hummingbird}, whereas later work uses a single static parameter shared across all comparison nodes~\cite{pigeon}.
Our prototype takes a middle-ground approach. We run a small calibration set (128--256 training inputs), record the input range of each comparison node, and use these ranges to select a parameter for each node.
Other variants used by existing frameworks can likewise be ported to \sys; once implemented at this stage, they can be shared across all compatible backends.

\subsection{\sys Compiler Backend}
\label{sec:backend}

\sys's backend lowers the IR produced by the earlier stages, which is still {runtime-agnostic}, into a graph executable by a target MPC runtime (Figure~\ref{fig:sys_overview}, rightmost).
We currently support backends for the following MPC runtimes:
(1) HPMPC~\cite{hpmpc}, an MPC library implementing the three-party, honest-majority Trio protocol. Due to its high performance, HPMPC has been used in several recent works~\cite{pigeon, truncation_sok, cryptdough};
and (2) CrypTen++~\cite{jinyu_mpc_cal}, which is an an accelerated version of the CrypTen~\cite{crypten} MPC library. 
CrypTen/CrypTen++ uses a two-party MPC protocol with a trusted third party, which is slower than HPMPC but tolerates stronger collusion assumptions (2-out-of-2 vs. 2-out-of-3). Both CrypTen and CrypTen++ have been used in several works~\cite{mpcformer, mpc_transformer, mpcpipe, pengzhi_ispass, hummingbird}.
Additionally, we provide an experimental FHE backend for
(3) Orion~\cite{orion}, an FHE library based on CKKS~\cite{ckks}, which we evaluate in Section~\ref{sec:eval_fhe}. 

The translation done by the MPC backends are simple. 
As Section~\ref{sec:to_int} already marked operators that takes in secret inputs (operators with the $_{SEC}$ subscripts), the backend only needs to replace them with the corresponding MPC kernels.
Because CKKS uses a different rescaling procedure from MPC (Section~\ref{sec:to_int}), we reuse both Orion's rescaling stage and its backend, thus sharing only \sys's frontend and operator-approximation stages.

\subsubsection{Additional Optimizations}
\label{sec:backend_additional_opts}
You can also apply additional backend-specific optimizations at this stage.
In particular, HPMPC converts arithmetic shares to binary shares during truncation using a binary adder, which is a standard technique in MPC~\cite{gmw, crypten, hpmpc}. We find that the best-performing adder depends on the tensor shape and assumed network conditions: ripple-carry and parallel-prefix adders perform better under different settings. However, the original HPMPC implementation always uses a ripple-carry adder for truncation.
Based on this observation, our HPMPC backend selects the binary adder according to the input shape and assumed network conditions.
This optimization is enabled by default in all of our evaluation results, including the baselines, except in Section~\ref{sec:eval_backend_opt}, where we evaluate its effect separately.

\subsection{Security Analysis}
\label{sec:security}

\sys does not introduce any additional security issues compared to the underlying MPC runtime.
\sys decomposes operators into the supported set (using different degrees of approximations) and maps each operator to the existing MPC kernel.
These kernels are secure on their own, and their security does not degrade under composition in the semi-honest MPC threat model. Thus, as long as each kernel is secure (ensuring correct implementation is an orthogonal problem to this paper), the entire execution is secure.
\sys's optimizations rely only on representative validation data that is considered public, and \sys's behavior does not depend on any observed data values during deployment.

\section{Evaluation}
\label{sec:eval}

In the evaluation, we aim to answer the following questions: (1) How much benefit does our approximation auto-tuner give to different MPC runtimes? (2) How does the benefit change with batch size, model encryption, network speed, and data distribution shift? (3) Does \sys improve programmability/usability? (4) Can the benefit of \sys extend to FHE?

\subsection{Evaluation Setup (MPC)}
\label{sec:eval_setup_mpc}

\subsubsection{MPC Runtimes}
We evaluate \sys in three MPC settings that differ in the number of parties and the protocols used:
(1) \textbf{HPMPC with $\bm{\mathit{TE}_{\{1\}}}$} uses Trio, a three-party MPC protocol from the HPMPC~\cite{hpmpc} library, with exact truncation (denoted $\mathit{TE}_{\{1\}}$ in prior work~\cite{truncation_sok}) for truncation operations. Exact truncation is considered more secure, but is more expensive than the stochastic truncation protocol described next;
(2) \textbf{HPMPC with $\bm{\mathit{TS}_{\{1\}}}$} uses the same Trio protocol and HPMPC implementation, but replaces exact truncation with a more efficient stochastic truncation protocol (denoted $\mathit{TS}_{\{1\}}$ in~\cite{truncation_sok}). The security of stochastic truncation has been debated: some works argue that it is insecure because it leaks more information than permitted by ideal-world simulation~\cite{truncation_broken}, while others argue that the additional leakage is benign~\cite{curl, truncation_sok}. This setting is applicable when the security guarantees of stochastic truncation are considered acceptable; and
(3) \textbf{CrypTen++}~\cite{jinyu_mpc_cal}, which is an optimized variant of CrypTen~\cite{crypten} that implements a two-party MPC protocol with the additional assumption of a trusted third party. We replace CrypTen's local truncation with the stochastic truncation protocol ($\mathit{TS}_{\{1\}}$) described above, because local truncation in CrypTen introduces non-negligible errors as model sizes increase~\cite{cheetah, aby3}.
All three settings compute over the integer ring $\mathbb{Z}_{2^{64}}$ with a scaling factor $s=2^{16}$. Except for the additional experiments in Section~\ref{sec:eval_settings}, all setups encrypt only the inputs, not the weights.
Our evaluation results include both offline and online overheads.

\subsubsection{Latency Measurements}
All experiments use a batch size of four inputs to represent low-latency cases, except for Section~\ref{sec:eval_settings}, which repeats the evaluations with a batch size of 128 for high-throughput cases.
Because larger batch sizes further increase the already high latency of MPC inference, small batch sizes are commonly used~\cite{gazelle, securenn, aby3, piranha, sigma}.
We measure the number of communication rounds and bytes and project their latency under two different network setups (LAN and WAN).
Following prior work~\cite{hummingbird, cheetah}, the LAN setup assumes 0.3\,ms latency and 10\,Gbit/s bandwidth, and the WAN assumes 40\,ms latency and 352\,Mbit/s bandwidth.

\subsubsection{Models and Datasets}
We evaluated four models under various datasets: (1) BERT-base~\cite{bert} on three GLUE tasks~\cite{glue_sst2}, SST-2, QNLI, and MNLI, at sequence length 128; (2) ViT-B/16~\cite{vit} and EfficientNet-B0~\cite{efficientnet_b0} on ImageNet~\cite{imagenet}; and (3) GPT-2 small~\cite{gpt2} on WikiText-2~\cite{wikitext}.
For each GLUE task (SST-2, QNLI, MNLI), we fine-tune BERT-base on the training set, excluding 8,192 examples. Of the 8,192 examples, 6,144 form the tuning set and 2,048 form the held-out set used for choosing the best approximations. We report test accuracies on the entire validation sets for SST-2 and QNLI, and 4,096 matched validation examples for MNLI, all of which are data never used in any of the previous stages.
For ViT-B/16 and EfficientNet, we tune on 2,048 training-set images, select the best approximations on another 1,024 training-set images, and report accuracy on 4,096 validation-set images.
For GPT-2 small~\cite{gpt2}, we used a context-window length of 128 and reported the next-token accuracy. We tuned on 128 such windows from the training set, held out another 64 windows, and reported test accuracy on 64 windows from the test set.
Following prior work~\cite{mpcformer, pengzhi_ispass}, we assume the client computes BERT/GPT-2 embeddings locally, and only run the rest with MPC.
Crucially, the tuning set (used to guide the auto-tuner), the held-out set (used to select the best knob values), and the test set (used to report test accuracy) are distinct and non-overlapping in all of the tested setups.

\subsubsection{Approximation Knobs}
BERT, ViT, and GPT-2 have 61 tunable knobs, and EfficientNet-B0 has 65.
We use the $e^x$ approximation from Section~\ref{sec:charac_approx_detail}, which performs $t=8$ iterations and an additional clamping. Tuning the knob reduces the iteration count.
GELU, SiLU, and sigmoid polynomials follow the recipes in BOLT~\cite{bolt}. Reducing the knob reduces the polynomial degree and ultimately degenerates into a ReLU or step function.
Inverse square root and reciprocal functions use the Newton-Raphson method with iteration counts of 5 (inverse square root) and 10 (reciprocal), and we reduced these iteration counts as the knob.

\subsection{Evaluation Setup (FHE)}
\label{sec:eval_setup_fhe}

Section~\ref{sec:eval_fhe} additionally evaluates the effect of \sys on a CKKS FHE runtime, Orion~\cite{orion}. Because FHE takes much longer to run, we evaluated only smaller CNN models: ResNet-20~\cite{resnet}, MiniONN~\cite{minionn}, and VGG-16~\cite{vgg} on CIFAR-10~\cite{cifar10} with a batch size of one.
Unlike MPC, which supports comparison and does not require operator approximation to run these models (hence their omission from the MPC experiments), CKKS FHE does not natively support comparison and must approximate ReLU with high-degree polynomials, so it can benefit from our shared auto-tuner stage. 
We used the polynomial degrees of each ReLU as the tuning knobs.
We tune the approximations on 2,000 validation-set images and hold out the remaining 3,000 validation-set images for choosing the best approximations. Because of FHE's performance limitations, we report simulated test accuracy on the 10,000 test-set images using Orion's~\cite{orion} simulator.

\subsection{MPC Performance Improvement}
\label{sec:eval_mpc}

\begin{figure}[t]
    \centering
    \includegraphics[width=\textwidth]{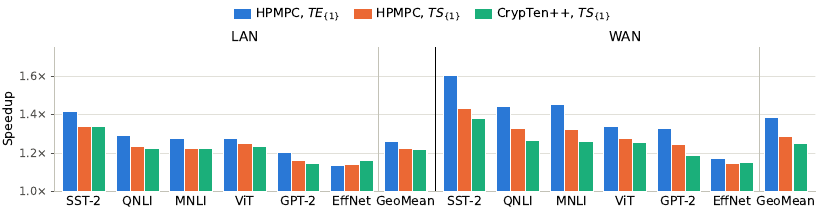}
    \caption{
    Speedup over the baseline (no tuning) with light tuning (quality criteria: held-out KL $\leq 0.005$). The test accuracy of all setups in this plot is within the noise (1 std) of floating-point accuracy.
    }
    \label{fig:main_bars}
\end{figure}

\subsubsection{Improvements Under Light Tuning}

Figure~\ref{fig:main_bars} shows the performance improvement achieved by the auto-tuner for different MPC runtimes while maintaining a KL divergence from the float model of at most 0.005 during auto-tuning. Under these settings, the test accuracy of all setups is within the noise (1 standard deviation) of the float model.
Without any measurable accuracy degradation, our auto-tuner gives \textbf{1.14--1.42$\times$ speedup on LAN} and \textbf{1.14--1.60$\times$ speedup on WAN}.
WAN generally benefits more because fewer communication rounds from approximations reduce the disproportionately slow round-trip latency of WAN.
Between MPC runtimes, HPMPC with $\mathit{TE}_{\{1\}}$ gets the most speedup (1.14--1.60$\times$). This is again because exact truncation incurs more communication than stochastic truncation ($\mathit{TS}_{\{1\}}$), and each truncation removal reduces communication more proportionally for exact truncation.
EfficientNet-B0 benefits the least because it has no iterative approximations (\emph{e.g.}, Newton-Raphson), which generally incur more sequential communication rounds than polynomial-based approximations.
Overall, the results show that the auto-tuner works across MPC runtimes and can bring speedup without accuracy penalties.

\subsubsection{Improvements Under More Aggressive Approximations}

\begin{figure}[t]
    \centering
    \includegraphics[width=\textwidth]{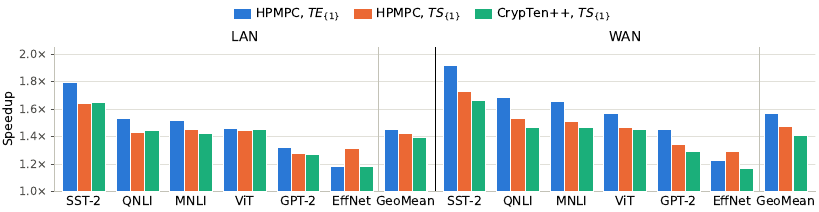}
    \caption{
    Speedup over the baseline (no tuning) with aggressive tuning (quality criteria: held-out KL $\leq 0.05$). The test accuracy of all setups degrades by at most -2.9\% from the floating-point accuracy.
    }
    \label{fig:main_bars_kl005}
\end{figure}

Figure~\ref{fig:main_bars_kl005} shows the performance improvement when slight accuracy degradation is tolerated, allowing more aggressive approximations. Using KL divergence $\leq 0.05$ as the quality criterion for the auto-tuner leads to a median 1\% to at most 2.9\% accuracy degradation in the setups we studied. In return, it achieves higher speedup of \textbf{1.18--1.79$\times$ in LAN} and \textbf{1.17--1.92$\times$ in WAN}.
While the gains are larger, the overall trend is similar---exact truncation gains more, and EfficienNet-B0 gains the least. Overall, the results show that \sys's auto-tuner can successfully trade off accuracy with additional performance across various MPC runtimes.

\subsubsection{Accuracy vs. Performance}

\begin{figure}[t]
    \centering
    \includegraphics[width=\textwidth]{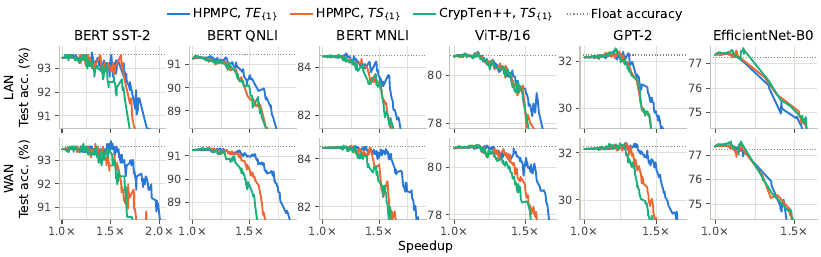}
    \caption{
    Test accuracy against speedup for the three MPC runtimes. 
    }
    \label{fig:main_frontier}
\end{figure}

Figure~\ref{fig:main_frontier} plots how the auto-tuner trades off test accuracy with performance as we progressively relax the quality constraint. The test accuracy trend is a bit noisy, but the general trend holds: our auto-tuner can gracefully trade off accuracy with additional performance across MPC runtimes, models, datasets, and network setups.
Graphs for datasets like SST-2 are noisier mainly because their test sets are small (their labeled validation sets, which we use as a test set, have only 872 samples). 

\subsubsection{Auto-tuner Cost}
One tuning run takes 180 steps for EfficientNet-B0 and 602--662 for BERT-base, ViT-B/16, and GPT-2. This represents 1--6.5 hours on one NVIDIA RTX A5000 GPU, a one-time cost per model and target. The search can also be parallelized for further speedup.

\subsection{Ablation Studies}
\label{sec:eval_settings}

\begin{figure}[t]
    \centering
    \includegraphics[width=\textwidth]{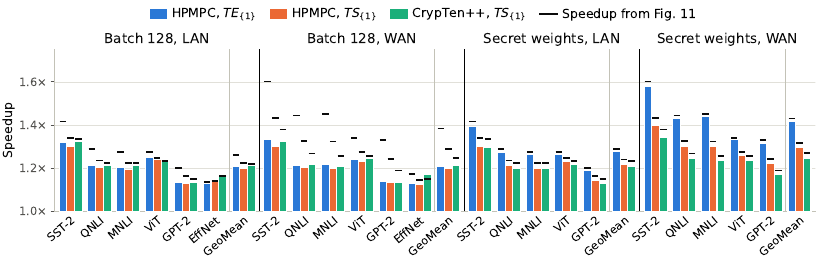}
    \caption{
    When the batch size increases (left) or the weights are encrypted (right), the speedup reduces from what was shown in Figure~\ref{fig:main_bars}.
    Black ticks show the speedup from Figure~\ref{fig:main_bars}, and the gap between the ticks and the bar shows how much speedup was lost with larger batch sizes or model encryption. 
    }
    \label{fig:ablation}
\end{figure}

\subsubsection{Larger Batch Size}
Figure~\ref{fig:ablation}(left) shows how the speedup changes relative to Figure~\ref{fig:main_bars} as we increase the batch size from 4 to 128. As the batch size increases, the communication round-trip latency is amortized anyways, and the benefit of \sys's auto-tuning decreases (from 1.14--1.42$\times$ to 1.13--1.33$\times$ on LAN, and from 1.14--1.60$\times$ to 1.13--1.32$\times$ on WAN). However, \sys's auto-tuner remains useful, improving performance by 20--21\% on average and up to 32\%.

\subsubsection{Encrypting Weights}
So far, our results encrypted only inputs, not weights. If we additionally encrypt weights, Conv2d and Linear layers become secret-by-secret multiplication instead of secret-by-public, and incur additional communications that our auto-tuner cannot improve. Figure~\ref{fig:ablation}(right) shows how the benefit changes with encrypted weights. As most of the overhead comes from approximating nonlinear layers (Section~\ref{sec:charac_approximation}), the speedup stays similar with or without model encryption. With encryption, \sys's auto-tuner still achieves, under light tuning (near iso-accuracy), 1.13--1.39$\times$ speedup on LAN and 1.17--1.58$\times$ on WAN.

\subsubsection{Distribution Shift}
\begin{figure}[t]
    \centering
    \includegraphics[width=\textwidth]{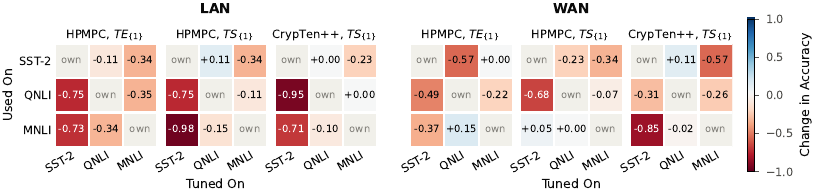}
    \caption{Reusing a tuned configuration on another GLUE task. Each cell is tuned for the task in its column to KL $\leq 0.005$ and evaluated on the task in its row. The reported numbers are changes in test accuracy compared to a setting tuned for the evaluated task with roughly the same latency.}
    \label{fig:transfer}
\end{figure}

We also study what happens when the input data distribution shifts after tuning the approximations. We evaluate a most extreme version of this: what happens if we tune the approximations for one GLUE dataset and use another GLUE dataset? Figure~\ref{fig:transfer} summarizes the result of running auto-tuning with KL divergence $\le 0.005$ as the criterion. 
Accuracy generally decreases when a knob configuration is used on a task it was not tuned for. The effect is not drastic and is within or close to noise in most cells.
However, approximations tuned for SST-2 did notably worse for QNLI and MNLI. This may be because SST-2 is an easier dataset than the other two~\cite{bert}, and the approximations might have been tuned more aggressively under the same budget.

\subsubsection{Network Changes}
Next, we study what happens when we use an approximation tuned for one network (\emph{e.g.}, LAN) for another network (\emph{e.g.}, WAN). This is an extreme version of studying the impact of network speed change during deployment, although no real network speed will change that radically (\emph{e.g.}, LAN to WAN). At KL divergence $< 0.005$, we observed that using approximations tuned for WAN on LAN changes overall latency by -1\% to +1.6\%. Similarly, using a setup tuned for LAN on WAN changes the latency by 0.5--3.8\%. We conclude that, while the auto-tuner accounts for LAN vs. WAN overhead, the dataset and workloads matter more in determining approximations.

\subsection{Performance Impact of Other Optimizations}
\label{sec:eval_extensibility}

While the paper's primary focus is on \sys's operator-approximation stage, \sys introduces additional optimizations at other compiler stages (float-to-int stage and backends).
Some have been explored in prior work (Section~\ref{sec:eval_hummingbird}), and their implementation on \sys is not novel in itself, but shows how \sys allows sharing common optimizations across multiple compatible backends.
Others are genuinely new to our knowledge (Section~\ref{sec:eval_backend_opt}) and is a small, orthogonal contribution.
We evaluate these additional optimizations.

\subsubsection{Additional Optimization in the Float-to-Int Stage (HummingBird~\cite{hummingbird})}
\label{sec:eval_hummingbird}

\begin{figure}
    \centering
    \begin{minipage}[t]{0.47\textwidth}
    \includegraphics[width=\textwidth]{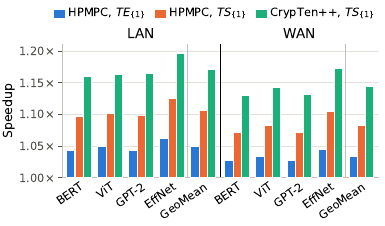}
    \caption{
        Speedup from applying the HummingBird~\cite{hummingbird} optimization (no auto-tuning applied). 
    }
    \label{fig:hummingbird}
    \end{minipage}%
    \hfill
    \begin{minipage}[t]{0.47\textwidth}
    \includegraphics[width=\textwidth]{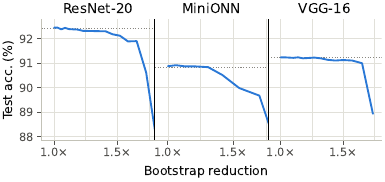}
    \caption{
    \sys's auto-tuner applied to CKKS FHE (Orion~\cite{orion}).
    }
    \label{fig:orion_frontier}
    \end{minipage}
\end{figure}

Additional optimizations can be added at the float-to-int stage of the compiler and shared with downstream backends. As an example, we implemented and evaluated the HummingBird~\cite{hummingbird} optimization from Section~\ref{sec:hummingbird}. 
Figure~\ref{fig:hummingbird} shows that our implementation of HummingBird alone improves performance by \textbf{1.04--1.20$\times$ on LAN} and \textbf{1.03--1.17$\times$ on WAN} on average. 
As the optimization saves more communicated bytes than rounds~\cite{hummingbird}, the benefit is slightly larger for LAN. Again, our prior results already contain this improvement even in their baseline numbers.

\subsubsection{Additional Optimization in the Backend}
\label{sec:eval_backend_opt}

We also evaluate the HPMPC-specific backend optimization explained in Section~\ref{sec:backend_additional_opts} on exact truncation ($\mathit{TE}_{\{1\}}$).
Our backend optimization, which chooses better adders based on the tensor shape and network speed, achieves \textbf{1.92$\times$ speedup on LAN} and \textbf{3.66$\times$ speedup on WAN} for BERT-base models. All of the prior results already include this optimization, even in the baselines.

\subsection{Programmability Benefits}
\label{sec:eval_extensibility_case_study}

We evaluate the practical benefit of using \sys over existing, fragmented frameworks.

\subsubsection{Supporting New Models}
\label{sec:adding_new_approx}

Both the original CrypTen~\cite{crypten} and HPMPC~\cite{hpmpc, pigeon} lacked support for the models we studied (Transformers and EfficientNet). Adding support for these models directly in these frameworks is not trivial because it requires changing many parts of their custom code. By adding these as a backend to \sys, \textbf{they can immediately support models like Transformers or EfficientNet}, by reusing the frontend and middle-end compiler stages. 
We argue that adding a new MPC runtime as a \sys backend provides an easier path to supporting a broad range of ML models than engineering each framework separately.

To illustrate why adding new model support to existing frameworks can be nontrivial, we present a case study on what it took to add Transformer support to CrypTen directly (what the CrypTen++~\cite{jinyu_mpc_cal} authors had to go through). 
To add LayerNorm support, Liu et al.~\cite{jinyu_mpc_cal} followed the maintainer's advice~\cite{crypten_layernorm} and modified two files in the CrypTen codebase (\texttt{nn/module.py} and \texttt{gradients.py}), adding 46 lines of code directly. Without the maintainer's help, this modification would not have been straightforward, and this effort doesn't transfer to other frameworks.
To add efficient GELU support, CrypTen's frontend needs to be modified, because its frontend immediately decomposes GELU into $\text{GELU}(x)=x\times\text{erf}(x)$ (erf is the Gauss error function).
This makes it hard to use faster approximations from BOLT~\cite{bolt}, which must be applied before GELU is decomposed.

For \sys, adding new support for LayerNorm or GELU is only 3 to 9 lines of code via our pass-writing interface, excluding the decorators. Our \sys prototype has \textbf{31 approximations} for \textbf{17 operators}. Excluding the decorators, the pass body is only \textbf{5.87 lines} on average. Most importantly, these are backend-agnostic and benefit multiple MPC backends.

\subsubsection{Writing New Optimizations}

In addition, \sys is built as an extension to PyTorch 2's compiler, and users can write additional optimizations with its popular pass-writing interface~\cite{export_ir_pass_tutorial}. This provides practical usability benefits for PyTorch users, unlike many existing frameworks that are implemented from scratch as custom C++ systems~\cite{spu, pigeon, hpmpc}. 
For example, implementing HummingBird in our float-to-int stage added only 43 lines of export IR pass code in Python.

\subsection{Generality Beyond MPC: an FHE Backend}
\label{sec:eval_fhe}

\begin{figure}
    \centering
    \includegraphics[width=0.95\linewidth]{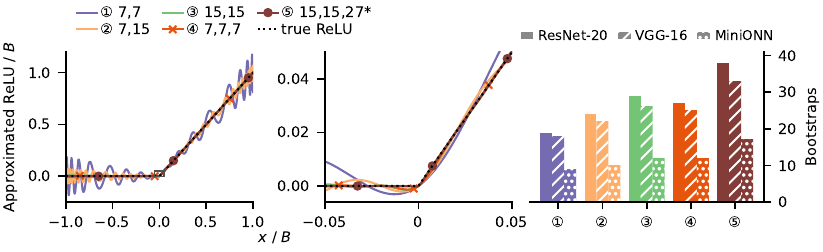}
    \caption{Various ReLU approximations from the auto-tuner (left), the area near zero zoomed-in (middle), and their relative performance (right). *Default of Orion~\cite{orion}.}
    \label{fig:fhe_relu_approx}
\end{figure}

Figure~\ref{fig:orion_frontier} shows the performance--accuracy trade-off obtained by reusing \sys's approximation auto-tuner with a CKKS FHE backend. We add Orion~\cite{orion} as a \sys backend and reuse \sys's frontend and approximation stages.
\sys's auto-tuner extends effectively to the CKKS FHE setting, reducing the number of bootstrapping operations by \textbf{1.46$\times$} at near-iso-accuracy on ResNet-20.
While this improvement is smaller than the 1.64$\times$ reduction reported by recent work that incorporates a tailored search strategy followed by post-search fine-tuning~\cite{autofhe}, we find the result meaningful because \sys uses a general-purpose auto-tuner and requires no fine-tuning.
Figure~\ref{fig:fhe_relu_approx} shows several ReLU approximations selected by our auto-tuner (left) and their relative performance (right). Each label denotes the sequence of polynomial degrees used in the composed approximation; for example, 15, 15, 27 indicates that the output of a degree-15 polynomial is fed into another degree-15 polynomial, whose output is then fed into a degree-27 polynomial.

\section{Additional Related Work}

\emph{\textbf{General MPC Compilers}}
General-purpose MPC compilers~\cite{silph, sok_mpc_compiler, emptoolkit, oblivc, oblivm, tinygarble, wysteria, sharemind, picco, aby, aby3, frigate, cbmc_gc, mp_spdz} differ from frameworks specific to those targeting ML inference in that they target an arbitrary (non-ML) program. 
These compilers usually target a smaller program (\emph{e.g.}, a function) and cannot scale to a large program, \emph{e.g.,} full Transformer inference code, and do not focus on approximating different ML operators in complex ML models. A recent general-purpose MPC framework, CryptDough~\cite{cryptdough}, has limited support for ML workloads, achieving competitive performance with Pigeon~\cite{pigeon} on small CNNs. However, it lacks many ML-specific optimizations (Pigeon still outperforms it under WAN) and, as other existing frameworks, does not support approximating operators needed for more complex ML models.
These are orthogonal to MPC-based ML frameworks like \sys, and MPC-based ML frameworks can use general MPC compilers to implement their lower-level MPC kernels.

\emph{\textbf{Additional Works on MPC-based ML.}}
There are several frameworks for MPC-based ML~\cite{crypten, cryptgpu, piranha, sigma, puma, falcon, trident, ariann, orca, hummingbird, gazelle, delphi, securenn, secureme, cheetah, bolt, sirnn, max_pir, cryptflow, cryptflow2, charmeleon, astra, blaze, flash, minionn, secureml, delphi, iron, pigeon, mpcpipe, hpmpc}.
They all implement unsupported ML operators with fixed, global approximations, and suffer from similar modularity/extensibility/debuggability issues.
The idea behind \sys applies to them as well.
Other works leveraged trusted hardware to accelerate MPC~\cite{stamp, ppmlac}, used quantization~\cite{ditto, quotient, xonn}, or built defenses for adversarial attacks~\cite{mpcdiff} for MPC-based ML systems. 

\emph{\textbf{Model Design for MPC-based ML.}}
Several works built tailored models for MPC-based ML~\cite{deepreduce, deepreshape, sphynx, mpcformer, salvit, mpcvit, secformer, snl, mpc_transformer, aespa, safenet, cryptonet, sisyphus, senet, truncformer, privit}.
These manually designed models usually require retraining with the approximated operators to achieve high accuracy, unlike \sys's auto-tuning, which does not require retraining.

\emph{\textbf{Approximating ReLU for FHE.}}
Many works address FHE's need to approximate ReLU, due to its key role in almost all ML tasks. Some~\cite{lola_fhe, smart_paf, paper_fhe, cryptonets, chou2018fastercryptonetsleveragingsparsity} replace ReLU with low-degree polynomials and design training methods to work around these activation functions' undesirable characteristics. Others~\cite{autofhe, ola_fhe} tuned their approximation based on sample inputs, customizing the approximation's degree and coefficients, and used post-approximation training to recover lost accuracy. While \sys mainly targets MPC runtimes, Section~\ref{sec:eval_fhe} shows it can achieve comparable benefits to \cite{autofhe} for FHE (achieving $\sim$18\% less, but \sys did not use auto-tuning).

\emph{\textbf{Alternatives for Private ML.}}
Trusted execution environment (TEE)~\cite{sgx, trustzone, amd-sev, keystone, aegis, xom, sp, hopper_confidential, graphcore_tee} is secure hardware that is practical and fast, but may have bugs and side channels that leak data~\cite{tee_bugs, coin_attack, tee_sidechannel}. 
Instance encoding~\cite{instahide_broken} adds statistical noise to hide the original data during ML training/inference. Some recent works have developed limited theoretical guarantees~\cite{liyue18, liyue19, maeng_fil2, pac_privacy, pac_privacy_followup}, but their privacy is much weaker than MPC.

\emph{\textbf{Neural Architecture Search (NAS).}} NAS~\cite{fbnet, mnasnet, micronet, bignasnet, oboyle_nas, proxylessnas, onceforall} automatically finds a model architecture for a given purpose, usually to lower the computational burden, while maintaining accuracy. Our approximation auto-tuner is conceptually similar to NAS, in that it automatically finds an ML model (by tuning the approximations) to maximize performance while roughly keeping the output quality.
However, NAS is usually done during training, while our approach works with a model that is already trained.

\section{Conclusion}

Existing MPC-based ML frameworks rely on a fixed set of transformations to map ML workloads onto MPC runtimes. These heuristically chosen transformations can hurt accuracy, performance, and usability.
In particular, globally fixed approximations for unsupported ML operators can introduce substantial overhead.
We design \sys, a multi-stage compiler with approximation auto-tuning that selects a set of approximations to maximize performance while preserving model quality.
Our evaluation shows that \sys improves performance across multiple MPC runtimes and also benefits an FHE runtime, demonstrating that multiple backends can effectively share the same frontend and middle-end.

\section*{AI Use}
ChatGPT and Claude Code were used to assist with code development, experiment execution, plot generation, and manuscript preparation. All AI-generated outputs were carefully reviewed and verified by the authors.

\section*{Acknowledgements}

This work was partly supported by the U.S. National Science Foundation under award No. CNS-2349610 and CCF-2529883. Any opinions, findings, and conclusions or recommendations expressed in this material are those of the author(s) and do not necessarily reflect the views of the National Science Foundation.

\bibliographystyle{ACM-Reference-Format}
\bibliography{refs}


\begin{thebibliography}{152}

%%% the Web address.

\ifx \showCODEN    \undefined \def \showCODEN     #1{\unskip}     \fi
\ifx \showISBNx    \undefined \def \showISBNx     #1{\unskip}     \fi
\ifx \showISBNxiii \undefined \def \showISBNxiii  #1{\unskip}     \fi
\ifx \showISSN     \undefined \def \showISSN      #1{\unskip}     \fi
\ifx \showLCCN     \undefined \def \showLCCN      #1{\unskip}     \fi
\ifx \shownote     \undefined \def \shownote      #1{#1}          \fi
\ifx \showarticletitle \undefined \def \showarticletitle #1{#1}   \fi
\ifx \showURL      \undefined \def \showURL       {\relax}        \fi
% invisible to TeX
\providecommand\bibfield[2]{#2}
\providecommand\bibinfo[2]{#2}
\providecommand\natexlab[1]{#1}
\providecommand\showeprint[2][]{arXiv:#2}

\bibitem[Agrawal et~al\mbox{.}(2019)]%
        {quotient}
\bibfield{author}{\bibinfo{person}{Nitin Agrawal}, \bibinfo{person}{Ali~Shahin Shamsabadi}, \bibinfo{person}{Matt~J. Kusner}, {and} \bibinfo{person}{Adri{\`{a}} Gasc{\'{o}}n}.} \bibinfo{year}{2019}\natexlab{}.
\newblock \showarticletitle{{QUOTIENT:} Two-Party Secure Neural Network Training and Prediction}. In \bibinfo{booktitle}{\emph{Proceedings of the 2019 {ACM} {SIGSAC} Conference on Computer and Communications Security, {CCS} 2019, London, UK, November 11-15, 2019}}. \bibinfo{publisher}{{ACM}}, \bibinfo{pages}{1231--1247}.
\newblock
\href{https://doi.org/10.1145/3319535.3339819}{doi:\nolinkurl{10.1145/3319535.3339819}}


\bibitem[Amazon(2023)]%
        {alexa}
\bibfield{author}{\bibinfo{person}{Amazon}.} \bibinfo{year}{2023}\natexlab{}.
\newblock \bibinfo{title}{Amazon Echo \& Alexa Devices}.
\newblock \bibinfo{howpublished}{\url{https://www.amazon.com/smart-home-devices/b?ie=UTF8&node=9818047011}}.
\newblock


\bibitem[{AMD}(2023)]%
        {amd-sev}
\bibfield{author}{\bibinfo{person}{{AMD}}.} \bibinfo{year}{2023}\natexlab{}.
\newblock \bibinfo{title}{{AMD Secure Encrypted Virtualization (SEV)}}.
\newblock \bibinfo{howpublished}{\url{https://www.amd.com/en/developer/sev.html}}.
\newblock


\bibitem[anakinxc(2025)]%
        {spu_hack}
\bibfield{author}{\bibinfo{person}{anakinxc}.} \bibinfo{year}{2025}\natexlab{}.
\newblock \bibinfo{title}{"hack\_softmax" (spu/examples/python/ml/flax\_llama7b/flax\_llama7b.py, Line 56)}.
\newblock \bibinfo{howpublished}{\url{https://github.com/secretflow/spu/blob/e86227631b115f5983aef8315adde387e0c0f63a/examples/python/ml/flax_llama7b/flax_llama7b.py\#L56}}.
\newblock


\bibitem[Ansel et~al\mbox{.}(2024)]%
        {pytorch2}
\bibfield{author}{\bibinfo{person}{Jason Ansel}, \bibinfo{person}{Edward~Z. Yang}, \bibinfo{person}{Horace He}, \bibinfo{person}{Natalia Gimelshein}, \bibinfo{person}{Animesh Jain}, \bibinfo{person}{Michael Voznesensky}, \bibinfo{person}{Bin Bao}, \bibinfo{person}{Peter Bell}, \bibinfo{person}{David Berard}, \bibinfo{person}{Evgeni Burovski}, \bibinfo{person}{Geeta Chauhan}, \bibinfo{person}{Anjali Chourdia}, \bibinfo{person}{Will Constable}, \bibinfo{person}{Alban Desmaison}, \bibinfo{person}{Zachary DeVito}, \bibinfo{person}{Elias Ellison}, \bibinfo{person}{Will Feng}, \bibinfo{person}{Jiong Gong}, \bibinfo{person}{Michael Gschwind}, \bibinfo{person}{Brian Hirsh}, \bibinfo{person}{Sherlock Huang}, \bibinfo{person}{Kshiteej Kalambarkar}, \bibinfo{person}{Laurent Kirsch}, \bibinfo{person}{Michael Lazos}, \bibinfo{person}{Mario Lezcano}, \bibinfo{person}{Yanbo Liang}, \bibinfo{person}{Jason Liang}, \bibinfo{person}{Yinghai Lu}, \bibinfo{person}{C.~K. Luk}, \bibinfo{person}{Bert Maher}, \bibinfo{person}{Yunjie Pan}, \bibinfo{person}{Christian Puhrsch}, \bibinfo{person}{Matthias Reso}, \bibinfo{person}{Mark Saroufim}, \bibinfo{person}{Marcos~Yukio Siraichi}, \bibinfo{person}{Helen Suk}, \bibinfo{person}{Shunting Zhang}, \bibinfo{person}{Michael Suo}, \bibinfo{person}{Phil Tillet}, \bibinfo{person}{Xu Zhao}, \bibinfo{person}{Eikan Wang}, \bibinfo{person}{Keren Zhou}, \bibinfo{person}{Richard Zou}, \bibinfo{person}{Xiaodong Wang}, \bibinfo{person}{Ajit Mathews}, \bibinfo{person}{William Wen}, \bibinfo{person}{Gregory Chanan}, \bibinfo{person}{Peng Wu}, {and} \bibinfo{person}{Soumith Chintala}.} \bibinfo{year}{2024}\natexlab{}.
\newblock \showarticletitle{PyTorch 2: Faster Machine Learning Through Dynamic Python Bytecode Transformation and Graph Compilation}. In \bibinfo{booktitle}{\emph{Proceedings of the 29th {ACM} International Conference on Architectural Support for Programming Languages and Operating Systems, Volume 2, {ASPLOS} 2024, La Jolla, CA, USA, 27 April 2024- 1 May 2024}}. \bibinfo{publisher}{{ACM}}, \bibinfo{pages}{929--947}.
\newblock
\href{https://doi.org/10.1145/3620665.3640366}{doi:\nolinkurl{10.1145/3620665.3640366}}


\bibitem[Ao and Boddeti(2024)]%
        {autofhe}
\bibfield{author}{\bibinfo{person}{Wei Ao} {and} \bibinfo{person}{Vishnu~Naresh Boddeti}.} \bibinfo{year}{2024}\natexlab{}.
\newblock \showarticletitle{{AutoFHE}: Automated Adaption of {CNNs} for Efficient Evaluation over {FHE}}. In \bibinfo{booktitle}{\emph{33rd USENIX Security Symposium (USENIX Security 24)}}. \bibinfo{publisher}{USENIX Association}, \bibinfo{address}{Philadelphia, PA}, \bibinfo{pages}{2173--2190}.
\newblock
\showISBNx{978-1-939133-44-1}
\urldef\tempurl%
\url{https://www.usenix.org/conference/usenixsecurity24/presentation/ao}
\showURL{%
\tempurl}


\bibitem[Arm(2023)]%
        {trustzone}
\bibfield{author}{\bibinfo{person}{Arm}.} \bibinfo{year}{2023}\natexlab{}.
\newblock \bibinfo{title}{TrustZone for Cortex-A}.
\newblock \bibinfo{howpublished}{\url{https://www.arm.com/technologies/trustzone-for-cortex-a}}.
\newblock


\bibitem[Bogdanov et~al\mbox{.}(2008)]%
        {sharemind}
\bibfield{author}{\bibinfo{person}{Dan Bogdanov}, \bibinfo{person}{Sven Laur}, {and} \bibinfo{person}{Jan Willemson}.} \bibinfo{year}{2008}\natexlab{}.
\newblock \showarticletitle{Sharemind: {A} Framework for Fast Privacy-Preserving Computations}. In \bibinfo{booktitle}{\emph{Computer Security - {ESORICS} 2008, 13th European Symposium on Research in Computer Security, M{\'{a}}laga, Spain, October 6-8, 2008. Proceedings}} \emph{(\bibinfo{series}{Lecture Notes in Computer Science}, Vol.~\bibinfo{volume}{5283})}. \bibinfo{publisher}{Springer}, \bibinfo{pages}{192--206}.
\newblock
\href{https://doi.org/10.1007/978-3-540-88313-5\_13}{doi:\nolinkurl{10.1007/978-3-540-88313-5\_13}}


\bibitem[Brutzkus et~al\mbox{.}(2019)]%
        {lola_fhe}
\bibfield{author}{\bibinfo{person}{Alon Brutzkus}, \bibinfo{person}{Oren Elisha}, {and} \bibinfo{person}{Ran Gilad-Bachrach}.} \bibinfo{year}{2019}\natexlab{}.
\newblock \bibinfo{title}{Low Latency Privacy Preserving Inference}.
\newblock
\showeprint[arxiv]{1812.10659}~[cs.LG]
\urldef\tempurl%
\url{https://arxiv.org/abs/1812.10659}
\showURL{%
\tempurl}


\bibitem[Byali et~al\mbox{.}(2020)]%
        {flash}
\bibfield{author}{\bibinfo{person}{Megha Byali}, \bibinfo{person}{Harsh Chaudhari}, \bibinfo{person}{Arpita Patra}, {and} \bibinfo{person}{Ajith Suresh}.} \bibinfo{year}{2020}\natexlab{}.
\newblock \showarticletitle{{FLASH:} Fast and Robust Framework for Privacy-preserving Machine Learning}.
\newblock \bibinfo{journal}{\emph{Proc. Priv. Enhancing Technol.}} \bibinfo{volume}{2020}, \bibinfo{number}{2} (\bibinfo{year}{2020}), \bibinfo{pages}{459--480}.
\newblock
\href{https://doi.org/10.2478/popets-2020-0036}{doi:\nolinkurl{10.2478/popets-2020-0036}}


\bibitem[Cai et~al\mbox{.}(2020)]%
        {onceforall}
\bibfield{author}{\bibinfo{person}{Han Cai}, \bibinfo{person}{Chuang Gan}, \bibinfo{person}{Tianzhe Wang}, \bibinfo{person}{Zhekai Zhang}, {and} \bibinfo{person}{Song Han}.} \bibinfo{year}{2020}\natexlab{}.
\newblock \showarticletitle{Once-for-All: Train One Network and Specialize it for Efficient Deployment}. In \bibinfo{booktitle}{\emph{8th International Conference on Learning Representations, {ICLR} 2020, Addis Ababa, Ethiopia, April 26-30, 2020}}. \bibinfo{publisher}{OpenReview.net}.
\newblock
\urldef\tempurl%
\url{https://openreview.net/forum?id=HylxE1HKwS}
\showURL{%
\tempurl}


\bibitem[Cai et~al\mbox{.}(2019)]%
        {proxylessnas}
\bibfield{author}{\bibinfo{person}{Han Cai}, \bibinfo{person}{Ligeng Zhu}, {and} \bibinfo{person}{Song Han}.} \bibinfo{year}{2019}\natexlab{}.
\newblock \showarticletitle{Proxyless{NAS}: Direct Neural Architecture Search on Target Task and Hardware}. In \bibinfo{booktitle}{\emph{International Conference on Learning Representations}}.
\newblock
\urldef\tempurl%
\url{https://openreview.net/forum?id=HylVB3AqYm}
\showURL{%
\tempurl}


\bibitem[Carlini et~al\mbox{.}(2020)]%
        {instahide_broken}
\bibfield{author}{\bibinfo{person}{Nicholas Carlini}, \bibinfo{person}{Samuel Deng}, \bibinfo{person}{Sanjam Garg}, \bibinfo{person}{Somesh Jha}, \bibinfo{person}{Saeed Mahloujifar}, \bibinfo{person}{Mohammad Mahmoody}, \bibinfo{person}{Shuang Song}, \bibinfo{person}{Abhradeep Thakurta}, {and} \bibinfo{person}{Florian Tram{\`{e}}r}.} \bibinfo{year}{2020}\natexlab{}.
\newblock \showarticletitle{An Attack on InstaHide: Is Private Learning Possible with Instance Encoding?}
\newblock \bibinfo{journal}{\emph{CoRR}}  \bibinfo{volume}{abs/2011.05315} (\bibinfo{year}{2020}).
\newblock
\showeprint[arXiv]{2011.05315}
\urldef\tempurl%
\url{https://arxiv.org/abs/2011.05315}
\showURL{%
\tempurl}


\bibitem[Chaudhari et~al\mbox{.}(2019)]%
        {astra}
\bibfield{author}{\bibinfo{person}{Harsh Chaudhari}, \bibinfo{person}{Ashish Choudhury}, \bibinfo{person}{Arpita Patra}, {and} \bibinfo{person}{Ajith Suresh}.} \bibinfo{year}{2019}\natexlab{}.
\newblock \showarticletitle{ASTRA: High Throughput 3PC over Rings with Application to Secure Prediction}. In \bibinfo{booktitle}{\emph{Proceedings of the 2019 ACM SIGSAC Conference on Cloud Computing Security Workshop}} (London, United Kingdom) \emph{(\bibinfo{series}{CCSW'19})}. \bibinfo{publisher}{Association for Computing Machinery}, \bibinfo{address}{New York, NY, USA}, \bibinfo{pages}{81–92}.
\newblock
\showISBNx{9781450368261}
\href{https://doi.org/10.1145/3338466.3358922}{doi:\nolinkurl{10.1145/3338466.3358922}}


\bibitem[Chaudhari et~al\mbox{.}(2020)]%
        {trident}
\bibfield{author}{\bibinfo{person}{Harsh Chaudhari}, \bibinfo{person}{Rahul Rachuri}, {and} \bibinfo{person}{Ajith Suresh}.} \bibinfo{year}{2020}\natexlab{}.
\newblock \showarticletitle{Trident: Efficient 4PC Framework for Privacy Preserving Machine Learning}. In \bibinfo{booktitle}{\emph{27th Annual Network and Distributed System Security Symposium, {NDSS} 2020, San Diego, California, USA, February 23-26, 2020}}. \bibinfo{publisher}{The Internet Society}.
\newblock
\urldef\tempurl%
\url{https://www.ndss-symposium.org/ndss-paper/trident-efficient-4pc-framework-for-privacy-preserving-machine-learning/}
\showURL{%
\tempurl}


\bibitem[Chen et~al\mbox{.}(2023)]%
        {silph}
\bibfield{author}{\bibinfo{person}{Edward Chen}, \bibinfo{person}{Jinhao Zhu}, \bibinfo{person}{Alex Ozdemir}, \bibinfo{person}{Riad~S. Wahby}, \bibinfo{person}{Fraser Brown}, {and} \bibinfo{person}{Wenting Zheng}.} \bibinfo{year}{2023}\natexlab{}.
\newblock \showarticletitle{Silph: {A} Framework for Scalable and Accurate Generation of Hybrid {MPC} Protocols}. In \bibinfo{booktitle}{\emph{44th {IEEE} Symposium on Security and Privacy, {SP} 2023, San Francisco, CA, USA, May 21-25, 2023}}. \bibinfo{publisher}{{IEEE}}, \bibinfo{pages}{848--863}.
\newblock
\href{https://doi.org/10.1109/SP46215.2023.10179397}{doi:\nolinkurl{10.1109/SP46215.2023.10179397}}


\bibitem[Cheon et~al\mbox{.}(2017)]%
        {ckks}
\bibfield{author}{\bibinfo{person}{Jung~Hee Cheon}, \bibinfo{person}{Andrey Kim}, \bibinfo{person}{Miran Kim}, {and} \bibinfo{person}{Yongsoo Song}.} \bibinfo{year}{2017}\natexlab{}.
\newblock \showarticletitle{Homomorphic Encryption for Arithmetic of Approximate Numbers}. In \bibinfo{booktitle}{\emph{Advances in Cryptology -- {ASIACRYPT} 2017}} \emph{(\bibinfo{series}{Lecture Notes in Computer Science}, Vol.~\bibinfo{volume}{10624})}. \bibinfo{publisher}{Springer}, \bibinfo{pages}{409--437}.
\newblock
\href{https://doi.org/10.1007/978-3-319-70694-8_15}{doi:\nolinkurl{10.1007/978-3-319-70694-8_15}}


\bibitem[Cheon et~al\mbox{.}(2024)]%
        {dacapo}
\bibfield{author}{\bibinfo{person}{Seonyoung Cheon}, \bibinfo{person}{Yongwoo Lee}, \bibinfo{person}{Dongkwan Kim}, \bibinfo{person}{Ju~Min Lee}, \bibinfo{person}{Sunchul Jung}, \bibinfo{person}{Taekyung Kim}, \bibinfo{person}{Dongyoon Lee}, {and} \bibinfo{person}{Hanjun Kim}.} \bibinfo{year}{2024}\natexlab{}.
\newblock \showarticletitle{$\{$DaCapo$\}$: Automatic Bootstrapping Management for Efficient Fully Homomorphic Encryption}. In \bibinfo{booktitle}{\emph{33rd USENIX Security Symposium (USENIX Security 24)}}. \bibinfo{pages}{6993--7010}.
\newblock


\bibitem[Cheon et~al\mbox{.}(2025)]%
        {halo}
\bibfield{author}{\bibinfo{person}{Seonyoung Cheon}, \bibinfo{person}{Yongwoo Lee}, \bibinfo{person}{Hoyun Youm}, \bibinfo{person}{Dongkwan Kim}, \bibinfo{person}{Sungwoo Yun}, \bibinfo{person}{Kunmo Jeong}, \bibinfo{person}{Dongyoon Lee}, {and} \bibinfo{person}{Hanjun Kim}.} \bibinfo{year}{2025}\natexlab{}.
\newblock \showarticletitle{HALO: Loop-aware Bootstrapping Management for Fully Homomorphic Encryption}. In \bibinfo{booktitle}{\emph{Proceedings of the 30th ACM International Conference on Architectural Support for Programming Languages and Operating Systems, Volume 1}} (Rotterdam, Netherlands) \emph{(\bibinfo{series}{ASPLOS '25})}. \bibinfo{publisher}{Association for Computing Machinery}, \bibinfo{address}{New York, NY, USA}, \bibinfo{pages}{572–585}.
\newblock
\showISBNx{9798400706981}
\href{https://doi.org/10.1145/3669940.3707275}{doi:\nolinkurl{10.1145/3669940.3707275}}


\bibitem[Chhabra et~al\mbox{.}(2011)]%
        {secureme}
\bibfield{author}{\bibinfo{person}{Siddhartha Chhabra}, \bibinfo{person}{Brian Rogers}, \bibinfo{person}{Yan Solihin}, {and} \bibinfo{person}{Milos Prvulovic}.} \bibinfo{year}{2011}\natexlab{}.
\newblock \showarticletitle{SecureME: a hardware-software approach to full system security}. In \bibinfo{booktitle}{\emph{Proceedings of the 25th International Conference on Supercomputing, 2011, Tucson, AZ, USA, May 31 - June 04, 2011}}, \bibfield{editor}{\bibinfo{person}{David~K. Lowenthal}, \bibinfo{person}{Bronis~R. de~Supinski}, {and} \bibinfo{person}{Sally~A. McKee}} (Eds.). \bibinfo{publisher}{{ACM}}, \bibinfo{pages}{108--119}.
\newblock
\href{https://doi.org/10.1145/1995896.1995914}{doi:\nolinkurl{10.1145/1995896.1995914}}


\bibitem[Chielle et~al\mbox{.}(2026)]%
        {paper_fhe}
\bibfield{author}{\bibinfo{person}{Eduardo Chielle}, \bibinfo{person}{Manaar Alam}, \bibinfo{person}{Jinting Liu}, \bibinfo{person}{Jovan Kascelan}, {and} \bibinfo{person}{Michail Maniatakos}.} \bibinfo{year}{2026}\natexlab{}.
\newblock \bibinfo{title}{PAPER: Privacy-Preserving Convolutional Neural Networks using Low-Degree Polynomial Approximations and Structural Optimizations on Leveled FHE}.
\newblock
\showeprint[arxiv]{2509.22857}~[cs.CR]
\urldef\tempurl%
\url{https://arxiv.org/abs/2509.22857}
\showURL{%
\tempurl}


\bibitem[Cho et~al\mbox{.}(2022a)]%
        {sphynx}
\bibfield{author}{\bibinfo{person}{Minsu Cho}, \bibinfo{person}{Zahra Ghodsi}, \bibinfo{person}{Brandon Reagen}, \bibinfo{person}{Siddharth Garg}, {and} \bibinfo{person}{Chinmay Hegde}.} \bibinfo{year}{2022}\natexlab{a}.
\newblock \showarticletitle{Sphynx: {A} Deep Neural Network Design for Private Inference}.
\newblock \bibinfo{journal}{\emph{{IEEE} Secur. Priv.}} \bibinfo{volume}{20}, \bibinfo{number}{5} (\bibinfo{year}{2022}), \bibinfo{pages}{22--34}.
\newblock
\href{https://doi.org/10.1109/MSEC.2022.3165475}{doi:\nolinkurl{10.1109/MSEC.2022.3165475}}


\bibitem[Cho et~al\mbox{.}(2022b)]%
        {snl}
\bibfield{author}{\bibinfo{person}{Minsu Cho}, \bibinfo{person}{Ameya Joshi}, \bibinfo{person}{Brandon Reagen}, \bibinfo{person}{Siddharth Garg}, {and} \bibinfo{person}{Chinmay Hegde}.} \bibinfo{year}{2022}\natexlab{b}.
\newblock \showarticletitle{Selective Network Linearization for Efficient Private Inference}. In \bibinfo{booktitle}{\emph{International Conference on Machine Learning, {ICML} 2022, 17-23 July 2022, Baltimore, Maryland, {USA}}} \emph{(\bibinfo{series}{Proceedings of Machine Learning Research}, Vol.~\bibinfo{volume}{162})}, \bibfield{editor}{\bibinfo{person}{Kamalika Chaudhuri}, \bibinfo{person}{Stefanie Jegelka}, \bibinfo{person}{Le~Song}, \bibinfo{person}{Csaba Szepesv{\'{a}}ri}, \bibinfo{person}{Gang Niu}, {and} \bibinfo{person}{Sivan Sabato}} (Eds.). \bibinfo{publisher}{{PMLR}}, \bibinfo{pages}{3947--3961}.
\newblock
\urldef\tempurl%
\url{https://proceedings.mlr.press/v162/cho22a.html}
\showURL{%
\tempurl}


\bibitem[Choi et~al\mbox{.}(2025)]%
        {cheddar}
\bibfield{author}{\bibinfo{person}{Wonseok Choi}, \bibinfo{person}{Jongmin Kim}, {and} \bibinfo{person}{Jung~Ho Ahn}.} \bibinfo{year}{2025}\natexlab{}.
\newblock \showarticletitle{Cheddar: A Swift Fully Homomorphic Encryption Library Designed for GPU Architectures}.
\newblock \bibinfo{journal}{\emph{arXiv preprint arXiv:2407.13055}} (\bibinfo{year}{2025}).
\newblock


\bibitem[Chou et~al\mbox{.}(2018)]%
        {chou2018fastercryptonetsleveragingsparsity}
\bibfield{author}{\bibinfo{person}{Edward Chou}, \bibinfo{person}{Josh Beal}, \bibinfo{person}{Daniel Levy}, \bibinfo{person}{Serena Yeung}, \bibinfo{person}{Albert Haque}, {and} \bibinfo{person}{Li Fei-Fei}.} \bibinfo{year}{2018}\natexlab{}.
\newblock \bibinfo{title}{Faster CryptoNets: Leveraging Sparsity for Real-World Encrypted Inference}.
\newblock
\showeprint[arxiv]{1811.09953}~[cs.CR]
\urldef\tempurl%
\url{https://arxiv.org/abs/1811.09953}
\showURL{%
\tempurl}


\bibitem[De~Castro et~al\mbox{.}(2025)]%
        {encryptedllm}
\bibfield{author}{\bibinfo{person}{Leo De~Castro}, \bibinfo{person}{Daniel Escudero}, \bibinfo{person}{Adya Agrawal}, \bibinfo{person}{Antigoni Polychroniadou}, {and} \bibinfo{person}{Manuela Veloso}.} \bibinfo{year}{2025}\natexlab{}.
\newblock \showarticletitle{EncryptedLLM: Privacy-Preserving Large Language Model Inference via GPU-Accelerated Fully Homomorphic Encryption.}. In \bibinfo{booktitle}{\emph{ICML}}.
\newblock


\bibitem[Demmler et~al\mbox{.}(2015)]%
        {aby}
\bibfield{author}{\bibinfo{person}{Daniel Demmler}, \bibinfo{person}{Thomas Schneider}, {and} \bibinfo{person}{Michael Zohner}.} \bibinfo{year}{2015}\natexlab{}.
\newblock \showarticletitle{{ABY} - {A} Framework for Efficient Mixed-Protocol Secure Two-Party Computation}. In \bibinfo{booktitle}{\emph{22nd Annual Network and Distributed System Security Symposium, {NDSS} 2015, San Diego, California, USA, February 8-11, 2015}}. \bibinfo{publisher}{The Internet Society}.
\newblock
\urldef\tempurl%
\url{https://www.ndss-symposium.org/ndss2015/aby---framework-efficient-mixed-protocol-secure-two-party-computation}
\showURL{%
\tempurl}


\bibitem[Deng et~al\mbox{.}(2009)]%
        {imagenet}
\bibfield{author}{\bibinfo{person}{Jia Deng}, \bibinfo{person}{Wei Dong}, \bibinfo{person}{Richard Socher}, \bibinfo{person}{Li{-}Jia Li}, \bibinfo{person}{Kai Li}, {and} \bibinfo{person}{Li Fei{-}Fei}.} \bibinfo{year}{2009}\natexlab{}.
\newblock \showarticletitle{ImageNet: {A} large-scale hierarchical image database}. In \bibinfo{booktitle}{\emph{2009 {IEEE} Computer Society Conference on Computer Vision and Pattern Recognition {(CVPR} 2009), 20-25 June 2009, Miami, Florida, {USA}}}. \bibinfo{publisher}{{IEEE} Computer Society}, \bibinfo{pages}{248--255}.
\newblock
\href{https://doi.org/10.1109/CVPR.2009.5206848}{doi:\nolinkurl{10.1109/CVPR.2009.5206848}}


\bibitem[Devlin et~al\mbox{.}(2019)]%
        {bert}
\bibfield{author}{\bibinfo{person}{Jacob Devlin}, \bibinfo{person}{Ming{-}Wei Chang}, \bibinfo{person}{Kenton Lee}, {and} \bibinfo{person}{Kristina Toutanova}.} \bibinfo{year}{2019}\natexlab{}.
\newblock \showarticletitle{{BERT:} Pre-training of Deep Bidirectional Transformers for Language Understanding}. In \bibinfo{booktitle}{\emph{Proceedings of the 2019 Conference of the North American Chapter of the Association for Computational Linguistics: Human Language Technologies, {NAACL-HLT} 2019, Minneapolis, MN, USA, June 2-7, 2019, Volume 1 (Long and Short Papers)}}. \bibinfo{publisher}{Association for Computational Linguistics}, \bibinfo{pages}{4171--4186}.
\newblock
\href{https://doi.org/10.18653/v1/n19-1423}{doi:\nolinkurl{10.18653/v1/n19-1423}}


\bibitem[Dhyani et~al\mbox{.}(2023)]%
        {privit}
\bibfield{author}{\bibinfo{person}{Naren Dhyani}, \bibinfo{person}{Jianqiao Mo}, \bibinfo{person}{Minsu Cho}, \bibinfo{person}{Ameya Joshi}, \bibinfo{person}{Siddharth Garg}, \bibinfo{person}{Brandon Reagen}, {and} \bibinfo{person}{Chinmay Hegde}.} \bibinfo{year}{2023}\natexlab{}.
\newblock \showarticletitle{PriViT: Vision Transformers for Fast Private Inference}.
\newblock \bibinfo{journal}{\emph{CoRR}}  \bibinfo{volume}{abs/2310.04604} (\bibinfo{year}{2023}).
\newblock
\showeprint[arXiv]{2310.04604}
\href{https://doi.org/10.48550/ARXIV.2310.04604}{doi:\nolinkurl{10.48550/ARXIV.2310.04604}}


\bibitem[Dong et~al\mbox{.}(2023)]%
        {puma}
\bibfield{author}{\bibinfo{person}{Ye Dong}, \bibinfo{person}{Wen{-}jie Lu}, \bibinfo{person}{Yancheng Zheng}, \bibinfo{person}{Haoqi Wu}, \bibinfo{person}{Derun Zhao}, \bibinfo{person}{Jin Tan}, \bibinfo{person}{Zhicong Huang}, \bibinfo{person}{Cheng Hong}, \bibinfo{person}{Tao Wei}, {and} \bibinfo{person}{Wenguang Chen}.} \bibinfo{year}{2023}\natexlab{}.
\newblock \showarticletitle{{PUMA:} Secure Inference of LLaMA-7B in Five Minutes}.
\newblock \bibinfo{journal}{\emph{CoRR}}  \bibinfo{volume}{abs/2307.12533} (\bibinfo{year}{2023}).
\newblock
\showeprint[arXiv]{2307.12533}
\href{https://doi.org/10.48550/ARXIV.2307.12533}{doi:\nolinkurl{10.48550/ARXIV.2307.12533}}


\bibitem[Dosovitskiy et~al\mbox{.}(2021)]%
        {vit}
\bibfield{author}{\bibinfo{person}{Alexey Dosovitskiy}, \bibinfo{person}{Lucas Beyer}, \bibinfo{person}{Alexander Kolesnikov}, \bibinfo{person}{Dirk Weissenborn}, \bibinfo{person}{Xiaohua Zhai}, \bibinfo{person}{Thomas Unterthiner}, \bibinfo{person}{Mostafa Dehghani}, \bibinfo{person}{Matthias Minderer}, \bibinfo{person}{Georg Heigold}, \bibinfo{person}{Sylvain Gelly}, \bibinfo{person}{Jakob Uszkoreit}, {and} \bibinfo{person}{Neil Houlsby}.} \bibinfo{year}{2021}\natexlab{}.
\newblock \showarticletitle{An Image is Worth 16x16 Words: Transformers for Image Recognition at Scale}. In \bibinfo{booktitle}{\emph{9th International Conference on Learning Representations, {ICLR} 2021, Virtual Event, Austria, May 3-7, 2021}}. \bibinfo{publisher}{OpenReview.net}.
\newblock
\urldef\tempurl%
\url{https://openreview.net/forum?id=YicbFdNTTy}
\showURL{%
\tempurl}


\bibitem[Dowlin et~al\mbox{.}(2016)]%
        {cryptonets}
\bibfield{author}{\bibinfo{person}{Nathan Dowlin}, \bibinfo{person}{Ran Gilad-Bachrach}, \bibinfo{person}{Kim Laine}, \bibinfo{person}{Kristin Lauter}, \bibinfo{person}{Michael Naehrig}, {and} \bibinfo{person}{John Wernsing}.} \bibinfo{year}{2016}\natexlab{}.
\newblock \showarticletitle{CryptoNets: applying neural networks to encrypted data with high throughput and accuracy}. In \bibinfo{booktitle}{\emph{Proceedings of the 33rd International Conference on International Conference on Machine Learning - Volume 48}} (New York, NY, USA) \emph{(\bibinfo{series}{ICML'16})}. \bibinfo{publisher}{JMLR.org}, \bibinfo{pages}{201–210}.
\newblock


\bibitem[Ebel et~al\mbox{.}(2025)]%
        {orion}
\bibfield{author}{\bibinfo{person}{Austin Ebel}, \bibinfo{person}{Karthik Garimella}, {and} \bibinfo{person}{Brandon Reagen}.} \bibinfo{year}{2025}\natexlab{}.
\newblock \showarticletitle{Orion: A Fully Homomorphic Encryption Framework for Deep Learning}. In \bibinfo{booktitle}{\emph{Proceedings of the 30th ACM International Conference on Architectural Support for Programming Languages and Operating Systems, Volume 2}} (Rotterdam, Netherlands) \emph{(\bibinfo{series}{ASPLOS '25})}. \bibinfo{publisher}{Association for Computing Machinery}, \bibinfo{address}{New York, NY, USA}, \bibinfo{pages}{734–749}.
\newblock
\showISBNx{9798400710797}
\href{https://doi.org/10.1145/3676641.3716008}{doi:\nolinkurl{10.1145/3676641.3716008}}


\bibitem[facebookresearch(2025)]%
        {crypten_conversion_bug}
\bibfield{author}{\bibinfo{person}{facebookresearch}.} \bibinfo{year}{2025}\natexlab{}.
\newblock \bibinfo{title}{CrypTen/crypten/mpc/primitives/arithmetic.py, Line 362}.
\newblock \bibinfo{howpublished}{\url{https://github.com/facebookresearch/CrypTen/blob/775868a02d6dac50774ce376a55b01fbd8bd85b6/crypten/mpc/primitives/arithmetic.py\#L362}}.
\newblock


\bibitem[Faisal et~al\mbox{.}(2026)]%
        {cryptdough}
\bibfield{author}{\bibinfo{person}{Muhammad Faisal}, \bibinfo{person}{Alessandra Lanz}, \bibinfo{person}{Sam Buxbaum}, \bibinfo{person}{Adam Godel}, \bibinfo{person}{Vasiliki Kalavri}, \bibinfo{person}{Mayank Varia}, {and} \bibinfo{person}{John Liagouris}.} \bibinfo{year}{2026}\natexlab{}.
\newblock \showarticletitle{CryptDough: A Unified Analytics Engine for Secure Multiparty Computation}. In \bibinfo{booktitle}{\emph{Proceedings of the ACM SIGOPS 32nd Symposium on Operating Systems Principles}} (Prague, Czech Republic) \emph{(\bibinfo{series}{SOSP '26})}. \bibinfo{publisher}{Association for Computing Machinery}, \bibinfo{address}{New York, NY, USA}, \bibinfo{pages}{145–166}.
\newblock
\showISBNx{9798400725852}
\href{https://doi.org/10.1145/3830418.3843909}{doi:\nolinkurl{10.1145/3830418.3843909}}


\bibitem[Fan(2018)]%
        {liyue18}
\bibfield{author}{\bibinfo{person}{Liyue Fan}.} \bibinfo{year}{2018}\natexlab{}.
\newblock \showarticletitle{Image Pixelization with Differential Privacy}. In \bibinfo{booktitle}{\emph{Data and Applications Security and Privacy {XXXII} - 32nd Annual {IFIP} {WG} 11.3 Conference, DBSec 2018, Bergamo, Italy, July 16-18, 2018, Proceedings}} \emph{(\bibinfo{series}{Lecture Notes in Computer Science}, Vol.~\bibinfo{volume}{10980})}, \bibfield{editor}{\bibinfo{person}{Florian Kerschbaum} {and} \bibinfo{person}{Stefano Paraboschi}} (Eds.). \bibinfo{publisher}{Springer}, \bibinfo{pages}{148--162}.
\newblock
\href{https://doi.org/10.1007/978-3-319-95729-6\_10}{doi:\nolinkurl{10.1007/978-3-319-95729-6\_10}}


\bibitem[Fan(2019)]%
        {liyue19}
\bibfield{author}{\bibinfo{person}{Liyue Fan}.} \bibinfo{year}{2019}\natexlab{}.
\newblock \showarticletitle{Differential privacy for image publication}. In \bibinfo{booktitle}{\emph{Theory and Practice of Differential Privacy (TPDP) Workshop}}, Vol.~\bibinfo{volume}{1}. \bibinfo{pages}{6}.
\newblock


\bibitem[Garimella et~al\mbox{.}(2021)]%
        {sisyphus}
\bibfield{author}{\bibinfo{person}{Karthik Garimella}, \bibinfo{person}{Nandan~Kumar Jha}, {and} \bibinfo{person}{Brandon Reagen}.} \bibinfo{year}{2021}\natexlab{}.
\newblock \showarticletitle{Sisyphus: {A} Cautionary Tale of Using Low-Degree Polynomial Activations in Privacy-Preserving Deep Learning}.
\newblock \bibinfo{journal}{\emph{CoRR}}  \bibinfo{volume}{abs/2107.12342} (\bibinfo{year}{2021}).
\newblock
\showeprint[arXiv]{2107.12342}
\urldef\tempurl%
\url{https://arxiv.org/abs/2107.12342}
\showURL{%
\tempurl}


\bibitem[Gilad-Bachrach et~al\mbox{.}(2016)]%
        {cryptonet}
\bibfield{author}{\bibinfo{person}{Ran Gilad-Bachrach}, \bibinfo{person}{Nathan Dowlin}, \bibinfo{person}{Kim Laine}, \bibinfo{person}{Kristin Lauter}, \bibinfo{person}{Michael Naehrig}, {and} \bibinfo{person}{John Wernsing}.} \bibinfo{year}{2016}\natexlab{}.
\newblock \showarticletitle{CryptoNets: Applying Neural Networks to Encrypted Data with High Throughput and Accuracy}. In \bibinfo{booktitle}{\emph{Proceedings of The 33rd International Conference on Machine Learning}} \emph{(\bibinfo{series}{Proceedings of Machine Learning Research}, Vol.~\bibinfo{volume}{48})}, \bibfield{editor}{\bibinfo{person}{Maria~Florina Balcan} {and} \bibinfo{person}{Kilian~Q. Weinberger}} (Eds.). \bibinfo{publisher}{PMLR}, \bibinfo{address}{New York, New York, USA}, \bibinfo{pages}{201--210}.
\newblock
\urldef\tempurl%
\url{https://proceedings.mlr.press/v48/gilad-bachrach16.html}
\showURL{%
\tempurl}


\bibitem[Goldreich et~al\mbox{.}(1987)]%
        {gmw}
\bibfield{author}{\bibinfo{person}{Oded Goldreich}, \bibinfo{person}{Silvio Micali}, {and} \bibinfo{person}{Avi Wigderson}.} \bibinfo{year}{1987}\natexlab{}.
\newblock \showarticletitle{How to Play any Mental Game or {A} Completeness Theorem for Protocols with Honest Majority}. In \bibinfo{booktitle}{\emph{Proceedings of the 19th Annual {ACM} Symposium on Theory of Computing, 1987, New York, New York, {USA}}}. \bibinfo{publisher}{{ACM}}, \bibinfo{pages}{218--229}.
\newblock
\href{https://doi.org/10.1145/28395.28420}{doi:\nolinkurl{10.1145/28395.28420}}


\bibitem[Gupta et~al\mbox{.}(2024)]%
        {sigma}
\bibfield{author}{\bibinfo{person}{Kanav Gupta}, \bibinfo{person}{Neha Jawalkar}, \bibinfo{person}{Ananta Mukherjee}, \bibinfo{person}{Nishanth Chandran}, \bibinfo{person}{Divya Gupta}, \bibinfo{person}{Ashish Panwar}, {and} \bibinfo{person}{Rahul Sharma}.} \bibinfo{year}{2024}\natexlab{}.
\newblock \showarticletitle{{SIGMA:} Secure {GPT} Inference with Function Secret Sharing}.
\newblock \bibinfo{journal}{\emph{Proc. Priv. Enhancing Technol.}} \bibinfo{volume}{2024}, \bibinfo{number}{4} (\bibinfo{year}{2024}), \bibinfo{pages}{61--79}.
\newblock
\href{https://doi.org/10.56553/POPETS-2024-0107}{doi:\nolinkurl{10.56553/POPETS-2024-0107}}


\bibitem[Hao et~al\mbox{.}(2022)]%
        {iron}
\bibfield{author}{\bibinfo{person}{Meng Hao}, \bibinfo{person}{Hongwei Li}, \bibinfo{person}{Hanxiao Chen}, \bibinfo{person}{Pengzhi Xing}, \bibinfo{person}{Guowen Xu}, {and} \bibinfo{person}{Tianwei Zhang}.} \bibinfo{year}{2022}\natexlab{}.
\newblock \showarticletitle{Iron: Private Inference on Transformers}. In \bibinfo{booktitle}{\emph{Advances in Neural Information Processing Systems 35: Annual Conference on Neural Information Processing Systems 2022, NeurIPS 2022, New Orleans, LA, USA, November 28 - December 9, 2022}}.
\newblock


\bibitem[Harth-Kitzerow et~al\mbox{.}(2025a)]%
        {truncation_sok}
\bibfield{author}{\bibinfo{person}{Christopher Harth-Kitzerow}, \bibinfo{person}{Ajith Suresh}, {and} \bibinfo{person}{Georg Carle}.} \bibinfo{year}{2025}\natexlab{a}.
\newblock \showarticletitle{SoK: Truncation Untangled: Scaling Fixed-Point Arithmetic for Privacy-Preserving Machine Learning to Large Models and Datasets}.
\newblock \bibinfo{journal}{\emph{Proceedings on Privacy Enhancing Technologies}} (\bibinfo{year}{2025}).
\newblock


\bibitem[Harth-Kitzerow et~al\mbox{.}(2025b)]%
        {hpmpc}
\bibfield{author}{\bibinfo{person}{Christopher Harth-Kitzerow}, \bibinfo{person}{Ajith Suresh}, \bibinfo{person}{Yongqin Wang}, \bibinfo{person}{Hossein Yalame}, \bibinfo{person}{Georg Carle}, {and} \bibinfo{person}{Murali Annavaram}.} \bibinfo{year}{2025}\natexlab{b}.
\newblock \showarticletitle{High-Throughput Secure Multiparty Computation with an Honest Majority in Various Network Settings}.
\newblock \bibinfo{journal}{\emph{Proceedings on Privacy Enhancing Technologies}} \bibinfo{volume}{2025}, \bibinfo{number}{1} (\bibinfo{year}{2025}), \bibinfo{pages}{250--272}.
\newblock
\urldef\tempurl%
\url{https://doi.org/10.56553/popets-2025-0015}
\showURL{%
\tempurl}


\bibitem[Harth{-}Kitzerow et~al\mbox{.}(2025)]%
        {pigeon}
\bibfield{author}{\bibinfo{person}{Christopher Harth{-}Kitzerow}, \bibinfo{person}{Yongqin Wang}, \bibinfo{person}{Rachit Rajat}, \bibinfo{person}{Georg Carle}, {and} \bibinfo{person}{Murali Annavaram}.} \bibinfo{year}{2025}\natexlab{}.
\newblock \showarticletitle{{PIGEON:} {A} High Throughput Framework for Private Inference of Neural Networks using Secure Multiparty Computation}.
\newblock \bibinfo{journal}{\emph{Proc. Priv. Enhancing Technol.}} \bibinfo{volume}{2025}, \bibinfo{number}{3} (\bibinfo{year}{2025}), \bibinfo{pages}{88--105}.
\newblock
\href{https://doi.org/10.56553/POPETS-2025-0090}{doi:\nolinkurl{10.56553/POPETS-2025-0090}}


\bibitem[Hastings et~al\mbox{.}(2019)]%
        {sok_mpc_compiler}
\bibfield{author}{\bibinfo{person}{Marcella Hastings}, \bibinfo{person}{Brett Hemenway}, \bibinfo{person}{Daniel Noble}, {and} \bibinfo{person}{Steve Zdancewic}.} \bibinfo{year}{2019}\natexlab{}.
\newblock \showarticletitle{SoK: General Purpose Compilers for Secure Multi-Party Computation}. In \bibinfo{booktitle}{\emph{2019 {IEEE} Symposium on Security and Privacy, {SP} 2019, San Francisco, CA, USA, May 19-23, 2019}}. \bibinfo{publisher}{{IEEE}}, \bibinfo{pages}{1220--1237}.
\newblock
\href{https://doi.org/10.1109/SP.2019.00028}{doi:\nolinkurl{10.1109/SP.2019.00028}}


\bibitem[He et~al\mbox{.}(2016)]%
        {resnet}
\bibfield{author}{\bibinfo{person}{Kaiming He}, \bibinfo{person}{Xiangyu Zhang}, \bibinfo{person}{Shaoqing Ren}, {and} \bibinfo{person}{Jian Sun}.} \bibinfo{year}{2016}\natexlab{}.
\newblock \showarticletitle{Deep Residual Learning for Image Recognition}. In \bibinfo{booktitle}{\emph{2016 {IEEE} Conference on Computer Vision and Pattern Recognition, {CVPR} 2016, Las Vegas, NV, USA, June 27-30, 2016}}. \bibinfo{publisher}{{IEEE} Computer Society}, \bibinfo{pages}{770--778}.
\newblock
\href{https://doi.org/10.1109/CVPR.2016.90}{doi:\nolinkurl{10.1109/CVPR.2016.90}}


\bibitem[Ho et~al\mbox{.}(2022)]%
        {xray}
\bibfield{author}{\bibinfo{person}{Trang-Thi Ho}, \bibinfo{person}{Khoa-Dang Tran}, {and} \bibinfo{person}{Yennun Huang}.} \bibinfo{year}{2022}\natexlab{}.
\newblock \showarticletitle{FedSGDCOVID: Federated SGD COVID-19 Detection under Local Differential Privacy Using Chest X-ray Images and Symptom Information}.
\newblock \bibinfo{journal}{\emph{Sensors}} \bibinfo{volume}{22}, \bibinfo{number}{10} (\bibinfo{year}{2022}), \bibinfo{pages}{3728}.
\newblock


\bibitem[Holzer et~al\mbox{.}(2012)]%
        {cbmc_gc}
\bibfield{author}{\bibinfo{person}{Andreas Holzer}, \bibinfo{person}{Martin Franz}, \bibinfo{person}{Stefan Katzenbeisser}, {and} \bibinfo{person}{Helmut Veith}.} \bibinfo{year}{2012}\natexlab{}.
\newblock \showarticletitle{Secure two-party computations in {ANSI} {C}}. In \bibinfo{booktitle}{\emph{the {ACM} Conference on Computer and Communications Security, CCS'12, Raleigh, NC, USA, October 16-18, 2012}}, \bibfield{editor}{\bibinfo{person}{Ting Yu}, \bibinfo{person}{George Danezis}, {and} \bibinfo{person}{Virgil~D. Gligor}} (Eds.). \bibinfo{publisher}{{ACM}}, \bibinfo{pages}{772--783}.
\newblock
\href{https://doi.org/10.1145/2382196.2382278}{doi:\nolinkurl{10.1145/2382196.2382278}}


\bibitem[Home(2023)]%
        {googlehome}
\bibfield{author}{\bibinfo{person}{Google Home}.} \bibinfo{year}{2023}\natexlab{}.
\newblock \bibinfo{title}{Brands you love, united with Google Home.}
\newblock \bibinfo{howpublished}{\url{https://home.google.com/explore-devices/}}.
\newblock


\bibitem[Huang et~al\mbox{.}(2022a)]%
        {stamp}
\bibfield{author}{\bibinfo{person}{Pengzhi Huang}, \bibinfo{person}{Thang Hoang}, \bibinfo{person}{Yueying Li}, \bibinfo{person}{Elaine Shi}, {and} \bibinfo{person}{G~Edward Suh}.} \bibinfo{year}{2022}\natexlab{a}.
\newblock \showarticletitle{STAMP: Lightweight TEE-Assisted MPC for Efficient Privacy-Preserving Machine Learning}.
\newblock \bibinfo{journal}{\emph{arXiv preprint arXiv:2210.10133}} (\bibinfo{year}{2022}).
\newblock


\bibitem[Huang et~al\mbox{.}(2026)]%
        {pengzhi_ispass}
\bibfield{author}{\bibinfo{person}{Pengzhi Huang}, \bibinfo{person}{Kiwan Maeng}, {and} \bibinfo{person}{G~Edward Suh}.} \bibinfo{year}{2026}\natexlab{}.
\newblock \showarticletitle{Beyond Latency: A System-Level Characterization of MPC and FHE for PPML}. In \bibinfo{booktitle}{\emph{International {IEEE} Symposium on Performance Analysis of Systems and Software, {ISPASS}}}. \bibinfo{publisher}{{IEEE}}.
\newblock


\bibitem[Huang et~al\mbox{.}(2022b)]%
        {cheetah}
\bibfield{author}{\bibinfo{person}{Zhicong Huang}, \bibinfo{person}{Wen{-}jie Lu}, \bibinfo{person}{Cheng Hong}, {and} \bibinfo{person}{Jiansheng Ding}.} \bibinfo{year}{2022}\natexlab{b}.
\newblock \showarticletitle{Cheetah: Lean and Fast Secure Two-Party Deep Neural Network Inference}. In \bibinfo{booktitle}{\emph{31st {USENIX} Security Symposium, {USENIX} Security 2022, Boston, MA, USA, August 10-12, 2022}}, \bibfield{editor}{\bibinfo{person}{Kevin R.~B. Butler} {and} \bibinfo{person}{Kurt Thomas}} (Eds.). \bibinfo{publisher}{{USENIX} Association}, \bibinfo{pages}{809--826}.
\newblock
\urldef\tempurl%
\url{https://www.usenix.org/conference/usenixsecurity22/presentation/huang-zhicong}
\showURL{%
\tempurl}


\bibitem[Intel(2023)]%
        {sgx}
\bibfield{author}{\bibinfo{person}{Intel}.} \bibinfo{year}{2023}\natexlab{}.
\newblock \bibinfo{title}{Intel® Software Guard Extensions}.
\newblock \bibinfo{howpublished}{\url{https://www.intel.com/content/www/us/en/developer/tools/software-guard-extensions/overview.html}}.
\newblock


\bibitem[Jawalkar et~al\mbox{.}(2024)]%
        {orca}
\bibfield{author}{\bibinfo{person}{Neha Jawalkar}, \bibinfo{person}{Kanav Gupta}, \bibinfo{person}{Arkaprava Basu}, \bibinfo{person}{Nishanth Chandran}, \bibinfo{person}{Divya Gupta}, {and} \bibinfo{person}{Rahul Sharma}.} \bibinfo{year}{2024}\natexlab{}.
\newblock \showarticletitle{Orca: FSS-based Secure Training and Inference with GPUs}. In \bibinfo{booktitle}{\emph{{IEEE} Symposium on Security and Privacy, {SP} 2024, San Francisco, CA, USA, May 19-23, 2024}}. \bibinfo{publisher}{{IEEE}}, \bibinfo{pages}{597--616}.
\newblock
\href{https://doi.org/10.1109/SP54263.2024.00063}{doi:\nolinkurl{10.1109/SP54263.2024.00063}}


\bibitem[Jayashankar et~al\mbox{.}(2025)]%
        {cinnamon}
\bibfield{author}{\bibinfo{person}{Siddharth Jayashankar}, \bibinfo{person}{Edward Chen}, \bibinfo{person}{Tom Tang}, \bibinfo{person}{Wenting Zheng}, {and} \bibinfo{person}{Dimitrios Skarlatos}.} \bibinfo{year}{2025}\natexlab{}.
\newblock \showarticletitle{Cinnamon: A Framework for Scale-Out Encrypted AI}. In \bibinfo{booktitle}{\emph{Proceedings of the 30th ACM International Conference on Architectural Support for Programming Languages and Operating Systems, Volume 1}} (Rotterdam, Netherlands) \emph{(\bibinfo{series}{ASPLOS '25})}. \bibinfo{publisher}{Association for Computing Machinery}, \bibinfo{address}{New York, NY, USA}, \bibinfo{pages}{133–150}.
\newblock
\showISBNx{9798400706981}
\href{https://doi.org/10.1145/3669940.3707260}{doi:\nolinkurl{10.1145/3669940.3707260}}


\bibitem[Jayashankar et~al\mbox{.}(2026)]%
        {cerium}
\bibfield{author}{\bibinfo{person}{Siddharth Jayashankar}, \bibinfo{person}{Joshua Kim}, \bibinfo{person}{Michael~B. Sullivan}, \bibinfo{person}{Wenting Zheng}, {and} \bibinfo{person}{Dimitrios Skarlatos}.} \bibinfo{year}{2026}\natexlab{}.
\newblock \showarticletitle{Cerium: A Multi-GPU Framework for Terabyte-Scale Encrypted Inference} \emph{(\bibinfo{series}{SOSP '26})}. \bibinfo{publisher}{Association for Computing Machinery}, \bibinfo{address}{New York, NY, USA}, \bibinfo{pages}{193–210}.
\newblock
\showISBNx{9798400725852}
\href{https://doi.org/10.1145/3830418.3843864}{doi:\nolinkurl{10.1145/3830418.3843864}}


\bibitem[Jha et~al\mbox{.}(2021)]%
        {deepreduce}
\bibfield{author}{\bibinfo{person}{Nandan~Kumar Jha}, \bibinfo{person}{Zahra Ghodsi}, \bibinfo{person}{Siddharth Garg}, {and} \bibinfo{person}{Brandon Reagen}.} \bibinfo{year}{2021}\natexlab{}.
\newblock \showarticletitle{DeepReDuce: ReLU Reduction for Fast Private Inference}. In \bibinfo{booktitle}{\emph{Proceedings of the 38th International Conference on Machine Learning, {ICML} 2021, 18-24 July 2021, Virtual Event}} \emph{(\bibinfo{series}{Proceedings of Machine Learning Research}, Vol.~\bibinfo{volume}{139})}, \bibfield{editor}{\bibinfo{person}{Marina Meila} {and} \bibinfo{person}{Tong Zhang}} (Eds.). \bibinfo{publisher}{{PMLR}}, \bibinfo{pages}{4839--4849}.
\newblock
\urldef\tempurl%
\url{http://proceedings.mlr.press/v139/jha21a.html}
\showURL{%
\tempurl}


\bibitem[Jha and Reagen(2024)]%
        {deepreshape}
\bibfield{author}{\bibinfo{person}{Nandan~Kumar Jha} {and} \bibinfo{person}{Brandon Reagen}.} \bibinfo{year}{2024}\natexlab{}.
\newblock \showarticletitle{DeepReShape: Redesigning Neural Networks for Efficient Private Inference}.
\newblock \bibinfo{journal}{\emph{Transactions on Machine Learning Research}} (\bibinfo{year}{2024}).
\newblock
\showISSN{2835-8856}
\urldef\tempurl%
\url{https://openreview.net/forum?id=iwCBWULItx}
\showURL{%
\tempurl}


\bibitem[Jung et~al\mbox{.}(2021)]%
        {100x}
\bibfield{author}{\bibinfo{person}{Wonkyung Jung}, \bibinfo{person}{Sangpyo Kim}, \bibinfo{person}{Jung~Ho Ahn}, \bibinfo{person}{Jung~Hee Cheon}, {and} \bibinfo{person}{Younho Lee}.} \bibinfo{year}{2021}\natexlab{}.
\newblock \showarticletitle{Over 100x Faster Bootstrapping in Fully Homomorphic Encryption through Memory-centric Optimization with GPUs}.
\newblock \bibinfo{journal}{\emph{{IACR} Trans. Cryptogr. Hardw. Embed. Syst.}} \bibinfo{volume}{2021}, \bibinfo{number}{4} (\bibinfo{year}{2021}), \bibinfo{pages}{114--148}.
\newblock
\href{https://doi.org/10.46586/tches.v2021.i4.114-148}{doi:\nolinkurl{10.46586/tches.v2021.i4.114-148}}


\bibitem[Juvekar et~al\mbox{.}(2018)]%
        {gazelle}
\bibfield{author}{\bibinfo{person}{Chiraag Juvekar}, \bibinfo{person}{Vinod Vaikuntanathan}, {and} \bibinfo{person}{Anantha~P. Chandrakasan}.} \bibinfo{year}{2018}\natexlab{}.
\newblock \showarticletitle{{GAZELLE:} {A} Low Latency Framework for Secure Neural Network Inference}. In \bibinfo{booktitle}{\emph{27th {USENIX} Security Symposium, {USENIX} Security 2018, Baltimore, MD, USA, August 15-17, 2018}}, \bibfield{editor}{\bibinfo{person}{William Enck} {and} \bibinfo{person}{Adrienne~Porter Felt}} (Eds.). \bibinfo{publisher}{{USENIX} Association}, \bibinfo{pages}{1651--1669}.
\newblock
\urldef\tempurl%
\url{https://www.usenix.org/conference/usenixsecurity18/presentation/juvekar}
\showURL{%
\tempurl}


\bibitem[Keller(2020)]%
        {mp_spdz}
\bibfield{author}{\bibinfo{person}{Marcel Keller}.} \bibinfo{year}{2020}\natexlab{}.
\newblock \showarticletitle{{MP-SPDZ:} {A} Versatile Framework for Multi-Party Computation}. In \bibinfo{booktitle}{\emph{{CCS} '20: 2020 {ACM} {SIGSAC} Conference on Computer and Communications Security, Virtual Event, USA, November 9-13, 2020}}, \bibfield{editor}{\bibinfo{person}{Jay Ligatti}, \bibinfo{person}{Xinming Ou}, \bibinfo{person}{Jonathan Katz}, {and} \bibinfo{person}{Giovanni Vigna}} (Eds.). \bibinfo{publisher}{{ACM}}, \bibinfo{pages}{1575--1590}.
\newblock
\href{https://doi.org/10.1145/3372297.3417872}{doi:\nolinkurl{10.1145/3372297.3417872}}


\bibitem[Khandaker et~al\mbox{.}(2020)]%
        {coin_attack}
\bibfield{author}{\bibinfo{person}{Mustakimur~Rahman Khandaker}, \bibinfo{person}{Yueqiang Cheng}, \bibinfo{person}{Zhi Wang}, {and} \bibinfo{person}{Tao Wei}.} \bibinfo{year}{2020}\natexlab{}.
\newblock \showarticletitle{{COIN} Attacks: On Insecurity of Enclave Untrusted Interfaces in {SGX}}. In \bibinfo{booktitle}{\emph{{ASPLOS} '20: Architectural Support for Programming Languages and Operating Systems, Lausanne, Switzerland, March 16-20, 2020}}, \bibfield{editor}{\bibinfo{person}{James~R. Larus}, \bibinfo{person}{Luis Ceze}, {and} \bibinfo{person}{Karin Strauss}} (Eds.). \bibinfo{publisher}{{ACM}}, \bibinfo{pages}{971--985}.
\newblock
\href{https://doi.org/10.1145/3373376.3378486}{doi:\nolinkurl{10.1145/3373376.3378486}}


\bibitem[Kim et~al\mbox{.}(2022b)]%
        {ark}
\bibfield{author}{\bibinfo{person}{Jongmin Kim}, \bibinfo{person}{Gwangho Lee}, \bibinfo{person}{Sangpyo Kim}, \bibinfo{person}{Gina Sohn}, \bibinfo{person}{Minsoo Rhu}, \bibinfo{person}{John Kim}, {and} \bibinfo{person}{Jung~Ho Ahn}.} \bibinfo{year}{2022}\natexlab{b}.
\newblock \showarticletitle{{ARK:} Fully Homomorphic Encryption Accelerator with Runtime Data Generation and Inter-Operation Key Reuse}. In \bibinfo{booktitle}{\emph{55th {IEEE/ACM} International Symposium on Microarchitecture, {MICRO} 2022, Chicago, IL, USA, October 1-5, 2022}}. \bibinfo{publisher}{{IEEE}}, \bibinfo{pages}{1237--1254}.
\newblock
\href{https://doi.org/10.1109/MICRO56248.2022.00086}{doi:\nolinkurl{10.1109/MICRO56248.2022.00086}}


\bibitem[Kim et~al\mbox{.}(2022a)]%
        {bts}
\bibfield{author}{\bibinfo{person}{Sangpyo Kim}, \bibinfo{person}{Jongmin Kim}, \bibinfo{person}{Michael~Jaemin Kim}, \bibinfo{person}{Wonkyung Jung}, \bibinfo{person}{John Kim}, \bibinfo{person}{Minsoo Rhu}, {and} \bibinfo{person}{Jung~Ho Ahn}.} \bibinfo{year}{2022}\natexlab{a}.
\newblock \showarticletitle{{BTS:} an accelerator for bootstrappable fully homomorphic encryption}. In \bibinfo{booktitle}{\emph{{ISCA} '22: The 49th Annual International Symposium on Computer Architecture, New York, New York, USA, June 18 - 22, 2022}}, \bibfield{editor}{\bibinfo{person}{Valentina Salapura}, \bibinfo{person}{Mohamed Zahran}, \bibinfo{person}{Fred Chong}, {and} \bibinfo{person}{Lingjia Tang}} (Eds.). \bibinfo{publisher}{{ACM}}, \bibinfo{pages}{711--725}.
\newblock
\href{https://doi.org/10.1145/3470496.3527415}{doi:\nolinkurl{10.1145/3470496.3527415}}


\bibitem[Knott et~al\mbox{.}(2021)]%
        {crypten}
\bibfield{author}{\bibinfo{person}{Brian Knott}, \bibinfo{person}{Shobha Venkataraman}, \bibinfo{person}{Awni~Y. Hannun}, \bibinfo{person}{Shubho Sengupta}, \bibinfo{person}{Mark Ibrahim}, {and} \bibinfo{person}{Laurens van~der Maaten}.} \bibinfo{year}{2021}\natexlab{}.
\newblock \showarticletitle{CrypTen: Secure Multi-Party Computation Meets Machine Learning}. In \bibinfo{booktitle}{\emph{Advances in Neural Information Processing Systems 34: Annual Conference on Neural Information Processing Systems 2021, NeurIPS 2021, December 6-14, 2021, virtual}}. \bibinfo{pages}{4961--4973}.
\newblock
\urldef\tempurl%
\url{https://proceedings.neurips.cc/paper/2021/hash/2754518221cfbc8d25c13a06a4cb8421-Abstract.html}
\showURL{%
\tempurl}


\bibitem[Krizhevsky et~al\mbox{.}(2009)]%
        {cifar10}
\bibfield{author}{\bibinfo{person}{Alex Krizhevsky}, \bibinfo{person}{Geoffrey Hinton}, {et~al\mbox{.}}} \bibinfo{year}{2009}\natexlab{}.
\newblock \showarticletitle{Learning multiple layers of features from tiny images}.
\newblock  (\bibinfo{year}{2009}).
\newblock


\bibitem[Kumar et~al\mbox{.}(2020)]%
        {cryptflow}
\bibfield{author}{\bibinfo{person}{Nishant Kumar}, \bibinfo{person}{Mayank Rathee}, \bibinfo{person}{Nishanth Chandran}, \bibinfo{person}{Divya Gupta}, \bibinfo{person}{Aseem Rastogi}, {and} \bibinfo{person}{Rahul Sharma}.} \bibinfo{year}{2020}\natexlab{}.
\newblock \showarticletitle{CrypTFlow: Secure TensorFlow Inference}. In \bibinfo{booktitle}{\emph{2020 {IEEE} Symposium on Security and Privacy, {SP} 2020, San Francisco, CA, USA, May 18-21, 2020}}. \bibinfo{publisher}{{IEEE}}, \bibinfo{pages}{336--353}.
\newblock
\href{https://doi.org/10.1109/SP40000.2020.00092}{doi:\nolinkurl{10.1109/SP40000.2020.00092}}


\bibitem[Kundu et~al\mbox{.}(2023)]%
        {senet}
\bibfield{author}{\bibinfo{person}{Souvik Kundu}, \bibinfo{person}{Shunlin Lu}, \bibinfo{person}{Yuke Zhang}, \bibinfo{person}{Jacqueline~Tiffany Liu}, {and} \bibinfo{person}{Peter~A. Beerel}.} \bibinfo{year}{2023}\natexlab{}.
\newblock \showarticletitle{Learning to Linearize Deep Neural Networks for Secure and Efficient Private Inference}. In \bibinfo{booktitle}{\emph{The Eleventh International Conference on Learning Representations, {ICLR} 2023, Kigali, Rwanda, May 1-5, 2023}}. \bibinfo{publisher}{OpenReview.net}.
\newblock
\urldef\tempurl%
\url{https://openreview.net/pdf?id=BGF9IeDfmlH}
\showURL{%
\tempurl}


\bibitem[Lam et~al\mbox{.}(2024)]%
        {max_pir}
\bibfield{author}{\bibinfo{person}{Maximilian Lam}, \bibinfo{person}{Jeff Johnson}, \bibinfo{person}{Wenjie Xiong}, \bibinfo{person}{Kiwan Maeng}, \bibinfo{person}{Udit Gupta}, \bibinfo{person}{Minsoo Rhu}, \bibinfo{person}{Hsien{-}Hsin~S. Lee}, \bibinfo{person}{Vijay~Janapa Reddi}, \bibinfo{person}{Gu{-}Yeon Wei}, \bibinfo{person}{David Brooks}, {and} \bibinfo{person}{G.~Edward Suh}.} \bibinfo{year}{2024}\natexlab{}.
\newblock \showarticletitle{GPU-based Private Information Retrieval for On-Device Machine Learning Inference}. In \bibinfo{booktitle}{\emph{Proceedings of the 29th {ACM} International Conference on Architectural Support for Programming Languages and Operating Systems, {ASPLOS} 2024}}. \bibinfo{publisher}{{ACM}}.
\newblock
\href{https://doi.org/10.48550/arXiv.2301.10904}{doi:\nolinkurl{10.48550/arXiv.2301.10904}}


\bibitem[Lee et~al\mbox{.}(2020)]%
        {keystone}
\bibfield{author}{\bibinfo{person}{Dayeol Lee}, \bibinfo{person}{David Kohlbrenner}, \bibinfo{person}{Shweta Shinde}, \bibinfo{person}{Krste Asanovic}, {and} \bibinfo{person}{Dawn Song}.} \bibinfo{year}{2020}\natexlab{}.
\newblock \showarticletitle{Keystone: an open framework for architecting trusted execution environments}. In \bibinfo{booktitle}{\emph{EuroSys '20: Fifteenth EuroSys Conference 2020, Heraklion, Greece, April 27-30, 2020}}, \bibfield{editor}{\bibinfo{person}{Angelos Bilas}, \bibinfo{person}{Kostas Magoutis}, \bibinfo{person}{Evangelos~P. Markatos}, \bibinfo{person}{Dejan Kostic}, {and} \bibinfo{person}{Margo~I. Seltzer}} (Eds.). \bibinfo{publisher}{{ACM}}, \bibinfo{pages}{38:1--38:16}.
\newblock
\href{https://doi.org/10.1145/3342195.3387532}{doi:\nolinkurl{10.1145/3342195.3387532}}


\bibitem[Lee et~al\mbox{.}(2025)]%
        {ola_fhe}
\bibfield{author}{\bibinfo{person}{Junghyun Lee}, \bibinfo{person}{Eunsang Lee}, \bibinfo{person}{Young-Sik Kim}, \bibinfo{person}{Yongwoo Lee}, \bibinfo{person}{Joon-Woo Lee}, \bibinfo{person}{Yongjune Kim}, {and} \bibinfo{person}{Jong-Seon No}.} \bibinfo{year}{2025}\natexlab{}.
\newblock \bibinfo{title}{Optimized Layerwise Approximation for Efficient Private Inference on Fully Homomorphic Encryption}.
\newblock
\showeprint[arxiv]{2310.10349}~[cs.CR]
\urldef\tempurl%
\url{https://arxiv.org/abs/2310.10349}
\showURL{%
\tempurl}


\bibitem[Lee et~al\mbox{.}(2005)]%
        {sp}
\bibfield{author}{\bibinfo{person}{Ruby~B. Lee}, \bibinfo{person}{Peter C.~S. Kwan}, \bibinfo{person}{John~Patrick McGregor}, \bibinfo{person}{Jeffrey~S. Dwoskin}, {and} \bibinfo{person}{Zhenghong Wang}.} \bibinfo{year}{2005}\natexlab{}.
\newblock \showarticletitle{Architecture for Protecting Critical Secrets in Microprocessors}. In \bibinfo{booktitle}{\emph{32st International Symposium on Computer Architecture {(ISCA} 2005), 4-8 June 2005, Madison, Wisconsin, {USA}}}. \bibinfo{publisher}{{IEEE} Computer Society}, \bibinfo{pages}{2--13}.
\newblock
\href{https://doi.org/10.1109/ISCA.2005.14}{doi:\nolinkurl{10.1109/ISCA.2005.14}}


\bibitem[Lee et~al\mbox{.}(2023)]%
        {elasm}
\bibfield{author}{\bibinfo{person}{Yongwoo Lee}, \bibinfo{person}{Seonyoung Cheon}, \bibinfo{person}{Dongkwan Kim}, \bibinfo{person}{Dongyoon Lee}, {and} \bibinfo{person}{Hanjun Kim}.} \bibinfo{year}{2023}\natexlab{}.
\newblock \showarticletitle{$\{$ELASM$\}$:$\{$Error-Latency-Aware$\}$ Scale Management for Fully Homomorphic Encryption}. In \bibinfo{booktitle}{\emph{32nd USENIX Security Symposium (USENIX Security 23)}}. \bibinfo{pages}{4697--4714}.
\newblock


\bibitem[Lee et~al\mbox{.}(2022)]%
        {hecate}
\bibfield{author}{\bibinfo{person}{Yongwoo Lee}, \bibinfo{person}{Seonyeong Heo}, \bibinfo{person}{Seonyoung Cheon}, \bibinfo{person}{Shinnung Jeong}, \bibinfo{person}{Changsu Kim}, \bibinfo{person}{Eunkyung Kim}, \bibinfo{person}{Dongyoon Lee}, {and} \bibinfo{person}{Hanjun Kim}.} \bibinfo{year}{2022}\natexlab{}.
\newblock \showarticletitle{HECATE: Performance-Aware Scale Optimization for Homomorphic Encryption Compiler}. In \bibinfo{booktitle}{\emph{2022 IEEE/ACM International Symposium on Code Generation and Optimization (CGO)}}. \bibinfo{pages}{193--204}.
\newblock
\href{https://doi.org/10.1109/CGO53902.2022.9741265}{doi:\nolinkurl{10.1109/CGO53902.2022.9741265}}


\bibitem[Li et~al\mbox{.}(2022)]%
        {mpcformer}
\bibfield{author}{\bibinfo{person}{Dacheng Li}, \bibinfo{person}{Rulin Shao}, \bibinfo{person}{Hongyi Wang}, \bibinfo{person}{Han Guo}, \bibinfo{person}{Eric~P. Xing}, {and} \bibinfo{person}{Hao Zhang}.} \bibinfo{year}{2022}\natexlab{}.
\newblock \showarticletitle{MPCFormer: fast, performant and private Transformer inference with {MPC}}.
\newblock \bibinfo{journal}{\emph{CoRR}}  \bibinfo{volume}{abs/2211.01452} (\bibinfo{year}{2022}).
\newblock
\showeprint[arXiv]{2211.01452}
\href{https://doi.org/10.48550/ARXIV.2211.01452}{doi:\nolinkurl{10.48550/ARXIV.2211.01452}}


\bibitem[Li et~al\mbox{.}(2021)]%
        {micronet}
\bibfield{author}{\bibinfo{person}{Yunsheng Li}, \bibinfo{person}{Yinpeng Chen}, \bibinfo{person}{Xiyang Dai}, \bibinfo{person}{Dongdong Chen}, \bibinfo{person}{Mengchen Liu}, \bibinfo{person}{Lu Yuan}, \bibinfo{person}{Zicheng Liu}, \bibinfo{person}{Lei Zhang}, {and} \bibinfo{person}{Nuno Vasconcelos}.} \bibinfo{year}{2021}\natexlab{}.
\newblock \showarticletitle{MicroNet: Improving Image Recognition with Extremely Low FLOPs}. In \bibinfo{booktitle}{\emph{2021 {IEEE/CVF} International Conference on Computer Vision, {ICCV} 2021, Montreal, QC, Canada, October 10-17, 2021}}. \bibinfo{publisher}{{IEEE}}, \bibinfo{pages}{458--467}.
\newblock
\href{https://doi.org/10.1109/ICCV48922.2021.00052}{doi:\nolinkurl{10.1109/ICCV48922.2021.00052}}


\bibitem[Li et~al\mbox{.}(2023)]%
        {truncation_broken}
\bibfield{author}{\bibinfo{person}{Yun Li}, \bibinfo{person}{Yufei Duan}, \bibinfo{person}{Zhicong Huang}, \bibinfo{person}{Cheng Hong}, \bibinfo{person}{Chao Zhang}, {and} \bibinfo{person}{Yifan Song}.} \bibinfo{year}{2023}\natexlab{}.
\newblock \showarticletitle{Efficient $\{$3PC$\}$ for binary circuits with application to $\{$Maliciously-Secure$\}$$\{$DNN$\}$ inference}. In \bibinfo{booktitle}{\emph{32nd USENIX Security Symposium (USENIX Security 23)}}. \bibinfo{pages}{5377--5394}.
\newblock


\bibitem[Lie et~al\mbox{.}(2003)]%
        {xom}
\bibfield{author}{\bibinfo{person}{David Lie}, \bibinfo{person}{John~C. Mitchell}, \bibinfo{person}{Chandramohan~A. Thekkath}, {and} \bibinfo{person}{Mark Horowitz}.} \bibinfo{year}{2003}\natexlab{}.
\newblock \showarticletitle{Specifying and Verifying Hardware for Tamper-Resistant Software}. In \bibinfo{booktitle}{\emph{2003 {IEEE} Symposium on Security and Privacy (S{\&}P 2003), 11-14 May 2003, Berkeley, CA, {USA}}}. \bibinfo{publisher}{{IEEE} Computer Society}, \bibinfo{pages}{166}.
\newblock
\href{https://doi.org/10.1109/SECPRI.2003.1199335}{doi:\nolinkurl{10.1109/SECPRI.2003.1199335}}


\bibitem[Liu et~al\mbox{.}(2015)]%
        {oblivm}
\bibfield{author}{\bibinfo{person}{Chang Liu}, \bibinfo{person}{Xiao~Shaun Wang}, \bibinfo{person}{Kartik Nayak}, \bibinfo{person}{Yan Huang}, {and} \bibinfo{person}{Elaine Shi}.} \bibinfo{year}{2015}\natexlab{}.
\newblock \showarticletitle{ObliVM: {A} Programming Framework for Secure Computation}. In \bibinfo{booktitle}{\emph{2015 {IEEE} Symposium on Security and Privacy, {SP} 2015, San Jose, CA, USA, May 17-21, 2015}}. \bibinfo{publisher}{{IEEE} Computer Society}, \bibinfo{pages}{359--376}.
\newblock
\href{https://doi.org/10.1109/SP.2015.29}{doi:\nolinkurl{10.1109/SP.2015.29}}


\bibitem[Liu et~al\mbox{.}(2017)]%
        {minionn}
\bibfield{author}{\bibinfo{person}{Jian Liu}, \bibinfo{person}{Mika Juuti}, \bibinfo{person}{Yao Lu}, {and} \bibinfo{person}{N. Asokan}.} \bibinfo{year}{2017}\natexlab{}.
\newblock \showarticletitle{Oblivious Neural Network Predictions via MiniONN Transformations}. In \bibinfo{booktitle}{\emph{Proceedings of the 2017 {ACM} {SIGSAC} Conference on Computer and Communications Security, {CCS} 2017, Dallas, TX, USA, October 30 - November 03, 2017}}, \bibfield{editor}{\bibinfo{person}{Bhavani Thuraisingham}, \bibinfo{person}{David Evans}, \bibinfo{person}{Tal Malkin}, {and} \bibinfo{person}{Dongyan Xu}} (Eds.). \bibinfo{publisher}{{ACM}}, \bibinfo{pages}{619--631}.
\newblock
\href{https://doi.org/10.1145/3133956.3134056}{doi:\nolinkurl{10.1145/3133956.3134056}}


\bibitem[Liu and Maeng(2025)]%
        {jinyu_mpc_cal}
\bibfield{author}{\bibinfo{person}{Jinyu Liu} {and} \bibinfo{person}{Kiwan Maeng}.} \bibinfo{year}{2025}\natexlab{}.
\newblock \showarticletitle{In-Depth Characterization of Machine Learning on an Optimized Multi-Party Computing Library}.
\newblock \bibinfo{journal}{\emph{IEEE Computer Architecture Letters}} \bibinfo{volume}{24}, \bibinfo{number}{2} (\bibinfo{year}{2025}), \bibinfo{pages}{341--344}.
\newblock
\href{https://doi.org/10.1109/LCA.2025.3624787}{doi:\nolinkurl{10.1109/LCA.2025.3624787}}


\bibitem[Liu et~al\mbox{.}(2021)]%
        {tee_bugs}
\bibfield{author}{\bibinfo{person}{Weijie Liu}, \bibinfo{person}{Hongbo Chen}, \bibinfo{person}{XiaoFeng Wang}, \bibinfo{person}{Zhi Li}, \bibinfo{person}{Danfeng Zhang}, \bibinfo{person}{Wenhao Wang}, {and} \bibinfo{person}{Haixu Tang}.} \bibinfo{year}{2021}\natexlab{}.
\newblock \showarticletitle{Understanding TEE containers, easy to use? Hard to trust}.
\newblock \bibinfo{journal}{\emph{arXiv preprint arXiv:2109.01923}} (\bibinfo{year}{2021}).
\newblock


\bibitem[Lou et~al\mbox{.}(2021)]%
        {safenet}
\bibfield{author}{\bibinfo{person}{Qian Lou}, \bibinfo{person}{Yilin Shen}, \bibinfo{person}{Hongxia Jin}, {and} \bibinfo{person}{Lei Jiang}.} \bibinfo{year}{2021}\natexlab{}.
\newblock \showarticletitle{SAFENet: {A} Secure, Accurate and Fast Neural Network Inference}. In \bibinfo{booktitle}{\emph{9th International Conference on Learning Representations, {ICLR} 2021, Virtual Event, Austria, May 3-7, 2021}}. \bibinfo{publisher}{OpenReview.net}.
\newblock
\urldef\tempurl%
\url{https://openreview.net/forum?id=Cz3dbFm5u-}
\showURL{%
\tempurl}


\bibitem[Lu et~al\mbox{.}(2025)]%
        {bumblebee}
\bibfield{author}{\bibinfo{person}{Wen{-}jie Lu}, \bibinfo{person}{Zhicong Huang}, \bibinfo{person}{Zhen Gu}, \bibinfo{person}{Jingyu Li}, \bibinfo{person}{Jian Liu}, \bibinfo{person}{Cheng Hong}, \bibinfo{person}{Kui Ren}, \bibinfo{person}{Tao Wei}, {and} \bibinfo{person}{Wenguang Chen}.} \bibinfo{year}{2025}\natexlab{}.
\newblock \showarticletitle{BumbleBee: Secure Two-party Inference Framework for Large Transformers}. In \bibinfo{booktitle}{\emph{32nd Annual Network and Distributed System Security Symposium, {NDSS} 2025, San Diego, California, USA, February 24-28, 2025}}. \bibinfo{publisher}{The Internet Society}.
\newblock
\urldef\tempurl%
\url{https://www.ndss-symposium.org/ndss-paper/bumblebee-secure-two-party-inference-framework-for-large-transformers/}
\showURL{%
\tempurl}


\bibitem[Luo et~al\mbox{.}(2024)]%
        {secformer}
\bibfield{author}{\bibinfo{person}{Jinglong Luo}, \bibinfo{person}{Yehong Zhang}, \bibinfo{person}{Jiaqi Zhang}, \bibinfo{person}{Xin Mu}, \bibinfo{person}{Hui Wang}, \bibinfo{person}{Yue Yu}, {and} \bibinfo{person}{Zenglin Xu}.} \bibinfo{year}{2024}\natexlab{}.
\newblock \showarticletitle{SecFormer: Towards Fast and Accurate Privacy-Preserving Inference for Large Language Models}.
\newblock \bibinfo{journal}{\emph{CoRR}}  \bibinfo{volume}{abs/2401.00793} (\bibinfo{year}{2024}).
\newblock
\showeprint[arXiv]{2401.00793}
\href{https://doi.org/10.48550/ARXIV.2401.00793}{doi:\nolinkurl{10.48550/ARXIV.2401.00793}}


\bibitem[Lv et~al\mbox{.}(2026)]%
        {smash}
\bibfield{author}{\bibinfo{person}{Yunlv Lv}, \bibinfo{person}{Rui Zhang}, \bibinfo{person}{Zhiyuan Zhang}, \bibinfo{person}{Ziyi Wan}, \bibinfo{person}{Lanxue Zhang}, \bibinfo{person}{Minhui Xue}, \bibinfo{person}{Jiangtao Li}, {and} \bibinfo{person}{Yanan Cao}.} \bibinfo{year}{2026}\natexlab{}.
\newblock \showarticletitle{{SMASH}: Scalable Maliciously Secure Hybrid Multi-party Computation Framework for Privacy-Preserving Large Language Models}. In \bibinfo{booktitle}{\emph{35th USENIX Security Symposium (USENIX Security 26)}}. \bibinfo{publisher}{USENIX Association}, \bibinfo{address}{Baltimore, MD, USA}, \bibinfo{pages}{1927--1946}.
\newblock
\urldef\tempurl%
\url{https://www.usenix.org/conference/usenixsecurity26/presentation/lv}
\showURL{%
\tempurl}


\bibitem[lvdmaaten(2022)]%
        {crypten_layernorm}
\bibfield{author}{\bibinfo{person}{lvdmaaten}.} \bibinfo{year}{2022}\natexlab{}.
\newblock \bibinfo{title}{comment on: How to directly define the LayerNorm module?}
\newblock \bibinfo{howpublished}{\url{https://github.com/facebookresearch/CrypTen/issues/399\#issuecomment-1229263703}}.
\newblock


\bibitem[Ma et~al\mbox{.}(2023)]%
        {spu}
\bibfield{author}{\bibinfo{person}{Junming Ma}, \bibinfo{person}{Yancheng Zheng}, \bibinfo{person}{Jun Feng}, \bibinfo{person}{Derun Zhao}, \bibinfo{person}{Haoqi Wu}, \bibinfo{person}{Wenjing Fang}, \bibinfo{person}{Jin Tan}, \bibinfo{person}{Chaofan Yu}, \bibinfo{person}{Benyu Zhang}, {and} \bibinfo{person}{Lei Wang}.} \bibinfo{year}{2023}\natexlab{}.
\newblock \showarticletitle{SecretFlow-SPU: {A} Performant and User-Friendly Framework for Privacy-Preserving Machine Learning}. In \bibinfo{booktitle}{\emph{Proceedings of the 2023 {USENIX} Annual Technical Conference, {USENIX} {ATC} 2023, Boston, MA, USA, July 10-12, 2023}}. \bibinfo{publisher}{{USENIX} Association}, \bibinfo{pages}{17--33}.
\newblock
\urldef\tempurl%
\url{https://www.usenix.org/conference/atc23/presentation/ma}
\showURL{%
\tempurl}


\bibitem[Maeng et~al\mbox{.}(2023)]%
        {maeng_fil2}
\bibfield{author}{\bibinfo{person}{Kiwan Maeng}, \bibinfo{person}{Chuan Guo}, \bibinfo{person}{Sanjay Kariyappa}, {and} \bibinfo{person}{G.~Edward Suh}.} \bibinfo{year}{2023}\natexlab{}.
\newblock \showarticletitle{Bounding the Invertibility of Privacy-preserving Instance Encoding using Fisher Information}. In \bibinfo{booktitle}{\emph{Advances in Neural Information Processing Systems 36: Annual Conference on Neural Information Processing Systems 2023, NeurIPS 2023, New Orleans, LA, USA, December 10 - 16, 2023}}.
\newblock
\urldef\tempurl%
\url{http://papers.nips.cc/paper\_files/paper/2023/hash/a344f7f474958cc0775be7e46bc94309-Abstract-Conference.html}
\showURL{%
\tempurl}


\bibitem[Maeng and Suh(2023)]%
        {hummingbird}
\bibfield{author}{\bibinfo{person}{Kiwan Maeng} {and} \bibinfo{person}{G.~Edward Suh}.} \bibinfo{year}{2023}\natexlab{}.
\newblock \bibinfo{title}{Approximating ReLU on a Reduced Ring for Efficient MPC-based Private Inference}.
\newblock
\showeprint[arxiv]{2309.04875}~[cs.LG]
\urldef\tempurl%
\url{https://arxiv.org/abs/2309.04875}
\showURL{%
\tempurl}


\bibitem[Merity et~al\mbox{.}(2017)]%
        {wikitext}
\bibfield{author}{\bibinfo{person}{Stephen Merity}, \bibinfo{person}{Caiming Xiong}, \bibinfo{person}{James Bradbury}, {and} \bibinfo{person}{Richard Socher}.} \bibinfo{year}{2017}\natexlab{}.
\newblock \showarticletitle{Pointer Sentinel Mixture Models}. In \bibinfo{booktitle}{\emph{International Conference on Learning Representations (ICLR)}}.
\newblock


\bibitem[Meta(2023)]%
        {fbportal}
\bibfield{author}{\bibinfo{person}{Meta}.} \bibinfo{year}{2023}\natexlab{}.
\newblock \bibinfo{title}{Meta Portal Go}.
\newblock \bibinfo{howpublished}{\url{https://www.meta.com/portal/products/portal-go/}}.
\newblock


\bibitem[Mishra et~al\mbox{.}(2020)]%
        {delphi}
\bibfield{author}{\bibinfo{person}{Pratyush Mishra}, \bibinfo{person}{Ryan Lehmkuhl}, \bibinfo{person}{Akshayaram Srinivasan}, \bibinfo{person}{Wenting Zheng}, {and} \bibinfo{person}{Raluca~Ada Popa}.} \bibinfo{year}{2020}\natexlab{}.
\newblock \showarticletitle{Delphi: {A} Cryptographic Inference Service for Neural Networks}. In \bibinfo{booktitle}{\emph{29th {USENIX} Security Symposium, {USENIX} Security 2020, August 12-14, 2020}}, \bibfield{editor}{\bibinfo{person}{Srdjan Capkun} {and} \bibinfo{person}{Franziska Roesner}} (Eds.). \bibinfo{publisher}{{USENIX} Association}, \bibinfo{pages}{2505--2522}.
\newblock
\urldef\tempurl%
\url{https://www.usenix.org/conference/usenixsecurity20/presentation/mishra}
\showURL{%
\tempurl}


\bibitem[Mohassel and Rindal(2018)]%
        {aby3}
\bibfield{author}{\bibinfo{person}{Payman Mohassel} {and} \bibinfo{person}{Peter Rindal}.} \bibinfo{year}{2018}\natexlab{}.
\newblock \showarticletitle{ABY\({}^{\mbox{3}}\): {A} Mixed Protocol Framework for Machine Learning}. In \bibinfo{booktitle}{\emph{Proceedings of the 2018 {ACM} {SIGSAC} Conference on Computer and Communications Security, {CCS} 2018, Toronto, ON, Canada, October 15-19, 2018}}, \bibfield{editor}{\bibinfo{person}{David Lie}, \bibinfo{person}{Mohammad Mannan}, \bibinfo{person}{Michael Backes}, {and} \bibinfo{person}{XiaoFeng Wang}} (Eds.). \bibinfo{publisher}{{ACM}}, \bibinfo{pages}{35--52}.
\newblock
\href{https://doi.org/10.1145/3243734.3243760}{doi:\nolinkurl{10.1145/3243734.3243760}}


\bibitem[Mohassel and Zhang(2017)]%
        {secureml}
\bibfield{author}{\bibinfo{person}{Payman Mohassel} {and} \bibinfo{person}{Yupeng Zhang}.} \bibinfo{year}{2017}\natexlab{}.
\newblock \showarticletitle{SecureML: {A} System for Scalable Privacy-Preserving Machine Learning}. In \bibinfo{booktitle}{\emph{2017 {IEEE} Symposium on Security and Privacy, {SP} 2017, San Jose, CA, USA, May 22-26, 2017}}. \bibinfo{publisher}{{IEEE} Computer Society}, \bibinfo{pages}{19--38}.
\newblock
\href{https://doi.org/10.1109/SP.2017.12}{doi:\nolinkurl{10.1109/SP.2017.12}}


\bibitem[Mood et~al\mbox{.}(2016)]%
        {frigate}
\bibfield{author}{\bibinfo{person}{Benjamin Mood}, \bibinfo{person}{Debayan Gupta}, \bibinfo{person}{Henry Carter}, \bibinfo{person}{Kevin R.~B. Butler}, {and} \bibinfo{person}{Patrick Traynor}.} \bibinfo{year}{2016}\natexlab{}.
\newblock \showarticletitle{Frigate: {A} Validated, Extensible, and Efficient Compiler and Interpreter for Secure Computation}. In \bibinfo{booktitle}{\emph{{IEEE} European Symposium on Security and Privacy, EuroS{\&}P 2016, Saarbr{\"{u}}cken, Germany, March 21-24, 2016}}. \bibinfo{publisher}{{IEEE}}, \bibinfo{pages}{112--127}.
\newblock
\href{https://doi.org/10.1109/EUROSP.2016.20}{doi:\nolinkurl{10.1109/EUROSP.2016.20}}


\bibitem[Moon et~al\mbox{.}(2024)]%
        {thor}
\bibfield{author}{\bibinfo{person}{Jungho Moon}, \bibinfo{person}{Dongwoo Yoo}, \bibinfo{person}{Xiaoqian Jiang}, {and} \bibinfo{person}{Miran Kim}.} \bibinfo{year}{2024}\natexlab{}.
\newblock \bibinfo{title}{{THOR}: Secure Transformer Inference with Homomorphic Encryption}.
\newblock \bibinfo{howpublished}{Cryptology {ePrint} Archive, Paper 2024/1881}.
\newblock
\href{https://doi.org/10.1145/3719027.3765150}{doi:\nolinkurl{10.1145/3719027.3765150}}


\bibitem[NVIDIA(2023)]%
        {hopper_confidential}
\bibfield{author}{\bibinfo{person}{NVIDIA}.} \bibinfo{year}{2023}\natexlab{}.
\newblock \bibinfo{title}{NVIDIA Confidential Computing}.
\newblock \bibinfo{howpublished}{\url{https://www.nvidia.com/en-us/data-center/solutions/confidential-computing/}}.
\newblock


\bibitem[Pang et~al\mbox{.}(2024a)]%
        {mpcdiff}
\bibfield{author}{\bibinfo{person}{Qi Pang}, \bibinfo{person}{Yuanyuan Yuan}, {and} \bibinfo{person}{Shuai Wang}.} \bibinfo{year}{2024}\natexlab{a}.
\newblock \showarticletitle{MPCDiff: Testing and Repairing MPC-Hardened Deep Learning Models}. In \bibinfo{booktitle}{\emph{31st Annual Network and Distributed System Security Symposium, {NDSS} 2024, San Diego, California, USA, February 26 - March 1, 2024}}. \bibinfo{publisher}{The Internet Society}.
\newblock
\urldef\tempurl%
\url{https://www.ndss-symposium.org/ndss-paper/mpcdiff-testing-and-repairing-mpc-hardened-deep-learning-models/}
\showURL{%
\tempurl}


\bibitem[Pang et~al\mbox{.}(2024b)]%
        {bolt}
\bibfield{author}{\bibinfo{person}{Qi Pang}, \bibinfo{person}{Jinhao Zhu}, \bibinfo{person}{Helen M{\"{o}}llering}, \bibinfo{person}{Wenting Zheng}, {and} \bibinfo{person}{Thomas Schneider}.} \bibinfo{year}{2024}\natexlab{b}.
\newblock \showarticletitle{{BOLT:} Privacy-Preserving, Accurate and Efficient Inference for Transformers}. In \bibinfo{booktitle}{\emph{{IEEE} Symposium on Security and Privacy, {SP} 2024, San Francisco, CA, USA, May 19-23, 2024}}. \bibinfo{publisher}{{IEEE}}, \bibinfo{pages}{4753--4771}.
\newblock
\href{https://doi.org/10.1109/SP54263.2024.00130}{doi:\nolinkurl{10.1109/SP54263.2024.00130}}


\bibitem[Park et~al\mbox{.}(2022)]%
        {aespa}
\bibfield{author}{\bibinfo{person}{Jaiyoung Park}, \bibinfo{person}{Michael~Jaemin Kim}, \bibinfo{person}{Wonkyung Jung}, {and} \bibinfo{person}{Jung~Ho Ahn}.} \bibinfo{year}{2022}\natexlab{}.
\newblock \showarticletitle{{AESPA:} Accuracy Preserving Low-degree Polynomial Activation for Fast Private Inference}.
\newblock \bibinfo{journal}{\emph{CoRR}}  \bibinfo{volume}{abs/2201.06699} (\bibinfo{year}{2022}).
\newblock
\showeprint[arXiv]{2201.06699}
\urldef\tempurl%
\url{https://arxiv.org/abs/2201.06699}
\showURL{%
\tempurl}


\bibitem[Patra and Suresh(2020)]%
        {blaze}
\bibfield{author}{\bibinfo{person}{Arpita Patra} {and} \bibinfo{person}{Ajith Suresh}.} \bibinfo{year}{2020}\natexlab{}.
\newblock \showarticletitle{{BLAZE:} Blazing Fast Privacy-Preserving Machine Learning}. In \bibinfo{booktitle}{\emph{27th Annual Network and Distributed System Security Symposium, {NDSS} 2020, San Diego, California, USA, February 23-26, 2020}}. \bibinfo{publisher}{The Internet Society}.
\newblock
\urldef\tempurl%
\url{https://www.ndss-symposium.org/ndss-paper/blaze-blazing-fast-privacy-preserving-machine-learning/}
\showURL{%
\tempurl}


\bibitem[Radford et~al\mbox{.}(2019)]%
        {gpt2}
\bibfield{author}{\bibinfo{person}{Alec Radford}, \bibinfo{person}{Jeffrey Wu}, \bibinfo{person}{Rewon Child}, \bibinfo{person}{David Luan}, \bibinfo{person}{Dario Amodei}, {and} \bibinfo{person}{Ilya Sutskever}.} \bibinfo{year}{2019}\natexlab{}.
\newblock \bibinfo{title}{Language Models are Unsupervised Multitask Learners}.
\newblock \bibinfo{howpublished}{OpenAI technical report}.
\newblock


\bibitem[Rastogi et~al\mbox{.}(2014)]%
        {wysteria}
\bibfield{author}{\bibinfo{person}{Aseem Rastogi}, \bibinfo{person}{Matthew~A. Hammer}, {and} \bibinfo{person}{Michael Hicks}.} \bibinfo{year}{2014}\natexlab{}.
\newblock \showarticletitle{Wysteria: {A} Programming Language for Generic, Mixed-Mode Multiparty Computations}. In \bibinfo{booktitle}{\emph{2014 {IEEE} Symposium on Security and Privacy, {SP} 2014, Berkeley, CA, USA, May 18-21, 2014}}. \bibinfo{publisher}{{IEEE} Computer Society}, \bibinfo{pages}{655--670}.
\newblock
\href{https://doi.org/10.1109/SP.2014.48}{doi:\nolinkurl{10.1109/SP.2014.48}}


\bibitem[Rathee et~al\mbox{.}(2021)]%
        {sirnn}
\bibfield{author}{\bibinfo{person}{Deevashwer Rathee}, \bibinfo{person}{Mayank Rathee}, \bibinfo{person}{Rahul Kranti~Kiran Goli}, \bibinfo{person}{Divya Gupta}, \bibinfo{person}{Rahul Sharma}, \bibinfo{person}{Nishanth Chandran}, {and} \bibinfo{person}{Aseem Rastogi}.} \bibinfo{year}{2021}\natexlab{}.
\newblock \showarticletitle{SiRnn: {A} Math Library for Secure {RNN} Inference}. In \bibinfo{booktitle}{\emph{42nd {IEEE} Symposium on Security and Privacy, {SP} 2021, San Francisco, CA, USA, 24-27 May 2021}}. \bibinfo{publisher}{{IEEE}}, \bibinfo{pages}{1003--1020}.
\newblock
\href{https://doi.org/10.1109/SP40001.2021.00086}{doi:\nolinkurl{10.1109/SP40001.2021.00086}}


\bibitem[Rathee et~al\mbox{.}(2020)]%
        {cryptflow2}
\bibfield{author}{\bibinfo{person}{Deevashwer Rathee}, \bibinfo{person}{Mayank Rathee}, \bibinfo{person}{Nishant Kumar}, \bibinfo{person}{Nishanth Chandran}, \bibinfo{person}{Divya Gupta}, \bibinfo{person}{Aseem Rastogi}, {and} \bibinfo{person}{Rahul Sharma}.} \bibinfo{year}{2020}\natexlab{}.
\newblock \showarticletitle{CrypTFlow2: Practical 2-Party Secure Inference}. In \bibinfo{booktitle}{\emph{{CCS} '20: 2020 {ACM} {SIGSAC} Conference on Computer and Communications Security, Virtual Event, USA, November 9-13, 2020}}, \bibfield{editor}{\bibinfo{person}{Jay Ligatti}, \bibinfo{person}{Xinming Ou}, \bibinfo{person}{Jonathan Katz}, {and} \bibinfo{person}{Giovanni Vigna}} (Eds.). \bibinfo{publisher}{{ACM}}, \bibinfo{pages}{325--342}.
\newblock
\href{https://doi.org/10.1145/3372297.3417274}{doi:\nolinkurl{10.1145/3372297.3417274}}


\bibitem[Reagen et~al\mbox{.}(2021)]%
        {cheetah_fhe}
\bibfield{author}{\bibinfo{person}{Brandon Reagen}, \bibinfo{person}{Woo-Seok Choi}, \bibinfo{person}{Yeongil Ko}, \bibinfo{person}{Vincent~T. Lee}, \bibinfo{person}{Hsien-Hsin~S. Lee}, \bibinfo{person}{Gu-Yeon Wei}, {and} \bibinfo{person}{David Brooks}.} \bibinfo{year}{2021}\natexlab{}.
\newblock \showarticletitle{Cheetah: Optimizing and Accelerating Homomorphic Encryption for Private Inference}. In \bibinfo{booktitle}{\emph{2021 IEEE International Symposium on High-Performance Computer Architecture (HPCA)}}. \bibinfo{pages}{26--39}.
\newblock
\href{https://doi.org/10.1109/HPCA51647.2021.00013}{doi:\nolinkurl{10.1109/HPCA51647.2021.00013}}


\bibitem[Research(2024)]%
        {crypten-conversion-bug}
\bibfield{author}{\bibinfo{person}{Facebook Research}.} \bibinfo{year}{2024}\natexlab{}.
\newblock \bibinfo{title}{CrypTen floating point conversion issue}.
\newblock \bibinfo{howpublished}{\url{https://github.com/facebookresearch/CrypTen/blob/775868a02d6dac50774ce376a55b01fbd8bd85b6/crypten/mpc/primitives/arithmetic.py\#L362}}.
\newblock


\bibitem[Riazi et~al\mbox{.}(2020)]%
        {heax}
\bibfield{author}{\bibinfo{person}{M.~Sadegh Riazi}, \bibinfo{person}{Kim Laine}, \bibinfo{person}{Blake Pelton}, {and} \bibinfo{person}{Wei Dai}.} \bibinfo{year}{2020}\natexlab{}.
\newblock \showarticletitle{{HEAX:} An Architecture for Computing on Encrypted Data}. In \bibinfo{booktitle}{\emph{{ASPLOS} '20: Architectural Support for Programming Languages and Operating Systems, Lausanne, Switzerland, March 16-20, 2020}}, \bibfield{editor}{\bibinfo{person}{James~R. Larus}, \bibinfo{person}{Luis Ceze}, {and} \bibinfo{person}{Karin Strauss}} (Eds.). \bibinfo{publisher}{{ACM}}, \bibinfo{pages}{1295--1309}.
\newblock
\href{https://doi.org/10.1145/3373376.3378523}{doi:\nolinkurl{10.1145/3373376.3378523}}


\bibitem[Riazi et~al\mbox{.}(2019)]%
        {xonn}
\bibfield{author}{\bibinfo{person}{M.~Sadegh Riazi}, \bibinfo{person}{Mohammad Samragh}, \bibinfo{person}{Hao Chen}, \bibinfo{person}{Kim Laine}, \bibinfo{person}{Kristin~E. Lauter}, {and} \bibinfo{person}{Farinaz Koushanfar}.} \bibinfo{year}{2019}\natexlab{}.
\newblock \showarticletitle{{XONN:} XNOR-based Oblivious Deep Neural Network Inference}. In \bibinfo{booktitle}{\emph{28th {USENIX} Security Symposium, {USENIX} Security 2019, Santa Clara, CA, USA, August 14-16, 2019}}. \bibinfo{publisher}{{USENIX} Association}, \bibinfo{pages}{1501--1518}.
\newblock
\urldef\tempurl%
\url{https://www.usenix.org/conference/usenixsecurity19/presentation/riazi}
\showURL{%
\tempurl}


\bibitem[Riazi et~al\mbox{.}(2018)]%
        {charmeleon}
\bibfield{author}{\bibinfo{person}{M.~Sadegh Riazi}, \bibinfo{person}{Christian Weinert}, \bibinfo{person}{Oleksandr Tkachenko}, \bibinfo{person}{Ebrahim~M. Songhori}, \bibinfo{person}{Thomas Schneider}, {and} \bibinfo{person}{Farinaz Koushanfar}.} \bibinfo{year}{2018}\natexlab{}.
\newblock \showarticletitle{Chameleon: {A} Hybrid Secure Computation Framework for Machine Learning Applications}. In \bibinfo{booktitle}{\emph{Proceedings of the 2018 on Asia Conference on Computer and Communications Security, AsiaCCS 2018, Incheon, Republic of Korea, June 04-08, 2018}}. \bibinfo{publisher}{{ACM}}, \bibinfo{pages}{707--721}.
\newblock
\href{https://doi.org/10.1145/3196494.3196522}{doi:\nolinkurl{10.1145/3196494.3196522}}


\bibitem[Ryffel et~al\mbox{.}(2022)]%
        {ariann}
\bibfield{author}{\bibinfo{person}{Th{\'{e}}o Ryffel}, \bibinfo{person}{Pierre Tholoniat}, \bibinfo{person}{David Pointcheval}, {and} \bibinfo{person}{Francis~R. Bach}.} \bibinfo{year}{2022}\natexlab{}.
\newblock \showarticletitle{AriaNN: Low-Interaction Privacy-Preserving Deep Learning via Function Secret Sharing}.
\newblock \bibinfo{journal}{\emph{Proc. Priv. Enhancing Technol.}} \bibinfo{volume}{2022}, \bibinfo{number}{1} (\bibinfo{year}{2022}), \bibinfo{pages}{291--316}.
\newblock
\href{https://doi.org/10.2478/POPETS-2022-0015}{doi:\nolinkurl{10.2478/POPETS-2022-0015}}


\bibitem[Samardzic et~al\mbox{.}(2021)]%
        {f1}
\bibfield{author}{\bibinfo{person}{Nikola Samardzic}, \bibinfo{person}{Axel Feldmann}, \bibinfo{person}{Aleksandar Krastev}, \bibinfo{person}{Srinivas Devadas}, \bibinfo{person}{Ronald~G. Dreslinski}, \bibinfo{person}{Christopher Peikert}, {and} \bibinfo{person}{Daniel S{\'{a}}nchez}.} \bibinfo{year}{2021}\natexlab{}.
\newblock \showarticletitle{{F1:} {A} Fast and Programmable Accelerator for Fully Homomorphic Encryption}. In \bibinfo{booktitle}{\emph{{MICRO} '21: 54th Annual {IEEE/ACM} International Symposium on Microarchitecture, Virtual Event, Greece, October 18-22, 2021}}. \bibinfo{publisher}{{ACM}}, \bibinfo{pages}{238--252}.
\newblock
\href{https://doi.org/10.1145/3466752.3480070}{doi:\nolinkurl{10.1145/3466752.3480070}}


\bibitem[Samardzic et~al\mbox{.}(2022)]%
        {craterlake}
\bibfield{author}{\bibinfo{person}{Nikola Samardzic}, \bibinfo{person}{Axel Feldmann}, \bibinfo{person}{Aleksandar Krastev}, \bibinfo{person}{Nathan Manohar}, \bibinfo{person}{Nicholas Genise}, \bibinfo{person}{Srinivas Devadas}, \bibinfo{person}{Karim Eldefrawy}, \bibinfo{person}{Chris Peikert}, {and} \bibinfo{person}{Daniel Sanchez}.} \bibinfo{year}{2022}\natexlab{}.
\newblock \showarticletitle{Craterlake: a hardware accelerator for efficient unbounded computation on encrypted data}. In \bibinfo{booktitle}{\emph{Proceedings of the 49th Annual International Symposium on Computer Architecture}}. \bibinfo{pages}{173--187}.
\newblock


\bibitem[Santos et~al\mbox{.}(2024)]%
        {curl}
\bibfield{author}{\bibinfo{person}{Manuel~B Santos}, \bibinfo{person}{Dimitris Mouris}, \bibinfo{person}{Mehmet Ugurbil}, \bibinfo{person}{Stanislaw Jarecki}, \bibinfo{person}{Jos{\'e} Reis}, \bibinfo{person}{Shubho Sengupta}, {and} \bibinfo{person}{Miguel de Vega}.} \bibinfo{year}{2024}\natexlab{}.
\newblock \showarticletitle{Curl: Private LLMs through Wavelet-Encoded Look-Up Tables.}. In \bibinfo{booktitle}{\emph{CAMLIS}}. \bibinfo{pages}{16--47}.
\newblock


\bibitem[Simonyan and Zisserman(2015)]%
        {vgg}
\bibfield{author}{\bibinfo{person}{Karen Simonyan} {and} \bibinfo{person}{Andrew Zisserman}.} \bibinfo{year}{2015}\natexlab{}.
\newblock \showarticletitle{Very Deep Convolutional Networks for Large-Scale Image Recognition}. In \bibinfo{booktitle}{\emph{3rd International Conference on Learning Representations, {ICLR} 2015, San Diego, CA, USA, May 7-9, 2015, Conference Track Proceedings}}.
\newblock
\urldef\tempurl%
\url{http://arxiv.org/abs/1409.1556}
\showURL{%
\tempurl}


\bibitem[Songhori et~al\mbox{.}(2015)]%
        {tinygarble}
\bibfield{author}{\bibinfo{person}{Ebrahim~M. Songhori}, \bibinfo{person}{Siam~U. Hussain}, \bibinfo{person}{Ahmad{-}Reza Sadeghi}, \bibinfo{person}{Thomas Schneider}, {and} \bibinfo{person}{Farinaz Koushanfar}.} \bibinfo{year}{2015}\natexlab{}.
\newblock \showarticletitle{TinyGarble: Highly Compressed and Scalable Sequential Garbled Circuits}. In \bibinfo{booktitle}{\emph{2015 {IEEE} Symposium on Security and Privacy, {SP} 2015, San Jose, CA, USA, May 17-21, 2015}}. \bibinfo{publisher}{{IEEE} Computer Society}, \bibinfo{pages}{411--428}.
\newblock
\href{https://doi.org/10.1109/SP.2015.32}{doi:\nolinkurl{10.1109/SP.2015.32}}


\bibitem[Suh et~al\mbox{.}(2003)]%
        {aegis}
\bibfield{author}{\bibinfo{person}{G.~Edward Suh}, \bibinfo{person}{Dwaine~E. Clarke}, \bibinfo{person}{Blaise Gassend}, \bibinfo{person}{Marten van Dijk}, {and} \bibinfo{person}{Srinivas Devadas}.} \bibinfo{year}{2003}\natexlab{}.
\newblock \showarticletitle{{AEGIS:} architecture for tamper-evident and tamper-resistant processing}. In \bibinfo{booktitle}{\emph{Proceedings of the 17th Annual International Conference on Supercomputing, {ICS} 2003, San Francisco, CA, USA, June 23-26, 2003}}, \bibfield{editor}{\bibinfo{person}{Utpal Banerjee}, \bibinfo{person}{Kyle~A. Gallivan}, {and} \bibinfo{person}{Antonio Gonz{\'{a}}lez}} (Eds.). \bibinfo{publisher}{{ACM}}, \bibinfo{pages}{160--171}.
\newblock
\href{https://doi.org/10.1145/782814.782838}{doi:\nolinkurl{10.1145/782814.782838}}


\bibitem[Tan et~al\mbox{.}(2019)]%
        {mnasnet}
\bibfield{author}{\bibinfo{person}{Mingxing Tan}, \bibinfo{person}{Bo Chen}, \bibinfo{person}{Ruoming Pang}, \bibinfo{person}{Vijay Vasudevan}, \bibinfo{person}{Mark Sandler}, \bibinfo{person}{Andrew Howard}, {and} \bibinfo{person}{Quoc~V. Le}.} \bibinfo{year}{2019}\natexlab{}.
\newblock \showarticletitle{MnasNet: Platform-Aware Neural Architecture Search for Mobile}. In \bibinfo{booktitle}{\emph{{IEEE} Conference on Computer Vision and Pattern Recognition, {CVPR} 2019, Long Beach, CA, USA, June 16-20, 2019}}. \bibinfo{publisher}{Computer Vision Foundation / {IEEE}}, \bibinfo{pages}{2820--2828}.
\newblock
\href{https://doi.org/10.1109/CVPR.2019.00293}{doi:\nolinkurl{10.1109/CVPR.2019.00293}}


\bibitem[Tan and Le(2019)]%
        {efficientnet_b0}
\bibfield{author}{\bibinfo{person}{Mingxing Tan} {and} \bibinfo{person}{Quoc~V. Le}.} \bibinfo{year}{2019}\natexlab{}.
\newblock \showarticletitle{{EfficientNet}: Rethinking Model Scaling for Convolutional Neural Networks}. In \bibinfo{booktitle}{\emph{Proceedings of the 36th International Conference on Machine Learning (ICML)}}. \bibinfo{pages}{6105--6114}.
\newblock


\bibitem[Tan et~al\mbox{.}(2021)]%
        {cryptgpu}
\bibfield{author}{\bibinfo{person}{Sijun Tan}, \bibinfo{person}{Brian Knott}, \bibinfo{person}{Yuan Tian}, {and} \bibinfo{person}{David~J. Wu}.} \bibinfo{year}{2021}\natexlab{}.
\newblock \showarticletitle{CryptGPU: Fast Privacy-Preserving Machine Learning on the {GPU}}. In \bibinfo{booktitle}{\emph{42nd {IEEE} Symposium on Security and Privacy, {SP} 2021, San Francisco, CA, USA, 24-27 May 2021}}. \bibinfo{publisher}{{IEEE}}, \bibinfo{pages}{1021--1038}.
\newblock
\href{https://doi.org/10.1109/SP40001.2021.00098}{doi:\nolinkurl{10.1109/SP40001.2021.00098}}


\bibitem[team(2024a)]%
        {export_ir_pass_tutorial}
\bibfield{author}{\bibinfo{person}{PyTorch team}.} \bibinfo{year}{2024}\natexlab{a}.
\newblock \bibinfo{title}{Custom Compiler Passes and Partitioners}.
\newblock \bibinfo{howpublished}{\url{https://pytorch.org/executorch/stable/compiler-custom-compiler-passes.html}}.
\newblock


\bibitem[team(2024b)]%
        {export_ir}
\bibfield{author}{\bibinfo{person}{PyTorch team}.} \bibinfo{year}{2024}\natexlab{b}.
\newblock \bibinfo{title}{torch.export IR Specification}.
\newblock \bibinfo{howpublished}{\url{https://pytorch.org/docs/stable/export.ir_spec.html}}.
\newblock


\bibitem[Tong et~al\mbox{.}(2024)]%
        {smart_paf}
\bibfield{author}{\bibinfo{person}{Jianming Tong}, \bibinfo{person}{Jingtian Dang}, \bibinfo{person}{Anupam Golder}, \bibinfo{person}{Callie Hao}, \bibinfo{person}{Arijit Raychowdhury}, {and} \bibinfo{person}{Tushar Krishna}.} \bibinfo{year}{2024}\natexlab{}.
\newblock \showarticletitle{Accurate Low-Degree Polynomial Approximation of Non-polynomial Operators for Fast Private Inference in Homomorphic Encryption}. In \bibinfo{booktitle}{\emph{Proceedings of Machine Learning and Systems (MLSys)}}.
\newblock
\urldef\tempurl%
\url{https://arxiv.org/abs/2404.03216}
\showURL{%
\tempurl}


\bibitem[Turner et~al\mbox{.}(2021)]%
        {oboyle_nas}
\bibfield{author}{\bibinfo{person}{Jack Turner}, \bibinfo{person}{Elliot~J. Crowley}, {and} \bibinfo{person}{Michael F.~P. O'Boyle}.} \bibinfo{year}{2021}\natexlab{}.
\newblock \showarticletitle{Neural architecture search as program transformation exploration}. In \bibinfo{booktitle}{\emph{{ASPLOS} '21: 26th {ACM} International Conference on Architectural Support for Programming Languages and Operating Systems, Virtual Event, USA, April 19-23, 2021}}. \bibinfo{publisher}{{ACM}}, \bibinfo{pages}{915--927}.
\newblock
\href{https://doi.org/10.1145/3445814.3446753}{doi:\nolinkurl{10.1145/3445814.3446753}}


\bibitem[Vaswani et~al\mbox{.}(2023)]%
        {graphcore_tee}
\bibfield{author}{\bibinfo{person}{Kapil Vaswani}, \bibinfo{person}{Stavros Volos}, \bibinfo{person}{Cedric Fournet}, \bibinfo{person}{Antonio~Nino Diaz}, \bibinfo{person}{Ken Gordon}, \bibinfo{person}{Balaji Vembu}, \bibinfo{person}{Sam Webster}, \bibinfo{person}{David Chisnall}, \bibinfo{person}{Saurabh Kulkarni}, \bibinfo{person}{Graham Cunningham}, \bibinfo{person}{Richard Osborne}, {and} \bibinfo{person}{Daniel Wilkinson}.} \bibinfo{year}{2023}\natexlab{}.
\newblock \showarticletitle{Confidential Computing within an {AI} Accelerator}. In \bibinfo{booktitle}{\emph{2023 USENIX Annual Technical Conference (USENIX ATC 23)}}. \bibinfo{publisher}{USENIX Association}, \bibinfo{address}{Boston, MA}, \bibinfo{pages}{501--518}.
\newblock
\showISBNx{978-1-939133-35-9}
\urldef\tempurl%
\url{https://www.usenix.org/conference/atc23/presentation/vaswani}
\showURL{%
\tempurl}


\bibitem[Wagh et~al\mbox{.}(2019)]%
        {securenn}
\bibfield{author}{\bibinfo{person}{Sameer Wagh}, \bibinfo{person}{Divya Gupta}, {and} \bibinfo{person}{Nishanth Chandran}.} \bibinfo{year}{2019}\natexlab{}.
\newblock \showarticletitle{SecureNN: 3-Party Secure Computation for Neural Network Training}.
\newblock \bibinfo{journal}{\emph{Proc. Priv. Enhancing Technol.}} \bibinfo{volume}{2019}, \bibinfo{number}{3} (\bibinfo{year}{2019}), \bibinfo{pages}{26--49}.
\newblock
\href{https://doi.org/10.2478/popets-2019-0035}{doi:\nolinkurl{10.2478/popets-2019-0035}}


\bibitem[Wagh et~al\mbox{.}(2021)]%
        {falcon}
\bibfield{author}{\bibinfo{person}{Sameer Wagh}, \bibinfo{person}{Shruti Tople}, \bibinfo{person}{Fabrice Benhamouda}, \bibinfo{person}{Eyal Kushilevitz}, \bibinfo{person}{Prateek Mittal}, {and} \bibinfo{person}{Tal Rabin}.} \bibinfo{year}{2021}\natexlab{}.
\newblock \showarticletitle{Falcon: Honest-Majority Maliciously Secure Framework for Private Deep Learning}.
\newblock \bibinfo{journal}{\emph{Proc. Priv. Enhancing Technol.}} \bibinfo{volume}{2021}, \bibinfo{number}{1} (\bibinfo{year}{2021}), \bibinfo{pages}{188--208}.
\newblock
\href{https://doi.org/10.2478/popets-2021-0011}{doi:\nolinkurl{10.2478/popets-2021-0011}}


\bibitem[Wang et~al\mbox{.}(2019)]%
        {glue_sst2}
\bibfield{author}{\bibinfo{person}{Alex Wang}, \bibinfo{person}{Amanpreet Singh}, \bibinfo{person}{Julian Michael}, \bibinfo{person}{Felix Hill}, \bibinfo{person}{Omer Levy}, {and} \bibinfo{person}{Samuel~R. Bowman}.} \bibinfo{year}{2019}\natexlab{}.
\newblock \showarticletitle{{GLUE:} {A} Multi-Task Benchmark and Analysis Platform for Natural Language Understanding}. In \bibinfo{booktitle}{\emph{7th International Conference on Learning Representations, {ICLR} 2019, New Orleans, LA, USA, May 6-9, 2019}}. \bibinfo{publisher}{OpenReview.net}.
\newblock
\urldef\tempurl%
\url{https://openreview.net/forum?id=rJ4km2R5t7}
\showURL{%
\tempurl}


\bibitem[Wang et~al\mbox{.}(2017)]%
        {tee_sidechannel}
\bibfield{author}{\bibinfo{person}{Wenhao Wang}, \bibinfo{person}{Guoxing Chen}, \bibinfo{person}{Xiaorui Pan}, \bibinfo{person}{Yinqian Zhang}, \bibinfo{person}{XiaoFeng Wang}, \bibinfo{person}{Vincent Bindschaedler}, \bibinfo{person}{Haixu Tang}, {and} \bibinfo{person}{Carl~A. Gunter}.} \bibinfo{year}{2017}\natexlab{}.
\newblock \showarticletitle{Leaky Cauldron on the Dark Land: Understanding Memory Side-Channel Hazards in {SGX}}. In \bibinfo{booktitle}{\emph{Proceedings of the 2017 {ACM} {SIGSAC} Conference on Computer and Communications Security, {CCS} 2017, Dallas, TX, USA, October 30 - November 03, 2017}}, \bibfield{editor}{\bibinfo{person}{Bhavani Thuraisingham}, \bibinfo{person}{David Evans}, \bibinfo{person}{Tal Malkin}, {and} \bibinfo{person}{Dongyan Xu}} (Eds.). \bibinfo{publisher}{{ACM}}, \bibinfo{pages}{2421--2434}.
\newblock
\href{https://doi.org/10.1145/3133956.3134038}{doi:\nolinkurl{10.1145/3133956.3134038}}


\bibitem[Wang et~al\mbox{.}(2016)]%
        {emptoolkit}
\bibfield{author}{\bibinfo{person}{Xiao Wang}, \bibinfo{person}{Alex~J. Malozemoff}, {and} \bibinfo{person}{Jonathan Katz}.} \bibinfo{year}{2016}\natexlab{}.
\newblock \bibinfo{title}{{EMP-toolkit: Efficient MultiParty computation toolkit}}.
\newblock \bibinfo{howpublished}{\url{https://github.com/emp-toolkit}}.
\newblock


\bibitem[Wang et~al\mbox{.}(2024)]%
        {mpcpipe}
\bibfield{author}{\bibinfo{person}{Yongqin Wang}, \bibinfo{person}{Rachit Rajat}, {and} \bibinfo{person}{Murali Annavaram}.} \bibinfo{year}{2024}\natexlab{}.
\newblock \showarticletitle{Mpc-pipe: An efficient pipeline scheme for semi-honest mpc machine learning}. In \bibinfo{booktitle}{\emph{Proceedings of the 29th ACM International Conference on Architectural Support for Programming Languages and Operating Systems, Volume 4}}. \bibinfo{pages}{203--219}.
\newblock


\bibitem[Wang et~al\mbox{.}(2022a)]%
        {mpc_transformer_workshop}
\bibfield{author}{\bibinfo{person}{Yongqin Wang}, \bibinfo{person}{G.~Edward Suh}, \bibinfo{person}{Wenjie Xiong}, \bibinfo{person}{Benjamin Lefaudeux}, \bibinfo{person}{Brian Knott}, \bibinfo{person}{Murali Annavaram}, {and} \bibinfo{person}{Hsien-Hsin~S. Lee}.} \bibinfo{year}{2022}\natexlab{a}.
\newblock \showarticletitle{Characterization of MPC-based Private Inference for Transformer-based Models}. In \bibinfo{booktitle}{\emph{2022 IEEE International Symposium on Performance Analysis of Systems and Software (ISPASS)}}. \bibinfo{pages}{187--197}.
\newblock
\href{https://doi.org/10.1109/ISPASS55109.2022.00025}{doi:\nolinkurl{10.1109/ISPASS55109.2022.00025}}


\bibitem[Wang et~al\mbox{.}(2022b)]%
        {mpc_transformer}
\bibfield{author}{\bibinfo{person}{Yongqin Wang}, \bibinfo{person}{G.~Edward Suh}, \bibinfo{person}{Wenjie Xiong}, \bibinfo{person}{Benjamin Lefaudeux}, \bibinfo{person}{Brian Knott}, \bibinfo{person}{Murali Annavaram}, {and} \bibinfo{person}{Hsien{-}Hsin~S. Lee}.} \bibinfo{year}{2022}\natexlab{b}.
\newblock \showarticletitle{Characterization of MPC-based Private Inference for Transformer-based Models}. In \bibinfo{booktitle}{\emph{International {IEEE} Symposium on Performance Analysis of Systems and Software, {ISPASS} 2022, Singapore, May 22-24, 2022}}. \bibinfo{publisher}{{IEEE}}, \bibinfo{pages}{187--197}.
\newblock
\href{https://doi.org/10.1109/ISPASS55109.2022.00025}{doi:\nolinkurl{10.1109/ISPASS55109.2022.00025}}


\bibitem[Watson et~al\mbox{.}(2022)]%
        {piranha}
\bibfield{author}{\bibinfo{person}{Jean-Luc Watson}, \bibinfo{person}{Sameer Wagh}, {and} \bibinfo{person}{Raluca~Ada Popa}.} \bibinfo{year}{2022}\natexlab{}.
\newblock \showarticletitle{Piranha: A {GPU} Platform for Secure Computation}. In \bibinfo{booktitle}{\emph{31st USENIX Security Symposium (USENIX Security 22)}}. \bibinfo{publisher}{USENIX Association}, \bibinfo{address}{Boston, MA}, \bibinfo{pages}{827--844}.
\newblock
\showISBNx{978-1-939133-31-1}
\urldef\tempurl%
\url{https://www.usenix.org/conference/usenixsecurity22/presentation/watson}
\showURL{%
\tempurl}


\bibitem[Wu et~al\mbox{.}(2019)]%
        {fbnet}
\bibfield{author}{\bibinfo{person}{Bichen Wu}, \bibinfo{person}{Xiaoliang Dai}, \bibinfo{person}{Peizhao Zhang}, \bibinfo{person}{Yanghan Wang}, \bibinfo{person}{Fei Sun}, \bibinfo{person}{Yiming Wu}, \bibinfo{person}{Yuandong Tian}, \bibinfo{person}{Peter Vajda}, \bibinfo{person}{Yangqing Jia}, {and} \bibinfo{person}{Kurt Keutzer}.} \bibinfo{year}{2019}\natexlab{}.
\newblock \showarticletitle{FBNet: Hardware-Aware Efficient ConvNet Design via Differentiable Neural Architecture Search}. In \bibinfo{booktitle}{\emph{{IEEE} Conference on Computer Vision and Pattern Recognition, {CVPR} 2019, Long Beach, CA, USA, June 16-20, 2019}}. \bibinfo{publisher}{Computer Vision Foundation / {IEEE}}, \bibinfo{pages}{10734--10742}.
\newblock
\href{https://doi.org/10.1109/CVPR.2019.01099}{doi:\nolinkurl{10.1109/CVPR.2019.01099}}


\bibitem[Wu et~al\mbox{.}(2024)]%
        {ditto}
\bibfield{author}{\bibinfo{person}{Haoqi Wu}, \bibinfo{person}{Wenjing Fang}, \bibinfo{person}{Yancheng Zheng}, \bibinfo{person}{Junming Ma}, \bibinfo{person}{Jin Tan}, {and} \bibinfo{person}{Lei Wang}.} \bibinfo{year}{2024}\natexlab{}.
\newblock \showarticletitle{Ditto: quantization-aware secure inference of transformers upon MPC}. In \bibinfo{booktitle}{\emph{Proceedings of the 41st International Conference on Machine Learning}} (Vienna, Austria) \emph{(\bibinfo{series}{ICML'24})}. \bibinfo{publisher}{JMLR.org}, Article \bibinfo{articleno}{2186}, \bibinfo{numpages}{20}~pages.
\newblock


\bibitem[Xiao and Devadas(2022)]%
        {pac_privacy}
\bibfield{author}{\bibinfo{person}{Hanshen Xiao} {and} \bibinfo{person}{Srinivas Devadas}.} \bibinfo{year}{2022}\natexlab{}.
\newblock \showarticletitle{PAC Security: Automatic Privacy Measurement and Control of Data Processing}.
\newblock \bibinfo{journal}{\emph{arXiv preprint arXiv:2210.03458}} (\bibinfo{year}{2022}).
\newblock


\bibitem[Xiao et~al\mbox{.}(2024)]%
        {pac_privacy_followup}
\bibfield{author}{\bibinfo{person}{Hanshen Xiao}, \bibinfo{person}{G.~Edward Suh}, {and} \bibinfo{person}{Srinivas Devadas}.} \bibinfo{year}{2024}\natexlab{}.
\newblock \showarticletitle{Formal Privacy Proof of Data Encoding: The Possibility and Impossibility of Learnable Encryption}. In \bibinfo{booktitle}{\emph{Proceedings of the 2024 on {ACM} {SIGSAC} Conference on Computer and Communications Security, {CCS} 2024, Salt Lake City, UT, USA, October 14-18, 2024}}. \bibinfo{publisher}{{ACM}}, \bibinfo{pages}{1834--1848}.
\newblock
\href{https://doi.org/10.1145/3658644.3670277}{doi:\nolinkurl{10.1145/3658644.3670277}}


\bibitem[Yates(2009)]%
        {fixed_point_book}
\bibfield{author}{\bibinfo{person}{Randy Yates}.} \bibinfo{year}{2009}\natexlab{}.
\newblock \showarticletitle{Fixed-point arithmetic: An introduction}.
\newblock \bibinfo{journal}{\emph{Digital Signal Labs}} \bibinfo{volume}{81}, \bibinfo{number}{83} (\bibinfo{year}{2009}), \bibinfo{pages}{198}.
\newblock


\bibitem[Yu et~al\mbox{.}(2024)]%
        {yu2024privacy}
\bibfield{author}{\bibinfo{person}{Da Yu}, \bibinfo{person}{Peter Kairouz}, \bibinfo{person}{Sewoong Oh}, {and} \bibinfo{person}{Zheng Xu}.} \bibinfo{year}{2024}\natexlab{}.
\newblock \showarticletitle{Privacy-preserving instructions for aligning large language models}.
\newblock \bibinfo{journal}{\emph{arXiv preprint arXiv:2402.13659}} (\bibinfo{year}{2024}).
\newblock


\bibitem[Yu et~al\mbox{.}(2020)]%
        {bignasnet}
\bibfield{author}{\bibinfo{person}{Jiahui Yu}, \bibinfo{person}{Pengchong Jin}, \bibinfo{person}{Hanxiao Liu}, \bibinfo{person}{Gabriel Bender}, \bibinfo{person}{Pieter{-}Jan Kindermans}, \bibinfo{person}{Mingxing Tan}, \bibinfo{person}{Thomas~S. Huang}, \bibinfo{person}{Xiaodan Song}, \bibinfo{person}{Ruoming Pang}, {and} \bibinfo{person}{Quoc Le}.} \bibinfo{year}{2020}\natexlab{}.
\newblock \showarticletitle{BigNAS: Scaling up Neural Architecture Search with Big Single-Stage Models}. In \bibinfo{booktitle}{\emph{Computer Vision - {ECCV} 2020 - 16th European Conference, Glasgow, UK, August 23-28, 2020, Proceedings, Part {VII}}} \emph{(\bibinfo{series}{Lecture Notes in Computer Science}, Vol.~\bibinfo{volume}{12352})}, \bibfield{editor}{\bibinfo{person}{Andrea Vedaldi}, \bibinfo{person}{Horst Bischof}, \bibinfo{person}{Thomas Brox}, {and} \bibinfo{person}{Jan{-}Michael Frahm}} (Eds.). \bibinfo{publisher}{Springer}, \bibinfo{pages}{702--717}.
\newblock
\href{https://doi.org/10.1007/978-3-030-58571-6\_41}{doi:\nolinkurl{10.1007/978-3-030-58571-6\_41}}


\bibitem[Yubeaton et~al\mbox{.}(2024)]%
        {truncformer}
\bibfield{author}{\bibinfo{person}{Patrick Yubeaton}, \bibinfo{person}{Jianqiao Mo}, \bibinfo{person}{Karthik Garimella}, \bibinfo{person}{Nandan~Kumar Jha}, \bibinfo{person}{Brandon Reagen}, \bibinfo{person}{Chinmay Hegde}, {and} \bibinfo{person}{Siddharth Garg}.} \bibinfo{year}{2024}\natexlab{}.
\newblock \showarticletitle{TruncFormer: Private {LLM} Inference Using Only Truncations}.
\newblock \bibinfo{journal}{\emph{CoRR}}  \bibinfo{volume}{abs/2412.01042} (\bibinfo{year}{2024}).
\newblock
\showeprint[arXiv]{2412.01042}
\href{https://doi.org/10.48550/ARXIV.2412.01042}{doi:\nolinkurl{10.48550/ARXIV.2412.01042}}


\bibitem[Zahur and Evans(2015)]%
        {oblivc}
\bibfield{author}{\bibinfo{person}{Samee Zahur} {and} \bibinfo{person}{David Evans}.} \bibinfo{year}{2015}\natexlab{}.
\newblock \showarticletitle{Obliv-C: {A} Language for Extensible Data-Oblivious Computation}.
\newblock \bibinfo{journal}{\emph{{IACR} Cryptol. ePrint Arch.}} (\bibinfo{year}{2015}), \bibinfo{pages}{1153}.
\newblock
\urldef\tempurl%
\url{http://eprint.iacr.org/2015/1153}
\showURL{%
\tempurl}


\bibitem[Zeng et~al\mbox{.}(2023)]%
        {mpcvit}
\bibfield{author}{\bibinfo{person}{Wenxuan Zeng}, \bibinfo{person}{Meng Li}, \bibinfo{person}{Wenjie Xiong}, \bibinfo{person}{Tong Tong}, \bibinfo{person}{Wen-jie Lu}, \bibinfo{person}{Jin Tan}, \bibinfo{person}{Runsheng Wang}, {and} \bibinfo{person}{Ru Huang}.} \bibinfo{year}{2023}\natexlab{}.
\newblock \showarticletitle{MPCViT: Searching for Accurate and Efficient MPC-Friendly Vision Transformer with Heterogeneous Attention}. In \bibinfo{booktitle}{\emph{Proceedings of the IEEE/CVF International Conference on Computer Vision}}. \bibinfo{pages}{5052--5063}.
\newblock


\bibitem[Zhang et~al\mbox{.}(2026)]%
        {cipherflow}
\bibfield{author}{\bibinfo{person}{Hedong Zhang}, \bibinfo{person}{Mengxin Zheng}, {and} \bibinfo{person}{Qian Lou}.} \bibinfo{year}{2026}\natexlab{}.
\newblock \showarticletitle{{CipherFlow}: Compiling {AI} Programs for Efficient Cryptographically Secure Inference}. In \bibinfo{booktitle}{\emph{Advances in Neural Information Processing Systems (NeurIPS)}}.
\newblock
\urldef\tempurl%
\url{https://loopleap-lab.github.io/CipherFlow/assets/cipherflow-paper.pdf}
\showURL{%
\tempurl}


\bibitem[Zhang et~al\mbox{.}(2025)]%
        {nexus}
\bibfield{author}{\bibinfo{person}{Jiawen Zhang}, \bibinfo{person}{Xinpeng Yang}, \bibinfo{person}{Lipeng He}, \bibinfo{person}{Kejia Chen}, \bibinfo{person}{Wen{-}jie Lu}, \bibinfo{person}{Yinghao Wang}, \bibinfo{person}{Xiaoyang Hou}, \bibinfo{person}{Jian Liu}, \bibinfo{person}{Kui Ren}, {and} \bibinfo{person}{Xiaohu Yang}.} \bibinfo{year}{2025}\natexlab{}.
\newblock \showarticletitle{Secure Transformer Inference Made Non-interactive}. In \bibinfo{booktitle}{\emph{32nd Annual Network and Distributed System Security Symposium, {NDSS} 2025, San Diego, California, USA, February 24-28, 2025}}. \bibinfo{publisher}{The Internet Society}.
\newblock
\urldef\tempurl%
\url{https://www.ndss-symposium.org/ndss-paper/secure-transformer-inference-made-non-interactive/}
\showURL{%
\tempurl}


\bibitem[Zhang et~al\mbox{.}(2023)]%
        {salvit}
\bibfield{author}{\bibinfo{person}{Yuke Zhang}, \bibinfo{person}{Dake Chen}, \bibinfo{person}{Souvik Kundu}, \bibinfo{person}{Chenghao Li}, {and} \bibinfo{person}{Peter~A Beerel}.} \bibinfo{year}{2023}\natexlab{}.
\newblock \showarticletitle{SAL-ViT: Towards Latency Efficient Private Inference on ViT using Selective Attention Search with a Learnable Softmax Approximation}. In \bibinfo{booktitle}{\emph{Proceedings of the IEEE/CVF International Conference on Computer Vision}}. \bibinfo{pages}{5116--5125}.
\newblock


\bibitem[Zhang et~al\mbox{.}(2013)]%
        {picco}
\bibfield{author}{\bibinfo{person}{Yihua Zhang}, \bibinfo{person}{Aaron Steele}, {and} \bibinfo{person}{Marina Blanton}.} \bibinfo{year}{2013}\natexlab{}.
\newblock \showarticletitle{{PICCO:} a general-purpose compiler for private distributed computation}. In \bibinfo{booktitle}{\emph{2013 {ACM} {SIGSAC} Conference on Computer and Communications Security, CCS'13, Berlin, Germany, November 4-8, 2013}}. \bibinfo{publisher}{{ACM}}, \bibinfo{pages}{813--826}.
\newblock
\href{https://doi.org/10.1145/2508859.2516752}{doi:\nolinkurl{10.1145/2508859.2516752}}


\bibitem[Zhou et~al\mbox{.}(2022)]%
        {ppmlac}
\bibfield{author}{\bibinfo{person}{Xing Zhou}, \bibinfo{person}{Zhilei Xu}, \bibinfo{person}{Cong Wang}, {and} \bibinfo{person}{Mingyu Gao}.} \bibinfo{year}{2022}\natexlab{}.
\newblock \showarticletitle{{PPMLAC:} high performance chipset architecture for secure multi-party computation}. In \bibinfo{booktitle}{\emph{{ISCA} '22: The 49th Annual International Symposium on Computer Architecture, New York, New York, USA, June 18 - 22, 2022}}. \bibinfo{publisher}{{ACM}}, \bibinfo{pages}{87--101}.
\newblock
\href{https://doi.org/10.1145/3470496.3527392}{doi:\nolinkurl{10.1145/3470496.3527392}}


\end{thebibliography}

\end{document}